\documentclass[a4paper,11pt]{article}
\pdfoutput=1
\usepackage{jheppub}

\usepackage[percent]{overpic}
\usepackage{natbib}
\usepackage{graphicx}
\usepackage{epsfig}
\usepackage{comment}
\usepackage{slashed}
\usepackage{bm}
\usepackage{amsmath}
\input{epsf}
\usepackage{psfrag}

\usepackage[usenames,dvipsnames]{color}

\newcommand{\beq}{\begin{eqnarray}}
\newcommand{\eeq}{\end{eqnarray}}
\newcommand{\Slash}[1]{{\ooalign{\hfil/\hfil\crcr$#1$}}}

\usepackage{amsmath,color,amsfonts}
\usepackage{amssymb,dsfont}
\usepackage{bm}% bold math
\usepackage{graphicx}
\usepackage{multirow}

\title{\boldmath Spin entanglement in $pp$ and $pA$ collisions: high-energy TMD effects and fragmentation probes}

\author[a,b]{Michael Fucilla,}

\affiliation[a]{National Centre for Nuclear Research, Pasteura 7, Warsaw 02-093, Poland}

\affiliation[b]{Institute of Nuclear Physics, Polish Academy of Sciences, Radzikowskiego 152, 31-342 Cracow, Poland}

\emailAdd{michael.fucilla@ncbj.gov.pl}
\emailAdd{michael.fucilla@ifj.edu.pl}

\abstract{Spin-spin correlations of a quark-antiquark pair produced in proton-proton or proton-Ion collisions are investigated in $k_T$-factorization within the shockwave formalism. We construct the full spin-density matrix of both inclusive and diffractive channels in the limit in which the two quarks are back-to-back. In the inclusive channel, due to linearly polarized gluons, we show that a $\cos (2 \phi)$-modulation of the concurrence, stabilizer R\'enyi entropy, and quantum discord emerges and depends on the interplay between six small-$x$ gluon TMDs. The modulation is predicted to be maximally strong and independent of the transverse momentum imbalance, ${\bm q}$, from a dilute (BFKL) perspective, while it is considerably suppressed by saturation effects. In diffraction, we observe that the pair is always entangled and exhibits the same properties found in diffractive DIS. We explain how these properties follow directly from the exclusive and eikonal nature of the Pomeron exchange. We trace this behavior to a spin-density matrix of rank at most two with maximally mixed marginals. For this class of states, entanglement and Bell nonlocality coincide even when the state is mixed, while the explicit form of the light-front helicity amplitudes ensures that both persist in the whole phase space. We suggest accessing the spin correlations in unpolarized hadron collisions through the Collins and dihadron fragmentation functions. We point out that, in inclusive production, for massless quarks, suitably chosen double-Collins asymmetries provide promising entanglement witnesses.}

\begin{document} 
\maketitle
\flushbottom

\section{Introduction}

Entanglement is a defining feature of quantum mechanics, and QCD offers many settings in which to study it. It has been investigated in hadronic wave functions and parton distributions~\cite{Kovner:2015hga,Kharzeev:2017qzs,Hentschinski:2023izh,Dumitru:2025bib,Golec-Biernat:2025hwa,Hentschinski:2026otq}, in high-energy scattering and evolution~\cite{Peschanski:2016hgk,Liu:2018gae,Ramos:2020kyc,Ouchen:2025tta,Hentschinski:2025pyq}, through partonic spin--orbit correlations~\cite{Bhattacharya:2024sno,Hatta:2024lbw,Agrawal:2025yoe}, and in observables proposed for hadronic and ultraperipheral collisions~\cite{Guo:2024jch,Brandenburg:2024ksp}. Among these settings, the spin correlations of particles produced in a collision offer a clean route to experimentally accessible quantum-information observables. Quantum information provides a framework for interpreting the spin correlations generated in high-energy collisions. The spin-density matrix of a particle pair encodes its polarizations and spin correlations, and allows one to investigate properties such as entanglement, nonstabilizerness, and quantum discord. At a collider, this state is inferred statistically from correlations among the particles produced in decays or fragmentation. Studies of $t\bar t$ production established how this approach could be used for quantum-state tomography and entanglement measurements at the Large Hadron Collider (LHC)~\cite{Afik:2020onf,Afik:2022kwm,Fabbrichesi:2021npl,Severi:2021cnj}\footnote{Theoretical studies have also begun to examine how radiation reduces fermion-pair entanglement, an important question for precision collider measurements~\cite{Aoude:2025ovu}.}. The top quark is particularly well suited to this program because it decays before hadronization, transferring its spin information to its decay products. ATLAS and CMS have observed spin entanglement in $t\bar t$ pairs near production threshold \cite{ATLAS:2023fsd,CMS:2024pts}. Subsequent measurements have extended the study to other kinematic regions and to a more detailed characterization of the spin-density matrix \cite{CMS:2024zkc,CMS:2025brx}. Analyses of these data have also addressed quantum properties beyond entanglement, including discord and nonstabilizerness~\cite{Afik:2026pxv}. More recently, ATLAS has reported strong evidence for entanglement between the two $Z$ bosons produced in $H\to ZZ^*$ decays~\cite{ATLAS:2026hye}. \\

In this work, we extend the collider study of entanglement in (two-qubit) quark-antiquark pair production in hadron-hadron collisions along two complementary directions. First, we investigate the sensitivity of quantum-information observables to the high-energy Transverse Momentum Dependent (TMD) QCD dynamics and gluon saturation. So far, these studies have been conducted only in the context of DIS, for future EIC measurements~\cite{Fucilla:2025kit,Hatta:2025obw,Fucilla:2026mkg,Fucilla:2026zhy}. We demonstrate that in proton-nucleus collisions there exists a pronounced sensitivity to linearly polarized gluons, which is reflected in a $\cos (2 \phi)$ modulation of concurrence, nonstabilizerness, and quantum discord. We show that a description of such modulation which neglects the effects of gluon saturation can lead to results that are in complete contradiction with the results obtained when saturation is taken into account. Therefore, quantum information quantities represent a potential new avenue for probing gluon saturation through quantum spin correlations in $pA$ collisions. We also discuss the case of diffractive events, which—though experimentally more challenging—exhibit the surprising quantum information properties already discovered in diffractive DIS~\cite{Fucilla:2025kit}, with saturation effects being the only factor that breaks the perfect symmetry between the two channels. We argue that these properties derive unequivocally from the combination of the exclusive and eikonal nature of the Pomeron interaction. We show how these latter constrain the spin-density matrix.

As a second, complementary direction, we propose to study the entanglement of quarks lighter than the top in hadronic collisions through spin-dependent fragmentation functions, building on the strategy developed for lepton--hadron collisions in Ref.~\cite{Fucilla:2026zhy}. Unlike the top quark, lighter quarks hadronize before decaying; decay-based spin measurements through heavy baryons are possible for charm and bottom~\cite{Kats:2023zxb}, but experimentally challenging. We instead use Collins and dihadron fragmentation functions to relate quark spin correlations to angular correlations among unpolarized hadrons in the final state. Double-Collins correlations in unpolarized hadronic collisions have been studied previously~\cite{Kang:2010vd}. Here we show that, for massless quarks, the spin coefficient of an appropriately selected double-Collins asymmetry equals the concurrence of the produced pair at small-$x$. This provides a way to probe light-quark spin entanglement in $pp$ ($pA$) collisions through unpolarized hadrons in the final state. \\

The paper is organized as follows. In Section II, we present the formulas needed to evaluate the concurrence, nonstabilizerness, and quantum discord. In Section III, we review the calculation of the $g \to Q\bar{Q}$ impact factor within the shockwave formalism. Sections IV and V are devoted to inclusive and diffractive $pA$ collisions, respectively. In Section VI, we discuss how spin correlations can be accessed through Collins and dihadron fragmentation functions. Finally, Section VII presents our conclusions and outlook. The Appendices provide additional details of the calculations, the operator definitions of the gluon TMDs relevant to our analysis, and their parametrization within the McLerran–Venugopalan (MV) model~\cite{McLerran:1993ni,McLerran:1993ka}.

\section{The spin density matrix and quantum information quantities}
\label{Sec:SpinDensityMatrix}
In this section, we present general formulas for computing the coefficients of the quark--antiquark spin-density matrix and the associated quantum information quantities, including entanglement, nonstabilizerness, and quantum discord. We collect the known expressions needed for the present analysis without proofs; detailed derivations with a notation closer to ours can be found in Refs.~\cite{Fucilla:2025kit,Fucilla:2026mkg,Fucilla:2026zhy}.

\begin{figure}[t]
    \centering
    \includegraphics[width=0.85\linewidth]{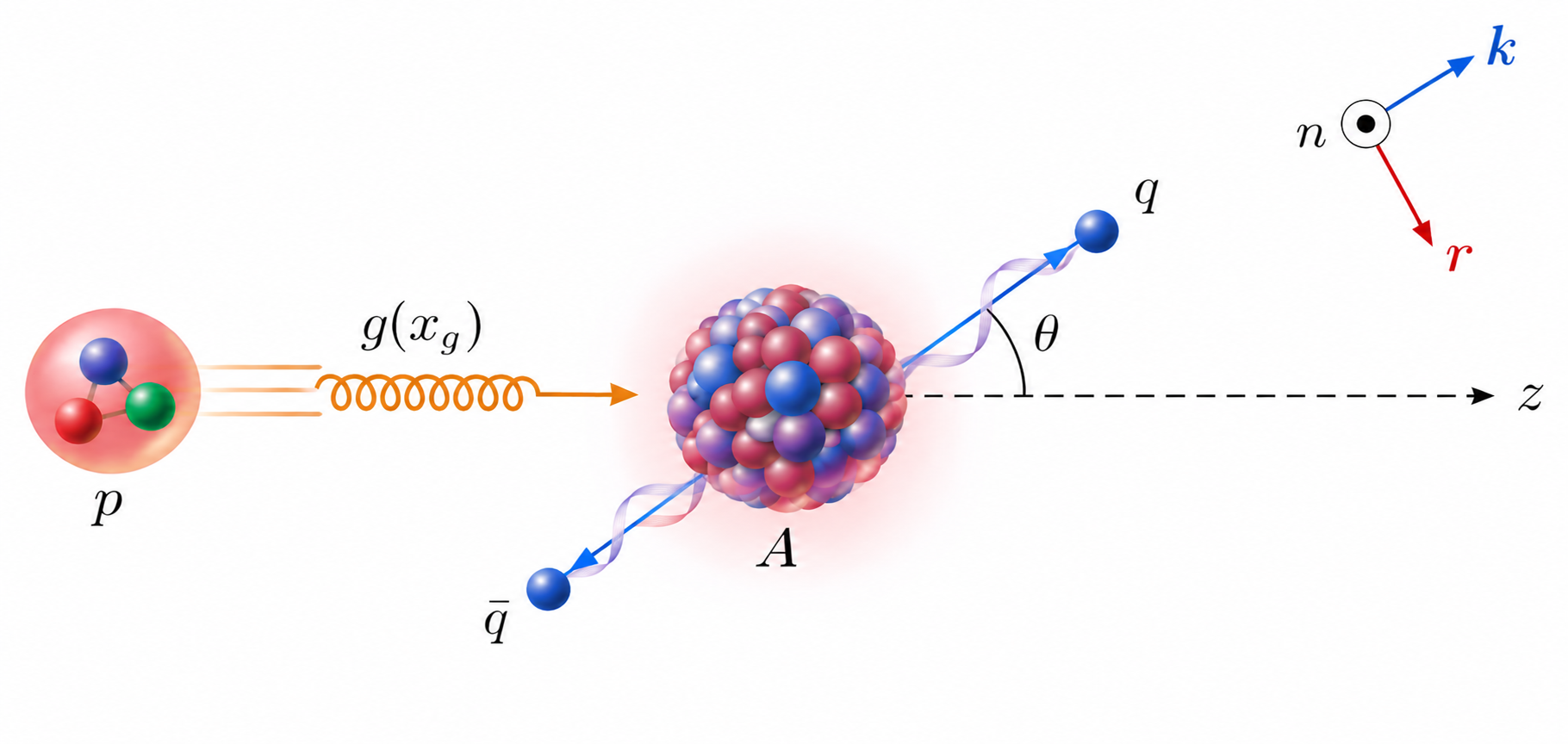}
    \caption{Schematic illustration of heavy quark–antiquark pair production in proton–nucleus collisions, initiated by a gluon carrying a longitudinal momentum fraction $x_g$ of the incoming proton. The collision axis defines the $z$ direction, while $\theta$ denotes the polar angle of the outgoing quark. The orthonormal basis $(\boldsymbol{k},\boldsymbol{r},\boldsymbol{n})$ is defined in the $q\bar q$ center-of-mass frame, with $\boldsymbol{k}$ aligned with the quark momentum, $\boldsymbol{r}$ lying in the scattering plane, and $\boldsymbol{n}$ normal to it.}
    \label{fig:Scattering_pACollision}
\end{figure}

\subsection{The spin-density matrix}

We describe the spin state of the quark--antiquark pair in its center-of-mass frame.\footnote{The production amplitudes in the shockwave formalism are evaluated in a different frame, the shockwave frame. In the back-to-back approximation adopted here, the pair center-of-mass frame is reached by a longitudinal boost along the collision axis, as discussed in Ref.~\cite{Fucilla:2025kit}.}
The scattering geometry of the considered process and the orthonormal spin basis $(\boldsymbol{k},\boldsymbol{r},\boldsymbol{n})$ are illustrated in Fig.~\ref{fig:Scattering_pACollision}, where $\theta$ is the angle
between the quark momentum and the collision axis. For the explicit spinor calculation, we choose the third spatial coordinate axis along the quark momentum and write
\begin{equation}
k^\mu=\frac{{\cal M}}{2}(1,0,0,\beta),
\qquad
k'^\mu=\frac{{\cal M}}{2}(1,0,0,-\beta),
\qquad
\beta=\sqrt{1-\frac{4m^2}{{\cal M}^2}},
\label{Eq:PairCMKinematics}
\end{equation}
where ${\cal M}^2=(k+k')^2$. The normalized spin-density matrix of this two-qubit system can be
decomposed as 
\begin{equation}
\rho
=
\frac{1}{4}
\left[
\mathbb{I}\otimes\mathbb{I}
+B^a\sigma^a\otimes\mathbb{I}
+\bar B^b\mathbb{I}\otimes\sigma^b
+C^{ab}\sigma^a\otimes\sigma^b
\right] \; ,
\label{Eq:DensityMatrixGeneral}
\end{equation}
where $\sigma^i$ denotes the $i$-th Pauli matrix. For a generic $q \bar q$-production amplitude
\begin{equation}
{\cal M}_{\alpha\alpha'}
=
\bar u_\alpha(k)\,\Gamma\,v_{\alpha'}(k') \; ,
\label{Eq:GenericQQbarAmplitude}
\end{equation}
we introduce the spin-open squared amplitude
\begin{equation}
{\cal R}_{\alpha\alpha';\beta\beta'}
=
{\rm Tr}\left[
\bar\Gamma\,
u_\beta(k)\bar u_\alpha(k)\,
\Gamma\,
v_{\alpha'}(k')\bar v_{\beta'}(k')
\right],
\qquad
\bar\Gamma\equiv\gamma^0\Gamma^\dagger\gamma^0 ,
\label{Eq:SpinOpenSquaredAmplitude}
\end{equation}
and its spin-summed normalization
\begin{equation}
{\cal A}
\equiv
\sum_{\alpha,\alpha'}
{\cal R}_{\alpha\alpha';\alpha\alpha'}
=
{\rm Tr}\left[
\bar\Gamma(\Slash{k}+m)
\Gamma(\Slash{k}'-m)
\right].
\label{Eq:SpinNormalization}
\end{equation}
Consequently,
\begin{equation}
{\cal R}_{\alpha\alpha';\beta\beta'}
=
{\cal A}\,
\xi_\alpha^\dagger\eta_{\alpha'}^\dagger\,
\rho\,
\xi_\beta\eta_{\beta'} .
\label{Eq:SpinOpenDensityMatrix}
\end{equation}
where $\xi$ and $\eta$ are the two-component spinors initially contained in the four-component Dirac spinors $u$ and $v$.
It can be shown~\cite{Fucilla:2026zhy} that the coefficients of the spin-density matrix can be conveniently computed in terms of four-dimensional quantities. They are given by the following master formulas~\cite{Fucilla:2026zhy}
\begin{gather}
    C^{ij} = \frac{1}{ {\cal A}} {\rm Tr}\left[\bar{\Gamma} \gamma^i\gamma_5(\Slash k+m)\Gamma \gamma^j\gamma_5(\Slash k'-m)\right] ,
    \label{Eq:Cij}
\end{gather}
\begin{gather}
    C^{ki} = \frac{1}{ {\cal A}} {\rm Tr} \left[\bar{\Gamma}\gamma_5\gamma^0\gamma^3(\Slash k+m)\Gamma \gamma^i\gamma_5(\Slash k'-m)\right] \; , 
    \label{Eq:Cki}
\end{gather}
\begin{gather}
C^{ik} =  - \frac{1}{ {\cal A}}  {\rm Tr}\left[\bar{\Gamma} \gamma^i\gamma_5(\Slash k+m)\Gamma \gamma_5\gamma^0\gamma^3(\Slash k'-m)\right]  ,
\label{Eq:Cik}
\end{gather}
\begin{gather}
    C^{kk} = - \frac{1}{ {\cal A}}  {\rm Tr}\left[\bar{\Gamma}\gamma_5\gamma^0\gamma^3(\Slash k+m)\Gamma \gamma_5\gamma^0\gamma^3(\Slash k'-m)\right]  \; ,
    \label{Eq:Ckk}
\end{gather} 
\begin{gather}
    B^i= \frac{1}{ {\cal A}}  {\rm Tr}\left[\bar{\Gamma}\gamma^i\gamma_5(\Slash k+m)\Gamma (\Slash k'-m)\right]   \; , \qquad
    B^k =  \frac{1}{ {\cal A}} {\rm Tr}\left[\bar{\Gamma}\gamma_5\gamma^0\gamma^3(\Slash k+m)\Gamma (\Slash k'-m)\right]  ,
\end{gather}
\begin{gather}
    \bar{B}^i=  \frac{1}{ {\cal A}}   {\rm Tr}\left[\bar{\Gamma}(\Slash k+m)\Gamma \gamma^i\gamma_5 (\Slash k'-m)\right]     \; , \qquad \bar{B}^k =-\frac{1}{{\cal A}} {\rm Tr}\left[\bar{\Gamma}(\Slash k+m)\Gamma \gamma_5\gamma^0\gamma^3(\Slash k'-m)\right]  \; ,
\end{gather}
where $i,j = \{ n,r \}$ and the Dirac matrices are considered in the chiral (Weyl) representation
\begin{equation}
 \gamma^\mu=
 \begin{pmatrix}
  0 & \; \sigma^\mu\\
  \bar\sigma^\mu&0
 \end{pmatrix} \; , \hspace{1 cm} \gamma_5=i\gamma^0\gamma^1\gamma^2\gamma^3 \; ,
 \label{eq:Weyl_gamma_matrices}
\end{equation}
with
\begin{equation}
 \sigma^\mu=( 1,\sigma^1,\sigma^2,\sigma^3),
 \qquad
 \bar \sigma^\mu=( 1,-\sigma^1,-\sigma^2,-\sigma^3),
 \label{eq:sigma_four_vectors}
\end{equation} 
For a parity-conserving production mechanism, the allowed components satisfy
\begin{equation}
B^r=B^k=\bar B^r=\bar B^k=0,
\qquad
C^{nr}=C^{rn}=C^{nk}=C^{kn}=0.
\label{Eq:ParityConstraints}
\end{equation}
For the process considered in this work, the normal polarizations
also vanish at leading order,
\begin{equation}
B^n=\bar B^n=0.
\label{Eq:LeadingOrderConstraint}
\end{equation}
In the following subsections, we give the formulas for the relevant quantum information quantities under the assumptions of parity conservation and vanishing normal polarizations, $B^n=\bar B^n=0$. Under these assumptions, the spin-density matrix is fully specified by the correlation coefficients $C^{ab}$. The resulting state thus has maximally mixed marginals\footnote{The marginals are the reduced spin-density matrices of the individual quark and antiquark, $\rho_q \equiv \operatorname{Tr}_{\bar q}\rho$ and $\rho_{\bar q} \equiv \operatorname{Tr}_q\rho$.}
\begin{gather}
\rho_q  = \rho_{\bar q} =\mathbb{I}/2 \; .
\label{Eq:MaximallyMixedMarginals}
\end{gather}

\subsection{Entanglement, Bell nonlocality, nonstabilizerness and quantum discord}
\label{Sec:QuantumInformationObservables}

We characterize the spin state of the pair through four complementary quantum information quantities: entanglement, Bell nonlocality, nonstabilizerness, and quantum discord.

\paragraph{Entanglement.} A bipartite state is entangled if it cannot be expressed as a mixture of product states. For the density matrices considered here, define
\begin{align}
\Delta_1
&=
R_1
-1+C_{nn},
\nonumber\\
\Delta_2
&=
R_2
-1-C_{nn}.
\label{Eq:DeltaEntanglement}
\end{align}
where
\begin{align}
R_1 &=
\sqrt{(C_{rr}-C_{kk})^2+(C_{rk}+C_{kr})^2},
\nonumber\\
R_2 &=
\sqrt{(C_{rr}+C_{kk})^2+(C_{rk}-C_{kr})^2} .
\end{align}
The Peres--Horodecki criterion~\cite{Peres:1996dw,Horodecki:1997vt} implies that the state is entangled whenever
$\Delta_1>0$ or $\Delta_2>0$~\cite{Afik:2020onf,Afik:2022kwm}. Defining the concurrence
\begin{equation}
{\cal C}[\rho]
=
\frac{1}{2}
\max\left\{
\Delta_1,\Delta_2,0
\right\},
\qquad
0\leq {\cal C}[\rho]\leq1 \; ,
\label{Eq:Concurrence}
\end{equation}
the condition for entanglement becomes ${\cal C}[\rho]>0$. \\

\paragraph{Bell nonlocality.} A more stringent form of quantum correlation is the Bell nonlocality. Bell nonlocality refers to correlations that cannot be reproduced
by a local hidden-variable model. This latter is signaled by the violation of the Bell-type Clauser–Horne–Shimony–Holt (CHSH) inequality~\cite{Clauser:1969ny,Horodecki:1995nsk} 
\begin{equation}
{\cal B}_{\rm max}
\equiv \max_{\{\boldsymbol n_i\}}
\left|
n_1^a C_{ab}\left(n_2^b+n_4^b\right)
+
n_3^a C_{ab}\left(n_2^b-n_4^b\right)
\right| \leq 2 ,
\qquad
|\boldsymbol n_i|=1 .
\label{Eq:CHSHParameter}
\end{equation}
If $\mu_1\geq\mu_2\geq\mu_3$ are the eigenvalues of
$C^{\mathcal T}C$, one has
\begin{equation}
{\cal B}_{\rm max}
=
2\sqrt{\mu_1+\mu_2}.
\label{Eq:MaximalCHSH}
\end{equation}
The Bell--CHSH bound is therefore violated when
\begin{equation}
\mu_1+\mu_2>1 .
\label{Eq:BellNonlocalityCondition}
\end{equation} \vspace{0.2 cm}

    \paragraph{Nonstabilizerness.} As an independent quantum-information characterization, we quantify
nonstabilizerness, also known as ``magic", which characterizes genuine quantum computational resources. To this scope, we use the stabilizer R\'enyi entropy~\cite{Gottesman:1998hu,Leone:2021rzd,White:2024nuc,Liu:2025qfl,Gargalionis:2026onv},
which in the present case becomes~\cite{Fucilla:2026mkg}
\begin{equation}
S_{\rm R}^{\rm stab}
=
-\ln\left[
\frac{
1+C_{nn}^4+C_{rr}^4+C_{kk}^4+C_{kr}^4+C_{rk}^4
}{
1+C_{nn}^2+C_{rr}^2+C_{kk}^2+C_{kr}^2+C_{rk}^2
}
\right].
\label{RenyiStabEntrop}
\end{equation} \vspace{0.2 cm}

\paragraph{Quantum discord.} The quantum discord is the difference between the total mutual information and the classical mutual information~\cite{Ollivier:2001fdq,Dakic:2010xfz}. This quantity should vanish from a classical perspective, but can be non-zero in quantum mechanics and captures non-classical correlations. Quantum discord captures a broader class of quantum correlations than entanglement: every entangled state has nonzero discord, while some separable states do as well. 

In our case, the single-spin polarizations vanish, and we thus have states with maximally mixed marginals. To obtain the entropic quantum discord, we use the fact that states with maximally mixed marginals can be brought
to Bell-diagonal form by local spin rotations~\cite{Lang:2010xtw,Han:2024ugl}. The four eigenvalues of the spin density matrix are
\begin{equation}
\lambda_{1,\pm}=\frac{1+C_{nn}\pm R_1}{4},
\qquad
\lambda_{2,\pm}=\frac{1-C_{nn}\pm R_2}{4},
\end{equation}
and the largest singular value of $C$ is
\begin{equation}
c_{\max}=
\max\left\{
|C_{nn}|,\frac{R_1+R_2}{2}
\right\}.
\end{equation}
Consequently, the discord is the same whether the local measurement
is performed on the quark or on the antiquark, and is given by~\cite{Lang:2010xtw}
\begin{equation}
D_{q|\bar q}(\rho)=D_{\bar q|q}(\rho)
=1+\sum_{i=1}^{2}\sum_{\eta=\pm}
\lambda_{i,\eta}\log_2\lambda_{i,\eta}
+h_2\!\left(\frac{1+c_{\max}}{2}\right)
\label{eq:discord_from_coefficients}
\end{equation}
where $h_2(x)=-x\log_2 x-(1-x)\log_2(1-x)$, with the convention
$0\log_2 0=0$. This expression applies both before and after
azimuthal averaging.

\section{The $g \rightarrow Q \bar Q$ amplitude in the CGC}

In this section we re-derive the amplitude for the $g \rightarrow Q \bar{Q}$ transition~\cite{Fujii:2006ab,Dominguez:2011wm,Marquet:2017xwy} in the shockwave formalism~\cite{McLerran:1993ka,Balitsky:1995ub}. This result is the building block for constructing and discussing the spin density matrix of both inclusive and diffractive hadroproduction of a heavy-quark pair.

\subsection{Conventions and kinematics}
Let us begin by fixing the convention used throughout the paper. We introduce the light-cone components of a generic four-vector, defined by
\begin{equation}
v^+\equiv v\cdot n_2=\frac{v^0+v^3}{\sqrt2},
\qquad
v^-\equiv v\cdot n_1=\frac{v^0-v^3}{\sqrt2}.
\label{eq:light_cone_definition}
\end{equation}
Accordingly,
\begin{equation}
v^\mu = v^+n_1^\mu+v^-n_2^\mu+v_\perp^\mu,
\qquad
v_\perp\cdot n_1=v_\perp\cdot n_2=0.
\label{eq:light_cone_decomposition}
\end{equation}
The corresponding light-cone Dirac matrices are
\begin{equation}
\gamma^\pm=\frac{\gamma^0\pm\gamma^3}{\sqrt2},
\qquad
\slashed n_1=\gamma^-,
\qquad
\slashed n_2=\gamma^+.
\label{eq:light_cone_gamma_matrices}
\end{equation}
\noindent The transverse metric and antisymmetric tensor associated with the
ordered pair $(n_1,n_2)$ are
\begin{align}
g_\perp^{\mu\nu}
\equiv
g^{\mu\nu}
-n_1^\mu n_2^\nu
-n_2^\mu n_1^\nu,
\qquad
\epsilon_\perp^{\mu\nu}
\equiv
\epsilon^{\mu\nu\rho\sigma}
n_{1\rho}n_{2\sigma}.
\label{eq:transverse_metric_and_epsilon}
\end{align}
For transverse indices $i,j=1,2$, these definitions give
\begin{equation}
g_\perp^{ij}=-\delta^{ij},
\qquad
\epsilon_\perp^{12}=+1.
\label{eq:transverse_components}
\end{equation}
Throughout the following, boldface transverse vectors are
two-dimensional Euclidean vectors.  In particular,
\begin{equation}
v_\perp\cdot w_\perp = -\bm v\cdot\bm w = -v^i w^i,
\qquad
\bm v^2=-v_\perp^2.
\label{eq:Minkowski_Euclidean_transverse_products}
\end{equation} 
The incoming gluon is considered to move in the $+$ direction,
\beq
 p_g^\mu=p_g^+n_1^\mu,
\eeq
The momenta of the produced quarks are decomposed as
\begin{equation}
    p_q^{\mu} = p_q^+ n_1^{\mu} + \frac{m^2 + {\bm p}_q^2}{2p_q^+} n_2^{\mu} + p_{q \perp}^{\mu} \; , \qquad  p_{\bar q}^{\mu} = p_{\bar q}^+ n_1^{\mu} + \frac{m^2 + {\bm p}_{\bar q}^2}{2 p_{\bar q}^+} n_2^{\mu} + p_{ {\bar q} \perp}^{\mu}
\end{equation}
where
\beq
  p_{  q}^+ =  z p_g^+,
 \qquad p_{\bar q}^+ =\bar z p_g^+,
 \qquad \bar z\equiv1-z \; .
\eeq

\subsection{Amplitude for the $g \rightarrow Q \bar{Q}$ transition in the shockwave formalism}
\begin{figure}
       \hspace{0.7 cm}\includegraphics[width=0.85\linewidth]{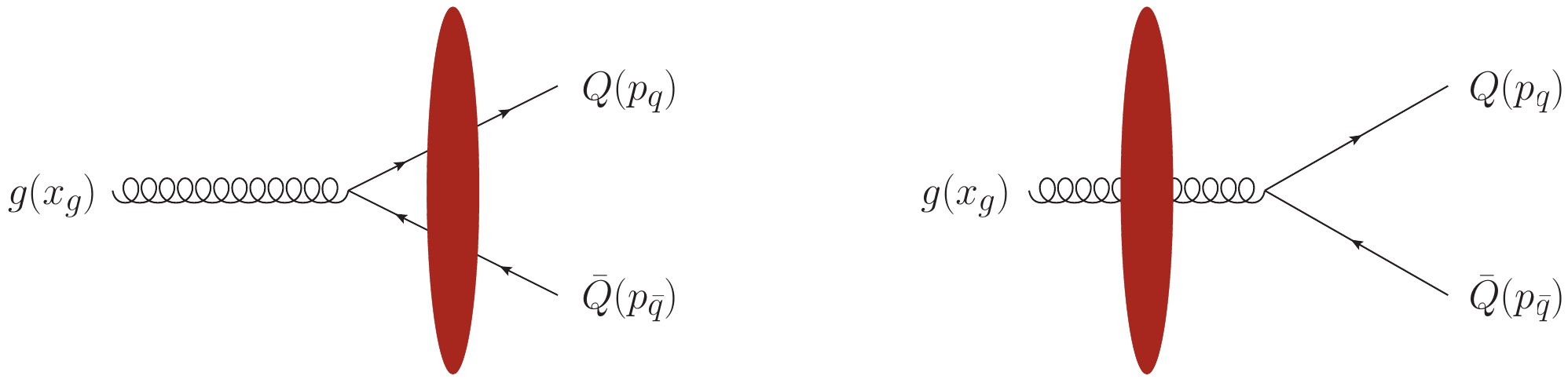}
    \caption{Diagram contributing to the $S$-matrix element for the dijet production in gluon-nucleus collisions.}
    \label{fig:Impact factor}
\end{figure}
The $S$-matrix element for the inclusive dijet production in gluon-nucleus collisions has two contributions. The first is analogous to the QED process $\gamma \rightarrow q \bar q$, and it occurs when the splitting of the collinear gluon into two quarks occurs before the shockwave; the associated $S$-matrix element reads
\begin{equation}
    S_{g \rightarrow q \bar q }^{\rm before} = - i g \varepsilon_{\mu} ( p_{g}) \int d^4 z \; \theta (-z^+) e^{-i p_{g} \cdot z} \; \bar{u}^{\rm shock}_{\alpha} (p_q, z) t^a \gamma^{\mu} v_{\alpha'}^{\rm shock} (p_{\bar{q}}, z) \; ,
    \label{Eq:DijetSMatrix1}
\end{equation}
where the effective quark/anti-quark spinors through the shockwave are
\begin{gather}
    \bar{u}^{\rm shock}_{\alpha} (p_q, z) = \theta (p_q^+) \int \frac{d^2 {\bm k}}{(2 \pi)^2} \int d^2 {\bm x}_{1} e^{-i {\bm x}_{1} \cdot ( {\bm p}_q - {\bm k}) } \mathcal{U}_F ( {\bm x}_{1} ) \nonumber \\ \times e^{i z^+ \frac{ {\bm k}^2 + m^2}{2 p_q^+} + i z^- p_q^+ - i {\bm z} \cdot {\bm k} } \bar{u}_{\alpha} (p_q) \gamma^+ \frac{ \gamma^- p_q^+ - \Slash{ \bm k} + m }{2 p_q^+} \; ,
\end{gather}
\begin{gather}
    v^{\rm shock}_{\alpha'} (p_{\bar{q}}, z) =  \theta (p_{\bar{q}}^+) \int \frac{d^2 {\bm p} }{(2 \pi)^2} \int d^2 {\bm x}_{2} e^{-i {\bm x}_{2} \cdot ( {\bm p}_{\bar q} - {\bm p}) } \mathcal{U}_F^{\dagger} ( {\bm x}_{2} ) \nonumber \\ \times e^{i z^+ \frac{ {\bm p}^2 + m^2}{2 p_{\bar{q}}^+} + i z^- p_{\bar{q}}^+ - i {\bm z} \cdot {\bm p} } \; \frac{ \gamma^- p_{\bar{q}}^+ - \Slash{ \bm p} - m }{2 p_{\bar{q}}^+} \gamma^+ v_{\alpha'} (p_{\bar{q}}) \; .
\end{gather}
and our convention for the Wilson line is\footnote{Here, the subscript $R$ labels the representation, and $T_R$ denotes the corresponding generator.}
\begin{equation}
\mathcal{U}_R ( \boldsymbol{x} ) = \mathcal{P} \exp\left[ i g \int dx^+  T_R \cdot A^-(x^+, x_\perp) \right] \; .
\end{equation}
The second contribution is the genuine non-Abelian diagram generated by the direct interaction of the collinear gluon with the shockwave; it reads
\begin{equation}
    S_{g \rightarrow q \bar q }^{\rm after} = - i g \; \bar{u}_{\alpha} (p_q) t^b \gamma^{\mu} v_{\alpha'} (p_{\bar{q}}) \int d^4 z \; \theta (z^+) e^{i (p_{q}+p_{\bar q}) \cdot z} \;  [\varepsilon_{\mu}^{\rm Shock} ( p_{g}, z)]^{ba} ,
    \label{Eq:DijetSMatrix2}
\end{equation}
where the effective gluon polarization vector through the shockwave is 
\begin{gather}
    [\varepsilon_{\mu}^{\rm Shock} ( p_{g}, z)]^{ba} = \varepsilon_{\perp \nu} (p_{g}) \int \frac{d^2 {\bm p} }{(2 \pi)^2} \int d^2 {\bm y} \; e^{i {\bm p} \cdot {\bm y} } [\mathcal{U}_A ( {\bm y} )]^{ba} \left[ g^{\nu}_{\mu  \perp} - \frac{n_{2 \mu} (p_{g \perp}^{\nu} - p_{\perp}^{\nu}) }{p_g^+} \right] \nonumber \\ \times e^{- i z^- p_g^+ - i z^+ \frac{ (\boldsymbol{p}_g - \boldsymbol{p})^2 }{ 2 p_g^+} + i{\bm z} \cdot ({\bm p}_g - {\bm p} ) }  \; .
\end{gather}
We start by computing the first contribution. Integrating over $z^+, z^-$ and $z_{\perp}$ in eq.~(\ref{Eq:DijetSMatrix1}), one can easily get
\begin{gather}
   \mathcal{M}_{g \rightarrow q \bar q }^{\rm before} = \frac{(-i) S_{g \rightarrow q \bar q }^{\rm before}  }{2 p_{g}^+ (2 \pi) \delta (p_{g}^+ - p_q^+ - p_{\bar{q}}^+ ) } \nonumber \\ = -i g  \int d^2 {\bm x}_{1}  \int d^2 {\bm x}_{2} e^{-i {\bm x}_{1} \cdot {\bm p}_{q} } e^{-i {\bm x}_{2} \cdot {\bm p}_{\bar{q} } } \Phi_{\alpha \alpha' \perp} ( {\bm x}_{12}) [\mathcal{U}_F ( {\bm x}_{1} ) t^a \mathcal{U}_F^{\dagger} ( {\bm x}_{2}) ] \; ,
   \label{Eq:MDijet1}
\end{gather}
where
\begin{gather}
    \Phi_{\alpha \alpha' \perp} ( {\bm x}_{12}) = \frac{ \varepsilon^i (p_{g}) }{ 2 p_{g}^+ } \int \frac{d^2 {\bm k} }{(2 \pi)^2} \frac{ e^{i {\bm k} \cdot {\bm x}_{12 } } }{ {\bm k}^2 + m^2 }  \bar{u}_{\alpha} (p_q) \left[ (1- 2 z) k^i + i \epsilon^{ij} k^j \gamma^5 - m \gamma^i  \right] \gamma^+  v_{\alpha'} (p_{\bar{q}}) \nonumber \\ 
    = \frac{ \varepsilon^i (p_{g}) }{ 2 p_{g}^+ } \int \frac{d^2 {\bm k} }{(2 \pi)^2} \frac{ e^{i {\bm k} \cdot {\bm x}_{12 } } }{ {\bm k}^2 
    + m^2 }  \bar{u}_{\alpha} (p_q) \left[ (1- 2 z) k^i - \frac{1}{2} [  \slashed{\bm k} , \gamma^i] - m \gamma^i  \right] \gamma^+  v_{\alpha'} (p_{\bar{q}}) \; .
    \label{Eq:IFmomRep}
\end{gather}
is the so-called splitting wavefunction. As we will show in a moment, the second contribution can be cast in a form that only modifies the color operator structure in eq.~(\ref{Eq:MDijet1}). Adding and subtracting a proper term to $\mathcal{M}_{g \rightarrow q \bar q }^{\rm before}$
\begin{gather}
   \mathcal{M}_{g \rightarrow q \bar q }^{\rm before}  = -i g  \int d^2 {\bm x}_{1}  \int d^2 {\bm x}_{2} e^{-i {\bm x}_{1} \cdot {\bm p}_{q} } e^{-i {\bm x}_{2} \cdot {\bm p}_{\bar{q} } } \Phi_{\alpha \alpha' \perp} ( {\bm x}_{12}) [\mathcal{U}_F ( {\bm x}_{1} ) t^a \mathcal{U}_F^{\dagger} ( {\bm x}_{2}) - t^b \mathcal{U}_A^{ba} (\boldsymbol{v}) ] \nonumber \\ - i g  \int d^2 {\bm x}_{1}  \int d^2 {\bm x}_{2} e^{-i {\bm x}_{1} \cdot {\bm p}_{q} } e^{-i {\bm x}_{2} \cdot {\bm p}_{\bar{q} } } \Phi_{\alpha \alpha' \perp} ( {\bm x}_{12}) [ t^b \mathcal{U}_A^{ba} (\boldsymbol{v}) ] \; ,
   \label{Eq:Mbefore}
\end{gather}
where $\boldsymbol{v} = z \boldsymbol{x}_1 + \bar z {\bm x}_2$, we want to show that the second term in the last line is exactly cancelled by the $\mathcal{M}_{g \rightarrow q \bar q }^{\rm after}$. Let us first observe that this term can be written as
\begin{gather}
   - i g  \Phi_{\alpha \alpha' \perp} ( {\bm P} ) \int d^2 \boldsymbol{v} e^{-i ({\bm p}_q + {\bm p}_{\bar q}) \cdot {\bm v} } [t^b \mathcal{U}_A^{ba} (\boldsymbol{v})]
\end{gather}
where we introduced $\boldsymbol{P} = \bar z {\bm p}_q - z {\bm p}_{\bar q} $ and we made explicit the frame choice ${\bm p}_{g}=0$. On the other hand, eq.~(\ref{Eq:DijetSMatrix2}) leads to
\begin{gather*}
    {\mathcal M}_{g \rightarrow q \bar q }^{\rm after} = i g \int d^2 \boldsymbol{v} e^{- i {\bm v} \cdot ({\bm p}_q + {\bm p}_{\bar q}) } [t^b \mathcal{U}_A^{ba} (\boldsymbol{v})]   \frac{ \varepsilon^i (p_{g}) }{ 2 p_{g}^+ }  \frac{\bar{u}_{\alpha} (p_q) \left[  \gamma^i - \frac{\gamma^+ (p_{q}^{i} + p_{\bar q}^{i} )}{ p_g^+}  \right]  v_{\alpha'} (p_{\bar{q}}) }{ \frac{{\bm p}_q^2 + m^2}{2 p_q^+} + \frac{{\bm p}_{\bar q}^2 + m^2}{2 p_{\bar q}^+} - \frac{ ({\bm p}_q + {\bm p}_{\bar q})^2 }{2 p_g^+} } \nonumber \\ = i g \int d^2 \boldsymbol{v} e^{- i {\bm v} \cdot ({\bm p}_q + {\bm p}_{\bar q}) } [t^b \mathcal{U}_A^{ba} (\boldsymbol{v})]   \frac{ \varepsilon^i (p_{g}) }{ 2 p_{g}^+ }  \frac{\bar{u}_{\alpha} (p_q) \left[ 2 z \bar z p_g^+ \gamma^i - 2 z \bar z  (p_{q}^{i} + p_{\bar q}^{i} )\gamma^+   \right]  v_{\alpha'} (p_{\bar{q}}) }{  {\bm P}^2
    + m^2 } \; .
\end{gather*}
We therefore need to show that
\begin{gather}
     \bar{u}_{\alpha} (p_q) \left[ 2 z \bar z p_g^+ \gamma^i - 2 z \bar z  (p_{q}^{i} + p_{\bar q}^{i} )\gamma^+   \right]  v_{\alpha'} (p_{\bar{q}}) \nonumber \\ = \bar{u}_{\alpha} (p_q) \left[ (1- 2 z) P^i - \frac{1}{2} [  \slashed{\bm P} , \gamma^i] - m \gamma^i  \right] \gamma^+  v_{\alpha'} (p_{\bar{q}}) \; .
     \label{Eq:DiracIden}
\end{gather}
To this end, we use the Dirac equations
\begin{gather}
    \bar{u}_{\alpha} (p_q) ( \slashed{p}_q-m) = 0 \; , \qquad ( \slashed{p}_{\bar q}+m) v_{\alpha'} (p_{\bar{q}}) = 0 \; ,
\end{gather}
to cast the first term on the first line of eq.~(\ref{Eq:DiracIden}) in the following form
\begin{gather}
     \bar{u}_{\alpha} (p_q)  2 z \bar z p_g^+ \gamma^i   v_{\alpha'} (p_{\bar{q}}) = \bar{u}_{\alpha} (p_q) \left[ -m \gamma^i \gamma^+ -  \bar z  p_{q}^j \gamma^j \gamma^i \gamma^+ - z p_{ \bar q}^j \gamma^i \gamma^j \gamma^+    \right] v_{\alpha'} (p_{\bar{q}})
\end{gather}
Using the last expression in the first line of eq.~(\ref{Eq:DiracIden}) and then moving to $P^i$ and
\begin{equation}
    q^i = p_{q}^{i} + p_{\bar q}^{i} \; ,
\end{equation}
is not difficult to prove eq.~(\ref{Eq:DiracIden}). We therefore obtain
\begin{equation}
   {\mathcal M}_{g \rightarrow q \bar q }^{\rm after}=  i g  \Phi_{\alpha \alpha' \perp} ( {\bm P} ) \int d^2 \boldsymbol{v} e^{-i ({\bm p}_q + {\bm p}_{\bar q}) \cdot {\bm v} } [t^b \mathcal{U}_A^{ba} ({\bm v})] \; .
   \label{Eq:Mafter}
\end{equation}
Summing eqs.~(\ref{Eq:Mbefore}) and~(\ref{Eq:Mafter}), we finally obtain 
\begin{equation}
    \mathcal{M}_{g \rightarrow q \bar q }  = -i g  \int d^2 {\bm x}_{1}  \int d^2 {\bm x}_{2} e^{-i {\bm x}_{1} \cdot {\bm p}_{q} } e^{-i {\bm x}_{2} \cdot {\bm p}_{\bar{q} } } \Phi_{\alpha \alpha' \perp} ( {\bm x}_{12}) [\mathcal{U}_F ( {\bm x}_{1} ) t^a \mathcal{U}_F^{\dagger} ( {\bm x}_{2}) - t^b \mathcal{U}_A^{ba} (\boldsymbol{v}) ] \; .
    \label{Eq:FullAmpInclusive}
\end{equation}
An important property of eq.~(\ref{Eq:FullAmpInclusive}) is its behavior in the limit of vanishing transverse separation, $\boldsymbol{u}\equiv\boldsymbol{x}_{12}\to 0$. In this limit, $\boldsymbol{x}_1=\boldsymbol{x}_2=\boldsymbol{v}$, and the fundamental Wilson lines satisfy
\begin{equation}
\mathcal{U}_F(\boldsymbol{v}) t^a 
\mathcal{U}_F^\dagger(\boldsymbol{v})
=
t^b \mathcal{U}_A^{ba}(\boldsymbol{v}) \; , 
\end{equation}
Consequently, the quantity in square brackets in eq.~(\ref{Eq:FullAmpInclusive}) vanishes in this limit. This cancellation is a manifestation of color coherence: a pointlike $q\bar q$ pair produced in a color-octet state carries the same total color charge and undergoes the same eikonal color rotation as its parent gluon. The target is therefore unable to distinguish whether the splitting occurred before or after the interaction with the shockwave. Only a finite transverse separation between the quark and the antiquark allows the target to resolve the internal structure of the pair.

\section{Inclusive proton-nucleus collisions}

\label{sec:IncluCollis}

We begin by discussing the inclusive proton-nucleus collisions. In this section, we derive the spin density matrix of the inclusive production of two heavy quarks in $pp$ or $pA$ collisions in the back-to-back limit. We perform the so-called twist expansion of shockwave/Color-Glass-Condensate results to obtain a TMD-factorized form~\cite{Dominguez:2011wm,Kotko:2015ura,vanHameren:2016ftb,Altinoluk:2019wyu,Altinoluk:2019fui,Boussarie:2021ybe,Bury:2020ndc,Altinoluk:2018uax,Altinoluk:2018byz,Altinoluk:2020qet,Caucal:2025mth,Caucal:2026xhj} (see Ref.~\cite{Caucal:2026yzl,Iancu:2026jlb} for recent reviews). Our derivation closely parallels that of the inclusive cross-section in Ref.~\cite{Marquet:2017xwy}, generalizing it to the spin density matrix. \\

At the leading order, the inclusive proton-nucleus cross-section for the process can be written in the hybrid high-energy and collinear factorization as 
\beq
 d\sigma^{pA\to Q\bar QX}
 =\int dx_p\,g(x_p,\mu^2)\,
 d\hat\sigma^{gA\to Q\bar QX}
 \big\vert_{p_g^+=x_p p_p^+,\,p_{g\perp}=0} \; ,
 \label{eq:hybrid_convolution}
\eeq
where the LO partonic cross section, $d\hat\sigma^{gA\to Q\bar QX}$, refers to the gluon-initiated channel
\beq
g(p_g,a,\lambda)+A\longrightarrow Q(p_q,i,\alpha)
 +\bar Q(p_{\bar q},j,\alpha')+X,
 \label{eq:process}
\eeq
In eq.~(\ref{eq:process}), $a$ and $\lambda$ are the color and physical transverse polarization of the incoming gluon, $i,j$ are the final fundamental color indices, and $\alpha,\alpha'$ are kept open. In the present work, we are interested in the fully differential cross-section (in the kinematical variables of the two quarks) and the gluon longitudinal fraction,
\begin{equation}
    x_g = \frac{p_g^+}{P_p^+} = \frac{p_q^+ + p_{\bar q}^+}{P_p^+} = \frac{m_{q \perp} e^{y_q} + m_{ \bar q \perp} e^{y_{\bar q}}}{\sqrt{s}} \; ,
\end{equation}
is therefore fixed by the kinematics of the outgoing particles. Therefore, we have
\begin{equation}
    \frac{d\sigma^{pA\to Q\bar QX}}{dy_q d y_{\bar q} d^2 {\bm p}_{q} d^2 {\bm p}_{\bar q } } = x_g \,g(x_g,\mu^2) \,  \frac{d\hat\sigma^{gA\to Q\bar QX}}{{dy_q d y_{\bar q} d^2 {\bm p}_{q} d^2 {\bm p}_{\bar q } }} \bigg \vert_{p_g^+=x_p p_p^+,\,p_{g\perp}=0} \; .
\end{equation}
Since the collinear gluon distribution $x_g g(x_g,\mu^2)$ multiplies all spin components by the same overall factor, it cancels upon normalization and therefore does not affect the spin-density matrix.

\subsection{The correlation limit}
We derive in this section the ``spin-open" partonic squared matrix element of the process. We start from the standard form of the CGC cross-section
\beq
  \left[ \frac{d\hat\sigma^{gA\to Q\bar QX}}{{d z d {\bar z} d^2 {\bm p}_{q} d^2 {\bm p}_{\bar q } }} \right]_{\alpha \alpha' \beta \beta'} = ||J|| \; 2 p_{g}^+ (2 \pi) \delta (p_{g}^+ - p_q^+ - p_{\bar{q}}^+) \frac{1}{2 (N_c^2-1)} \sum_{\rm hel. \; col.} \langle | \mathcal{M}_{g \rightarrow q \bar q }  |^2 \rangle \; , \nonumber \\
\eeq
where we average over the initial gluon color and helicity degrees of freedom and perform the CGC statistical average~\cite{Gelis:2010nm}. The Jacobian factor comes from the change of variables  
$d^6 {\rm P.S.} = ||J|| d z d \bar z d^2 {\bm p}_{q} d^2 {\bm p}_{\bar q }$, 
where $d^6 {\rm P.S.}$ is the standard two-particle phase space and
\begin{gather*}
 ||J|| = \frac{1}{4 z \bar{z} (2 \pi)^6} \; .
\end{gather*}
We easily obtain
\begin{gather}
  \left[ \frac{d\hat\sigma^{gA\to Q\bar QX}}{{d z  d^2 {\bm p}_{q} d^2 {\bm p}_{\bar q } }} \right]_{\alpha \alpha' \beta \beta'} =   \frac{   \alpha_s  }{ 2 } p_g^+ \int \frac{d^2 {\bm x}_{1}   d^2 {\bm x}_{2}  d^2 {\bm x}_{1'}  d^2 {\bm x}_{2'}}{(2 \pi)^8} \; e^{-i {\bm x}_{11'} \cdot {\bm p}_{q} - i {\bm x}_{22'} \cdot {\bm p}_{\bar{q} } }  \psi_{\beta \beta' \perp}^{*} (\boldsymbol{x}_{1'2'}) \psi_{\alpha \alpha' \perp} (\boldsymbol{x}_{12}) \nonumber \\ \times [S^{(4)}_{q \bar q \bar q q}    ( {\bm x}_1,{\bm x}_2, {\bm x}_{1'}, {\bm x}_{2'}) - S^{(3)}_{q g \bar q } ( {\bm v} , {\bm x}_{1'}, {\bm x}_{2'}) - S^{(3)}_{q g \bar q } ( {\bm v'} , {\bm x}_{2}, {\bm x}_{1}) + S^{(2)}_{gg} ( {\bm v} , {\bm v}') ] \; ,
   \label{Eq:GeneralInclusiveCross}
\end{gather}
where 
\begin{gather}
 S^{(4)}_{q \bar q \bar q q}    ( {\bm x}_1,{\bm x}_2, {\bm x}_{1'}, {\bm x}_{2'}) = \frac{2}{N_c^2-1} \left \langle {\rm Tr}_F \left( \mathcal{U}_F ( {\bm x}_{2'} ) t^a \mathcal{U}_F^{\dagger} ( {\bm x}_{1'}) \mathcal{U}_F ( {\bm x}_{1} ) t^a \mathcal{U}_F^{\dagger} ( {\bm x}_{2}) \right) \right \rangle \; ,
\end{gather}
\begin{gather}
     S^{(3)}_{q g \bar q } ( {\bm v} , {\bm x}_{1'}, {\bm x}_{2'}) = \frac{2}{N_c^2-1} \left \langle {\rm Tr}_F \left( \mathcal{U}_F ( {\bm x}_{2'} ) t^a \mathcal{U}_F^{\dagger} ( {\bm x}_{1'}) t^b \right) \mathcal{U}_A^{ba} ( {\bm v})  \right \rangle \; ,
\end{gather}
\begin{gather}
     S^{(3)}_{q g \bar q } ( {\bm v'} , {\bm x}_{2}, {\bm x}_{1}) = \frac{2}{N_c^2-1} \left \langle {\rm Tr}_F \left( \mathcal{U}_F^{\dagger} ( {\bm x}_{2} ) t^a \mathcal{U}_F ( {\bm x}_{1}) t^b \right) \mathcal{U}_A^{\dagger ba} ( {\bm v'})  \right \rangle \; , 
\end{gather}
\begin{gather}
   S^{(2)}_{gg} ( {\bm v} , {\bm v}') = \frac{1}{N_c^2-1} \left \langle  {\rm Tr}_A (\mathcal{U}_A^{\dagger} ( {\bm v'}) \mathcal{U}_A ( {\bm v}) ) \right \rangle \; ,
\end{gather}
and we have defined\footnote{We also left implicit the sum over polarization indices of the incoming gluon in the $\psi^{*} \psi$ product.}
\begin{gather}
    \psi_{\alpha \alpha' \perp} (\boldsymbol{x}_{12}) = \frac{(2 \pi)^2 }{ \sqrt{2 z \bar z  p_g^+ } }\Phi_{\alpha \alpha' \perp} ( {\bm x}_{12}) \; .
\end{gather}
Eq.~(\ref{Eq:GeneralInclusiveCross}) gives the full CGC cross-section. However, we are interested in the so-called correlation limit, i.e., the configuration in which the momenta of the individual quarks are much larger than their imbalance ${\bm q} = {\bm p}_{q} + {\bm p}_{\bar q}$. In this approximation
\begin{gather*}
  {\bm  P} \simeq {\bm p}_{q} \simeq - {\bm p}_{\bar q} \qquad {\rm and}  \qquad | {\bm P}| \gg |{\bm q}| \; \; ,
\end{gather*}
meaning that the quark and antiquark are almost back-to-back in the transverse plane. It is important to stress that, although in this leading-power (TMD) approximation, ${\bm P} \gg {\bm q}, Q_s$, saturation effects are still present when ${\bm q} \sim Q_s$. To extract the leading-power contribution to eq.~(\ref{Eq:GeneralInclusiveCross}), it is easier to write the cross-section in terms of variables that are conjugated to ${\bm P}$ and ${\bm q}$, rather than ${\bm p}_q$ and ${\bm p}_{\bar q}$. The correct transformation is
\beq
 {\bm u} = {\bm x}_1  - {\bm x}_{2} \; , \hspace{0.5 cm} {\bm u}' = {\bm x}_{1'} - {\bm x}_{2'} \; , \hspace{0.5 cm} {\bm v} = z {\bm x}_{1} + \bar{z} {\bm x}_{2} \; , \hspace{0.5 cm} {\bm v}' = z {\bm x}_{1'} + \bar{z} {\bm x}_{2'} \; ,
\eeq
whose inverse reads
\begin{gather*}
    {\bm x}_1 = {\bm v} + {\bar z} {\bm u} \; , \hspace{0.5 cm} {\bm x}_{1'} = {\bm v'} + {\bar z} {\bm u'} \; , \hspace{0.5 cm} {\bm x}_2 = {\bm v} - z {\bm u} \; , \hspace{0.5 cm} {\bm x}_{2'} = {\bm v}' - z {\bm u}' \; .
\end{gather*}
The full transformation has unit Jacobian, meaning
\begin{gather}
\int d^2 {\bm x}_{1} d^2 {\bm x}_{2} d^2 {\bm x}_{1'} d^2 {\bm x}_{2'} e^{-i {\bm x}_{11'} \cdot {\bm p}_{q} - i {\bm x}_{22'} \cdot {\bm p}_{\bar{q} } } \; ...  = \int d^2 {\bm u}  d^2 {\bm u}^\prime  d^2 {\bm v}  d^2 {\bm v}^\prime  e^{-i {\bm P} ({\bm u}-{\bm u}^\prime)  -i {\bm q}  ({\bm v}- {\bm v}^\prime) } \; ... \, .
\end{gather} 
The relevant quantity to expand in powers of $\boldsymbol{u}$ and
$\boldsymbol{u}'$ is
\begin{align}
S(\boldsymbol{v},\boldsymbol{v}',\boldsymbol{u},\boldsymbol{u}')
={}&
S^{(4)}_{q\bar q\bar q q}
(\boldsymbol{x}_1,\boldsymbol{x}_2,
 \boldsymbol{x}_{1'},\boldsymbol{x}_{2'})
-
S^{(3)}_{qg\bar q}
(\boldsymbol{v},\boldsymbol{x}_{1'},\boldsymbol{x}_{2'}) -
S^{(3)}_{qg\bar q}
(\boldsymbol{v}',\boldsymbol{x}_{2},\boldsymbol{x}_{1})
+
S^{(2)}_{gg}(\boldsymbol{v},\boldsymbol{v}') \; .
\label{Eq:SCombination}
\end{align}
As already explained, color coherence implies that every non-vanishing term in the Taylor expansion must
contain at least one power of $\boldsymbol{u}$ and one power of
$\boldsymbol{u}'$, i.e.
\begin{equation}
    S(\boldsymbol{v},\boldsymbol{v}',\boldsymbol{0},\boldsymbol{u}')
    =
    S(\boldsymbol{v},\boldsymbol{v}',\boldsymbol{u},\boldsymbol{0})
    =
    0 \; .
\label{Eq:ColorCoherenceS}
\end{equation}
The leading contribution is therefore of
$\mathcal{O}(uu')$. It is useful to make the cancellation manifest directly at the
amplitude level. Using
\begin{equation}
    \mathcal{U}_F(\boldsymbol{v})t^a
    \mathcal{U}_F^\dagger(\boldsymbol{v})
    =
    t^b\mathcal{U}_A^{ba}(\boldsymbol{v}),
\label{Eq:FundamentalAdjointRelation}
\end{equation}
we introduce
\begin{equation}
    \mathcal{C}^a(\boldsymbol{v},\boldsymbol{u})
    =
    \mathcal{U}_F(\boldsymbol{x}_1)t^a
    \mathcal{U}_F^\dagger(\boldsymbol{x}_2)
    -
    \mathcal{U}_F(\boldsymbol{v})t^a
    \mathcal{U}_F^\dagger(\boldsymbol{v}) \; .
\label{Eq:ColorAmplitudeDefinition}
\end{equation}
The target-dependent quantity in eq.~(\ref{Eq:SCombination}) can then
be written as
\begin{equation}
    S(\boldsymbol{v},\boldsymbol{v}',\boldsymbol{u},\boldsymbol{u}')
    =
    \frac{2}{N_c^2-1}
    \left\langle
    {\rm Tr}_F\left[
    \mathcal{C}^a(\boldsymbol{v},\boldsymbol{u})
    \mathcal{C}^{a\dagger}
    (\boldsymbol{v}',\boldsymbol{u}')
    \right]
    \right\rangle .
\label{Eq:SAmplitudeLevel}
\end{equation}
Equation~(\ref{Eq:ColorCoherenceS}) follows immediately from
\begin{equation}
    \mathcal{C}^a(\boldsymbol{v},\boldsymbol{0})=0 \; .
\end{equation}
Expanding the Wilson lines around $\boldsymbol{u}=\boldsymbol{0}$ gives
\begin{align}
    \mathcal{U}_F(\boldsymbol{x}_1)
    &=
    \mathcal{U}_F(\boldsymbol{v})
    +
    \bar z\,u^i
    \partial_i\mathcal{U}_F(\boldsymbol{v})
    +
    \mathcal{O}(u^2),
    \label{Eq:U1Expansion}
    \\
    \mathcal{U}_F^\dagger(\boldsymbol{x}_2)
    &=
    \mathcal{U}_F^\dagger(\boldsymbol{v})
    -
    z\,u^i
    \partial_i\mathcal{U}_F^\dagger(\boldsymbol{v})
    +
    \mathcal{O}(u^2).
    \label{Eq:U2Expansion}
\end{align}
It follows that
\begin{equation}
    \mathcal{C}^a(\boldsymbol{v},\boldsymbol{u})
    =
    u^i\mathcal{C}^a_i(\boldsymbol{v})
    +
    \mathcal{O}(u^2),
\label{Eq:ColorAmplitudeExpansion}
\end{equation}
where
\begin{equation}
    \mathcal{C}^a_i(\boldsymbol{v})
    =
    \bar z\,
    \big[\partial_i\mathcal{U}_F(\boldsymbol{v})\big]
    t^a\mathcal{U}_F^\dagger(\boldsymbol{v})
    -
    z\,
    \mathcal{U}_F(\boldsymbol{v})t^a
    \big[\partial_i\mathcal{U}_F^\dagger(\boldsymbol{v})\big].
\label{Eq:CiDefinition}
\end{equation}
Similarly,
\begin{equation}
    \mathcal{C}^{a\dagger}
    (\boldsymbol{v}',\boldsymbol{u}')
    =
    u^{\prime j}\,
    \overline{\mathcal{C}}^a_j(\boldsymbol{v}')
    +
    \mathcal{O}(u'^2),
\label{Eq:ConjugateColorAmplitudeExpansion}
\end{equation}
with
\begin{equation}
    \overline{\mathcal{C}}^a_j(\boldsymbol{v}')
    =
    \bar z\,
    \mathcal{U}_F(\boldsymbol{v}')t^a
    \big[\partial_j\mathcal{U}_F^\dagger(\boldsymbol{v}')\big]
    -
    z\,
    \big[\partial_j\mathcal{U}_F(\boldsymbol{v}')\big]
    t^a\mathcal{U}_F^\dagger(\boldsymbol{v}').
\label{Eq:CbarjDefinition}
\end{equation}
Substituting eqs.~(\ref{Eq:ColorAmplitudeExpansion}) and
(\ref{Eq:ConjugateColorAmplitudeExpansion}) into
eq.~(\ref{Eq:SAmplitudeLevel}), one obtains
\begin{equation}
    S(\boldsymbol{v},\boldsymbol{v}',\boldsymbol{u},\boldsymbol{u}')
    =
    u^i u^{\prime j}\,
    \mathcal{G}^{ij}(\boldsymbol{v},\boldsymbol{v}')
    +
    \mathcal{O}(u^2u',uu'^2),
\label{Eq:SFinalCorrelationLimit}
\end{equation}
where
\begin{equation}
    \mathcal{G}_{ij}(\boldsymbol{v},\boldsymbol{v}')
    =
    \frac{2}{N_c^2-1}
    \left\langle
    {\rm Tr}_F\left[
    \mathcal{C}^a_i(\boldsymbol{v})
    \overline{\mathcal{C}}^a_j(\boldsymbol{v}')
    \right]
    \right\rangle .
\label{Eq:GijAmplitudeRepresentation}
\end{equation}
Explicitly,
\begin{align}
\mathcal{G}_{ij}(\boldsymbol{v},\boldsymbol{v}')
={}&
\frac{2}{N_c^2-1}
\Bigg\langle
{\rm Tr}_F
\Bigg[
\bar z^2
\big(\partial_i\mathcal{U}_F(\boldsymbol{v})\big)t^a
\mathcal{U}_F^\dagger(\boldsymbol{v})
\mathcal{U}_F(\boldsymbol{v}')t^a
\big(\partial_j\mathcal{U}_F^\dagger(\boldsymbol{v}')\big)
\nonumber\\
&\hspace{0.4cm}
-z\bar z
\big(\partial_i\mathcal{U}_F(\boldsymbol{v})\big)t^a
\mathcal{U}_F^\dagger(\boldsymbol{v})
\big(\partial_j\mathcal{U}_F(\boldsymbol{v}')\big)t^a
\mathcal{U}_F^\dagger(\boldsymbol{v}')  \nonumber\\
&\hspace{0.4cm}
-z \bar z
\mathcal{U}_F(\boldsymbol{v})t^a
\big(\partial_i\mathcal{U}_F^\dagger(\boldsymbol{v})\big)
\mathcal{U}_F(\boldsymbol{v}')t^a
\big(\partial_j\mathcal{U}_F^\dagger(\boldsymbol{v}')\big)
\nonumber\\
&\hspace{0.4cm}
+z^2
\mathcal{U}_F(\boldsymbol{v})t^a
\big(\partial_i\mathcal{U}_F^\dagger(\boldsymbol{v})\big)
\big(\partial_j\mathcal{U}_F(\boldsymbol{v}')\big)t^a
\mathcal{U}_F^\dagger(\boldsymbol{v}')
\Bigg]
\Bigg\rangle .
\label{Eq:GijExpandedWilsonLines}
\end{align}
This quantity can be cast in the following form~\cite{Marquet:2017xwy}
\begin{gather}
 \mathcal{G}_{ij}(\boldsymbol{v},\boldsymbol{v}')
=
\frac{N_c}{2C_F}
\Bigg\{
\bar z^2
\left\langle
\big[
\partial_{v^i}\partial_{v^{\prime j}}
D(\boldsymbol{v},\boldsymbol{v}')
\big]
D(\boldsymbol{v}',\boldsymbol{v})
\right\rangle
+
z^2
\left\langle
D(\boldsymbol{v},\boldsymbol{v}')
\big[
\partial_{v^i}\partial_{v^{\prime j}}
D(\boldsymbol{v}',\boldsymbol{v})
\big]
\right\rangle
\nonumber\\
-
2z\bar z\,
{\rm Re}
\left\langle
\big[
\partial_{v^i}D(\boldsymbol{v},\boldsymbol{v}')
\big]
\big[
\partial_{v^{\prime j}}
D(\boldsymbol{v}',\boldsymbol{v})
\big]
\right\rangle \nonumber \\ 
+
\frac{1}{N_c^2}
\left\langle
\frac{1}{N_c}
{\rm Tr}_F\left[
\big(\partial_i\mathcal{U}_F(\boldsymbol{v})\big)
\mathcal{U}_F^\dagger(\boldsymbol{v})
\big(\partial_j\mathcal{U}_F(\boldsymbol{v}')\big)
\mathcal{U}_F^\dagger(\boldsymbol{v}')
\right]
\right\rangle
\Bigg\} \; ,
\label{Eq:GijFinal_MainTex}
\end{gather}
where
\begin{equation}
    D(\boldsymbol{x},\boldsymbol{y})
    =
    \frac{1}{N_c}
    {\rm Tr}_F\left[
    \mathcal{U}_F(\boldsymbol{x})
    \mathcal{U}_F^\dagger(\boldsymbol{y})
    \right] \; .
    \label{Eq:DipoleDefinition}
\end{equation}
is the dipole operator. Deriving eq.~(\ref{Eq:GijFinal_MainTex}) is far from being trivial\footnote{The comparison with Ref.~\cite{Marquet:2017xwy} is not easy at the intermediate step of the calculation, because we use a different convention for the Wilson lines. In particular, it should be noted that
\[
U_R^{\rm Ref.}(\boldsymbol{x})
=
\mathcal{U}_R^\dagger(\boldsymbol{x}) \; .
\]
}; for convenience of the reader more details are given in Appendix \ref{CorrLim&TMDs}.

\subsection{Spin density matrix of the inclusive dijet production in $pp$ or $pA$ with high-energy TMD effects}
Employing eq.~(\ref{Eq:SFinalCorrelationLimit}), we can finally write the leading-power ``spin-open" cross-section as 
\begin{gather*}
   \left[ \frac{d\sigma }{dz d^2 {\bm P}  d^2 {\bm q}} \right]_{\alpha\alpha'\beta\beta'} = \frac{   \alpha_s  }{ 2 (2 \pi)^4 } p_g^+  \int d^2 {\bm u} d^2 {\bm u'} e^{-i {\bm P}  ({\bm u} - {\bm u}')  } u^{i} u'^{ j} \psi^{  *}_{\beta\beta'\perp}( {\bm u}')  \psi_{\alpha\alpha'\perp}  ( {\bm u}) \nonumber \\ \times \int \frac{d^2 {\bm v} \;  d^2 {\bm v}'}{(2 \pi)^4} \mathcal{G}^{ij}( {\bm v} , {\bm v}') e^{-i {\bm q} ( {\bm v} -{\bm v}')}  
\end{gather*}
The integral over $u$ and $u'$ can be performed to obtain the hard spin-density matrix 
\begin{gather}
 \left[ \frac{\partial}{\partial  P^i } \psi_{\beta\beta'\perp}^{L/T(\lambda)*} ( {\bm P} ) \right]  \left[ \frac{\partial}{\partial P^j } \psi^{L/T(\lambda)}_{\alpha\alpha'\perp} ( {\bm P} ) \right] 
= \label{Eq:GeneralHardFactor} \\  \times \frac{1}{4} \xi_\alpha^\dagger \eta_{\alpha'}^\dagger  \left[\delta^{ij}\left(A^{(0)} \mathbb{I}_{4 \times 4}  +   \tilde{ C}^{(0)}_{ab} \sigma^a \otimes \sigma^b \right)   +\left(\frac{2P^i P^j}{{\bm P}^2}-\delta^{ij}\right)\left(A^{(2)}  \mathbb{I}_{4 \times 4} +   \tilde{C}^{(2)}_{ab} \sigma^a \otimes \sigma^b \right)\right]\xi_\beta \eta_{\beta'}  , \notag
\end{gather}
with the hard coefficients, calculable by means of Eqs.~(\ref{Eq:Cij})-(\ref{Eq:Ckk}). For the trace part, one has
\begin{gather}
    A^{(0)}
    =
    \frac{2(2\pi)^4}{p_g^+}
    \frac{D }
    {\left(|{\bm P}|^2+m^2\right)^4} \; ,
    \label{Eq:A0}
\end{gather}
\begin{gather}
    \tilde{C}_{nn}^{(0)}
    = -\frac{
        2z\bar z\left(|{\bm P}|^4+m^4\right)
    }{
        D 
    }  A^{(0)} \; ,
    \label{Eq:C0nn}
\end{gather}
\begin{gather}
    \tilde{C}_{rr}^{(0)}
    =
    -\frac{2z\bar z}{\Delta }
    \Big[
        m^6(1-2z)^2
        +m^4|{\bm P}|^2
        -m^2|{\bm P}|^4
        \left(1+4z\bar z\right)
        -|{\bm P}|^6
    \Big] A^{(0)} \; ,
    \label{Eq:C0rr}
\end{gather}
\begin{gather}
    \tilde{C}_{kk}^{(0)}
    =
    -\frac{1}{2\Delta }
    \Big\{
        m^6
        \left[
            (1-2z)^4+(1-2z)^2
        \right]
        +m^4|{\bm P}|^2
        \left[
            1+(1-2z)^2
        \right]
        \nonumber\\
        \hspace{2cm}
        +m^2|{\bm P}|^4
        \left[
            (1-2z)^4
            -5(1-2z)^2+2
        \right]
        -|{\bm P}|^6
        \left[
            1+(1-2z)^2
        \right]
    \Big\} A^{(0)} \; ,
    \label{Eq:C0kk}
\end{gather}
\begin{gather}
  \tilde{C}_{rk}^{(0)}
    =
    \tilde{C}_{kr}^{(0)}
    =
    -\frac{
        4z\bar z(1-2z)\,
        m\,|{\bm P}|^3
        \left(m^2+|{\bm P}|^2\right)
    }{
        \Delta 
    } A^{(0)} \; ,
    \label{Eq:C0rk}
\end{gather}
where, for compactness, we introduced
\begin{gather}
    D
    =
    \left(z^2+\bar z^2\right)
    \left(|{\bm P}|^4+m^4\right)
    +2m^2|{\bm P}|^2,
    \qquad
    \Delta
    =
    \left[m^2(1-2z)^2+|{\bm P}|^2\right]D \; .
\end{gather}
For the traceless part, we obtain 
\begin{gather}
    A^{(2)}
    =
    \frac{8(2\pi)^4}{p_g^+}
    \frac{z\bar z\,m^2|{\bm P}|^2}
    {\left(|{\bm P}|^2+m^2\right)^4} \; ,
    \label{Eq:A2}
\end{gather}

\begin{gather}
    \tilde C_{nn}^{(2)} = 
    \tilde{C}_{rr}^{(2)} =
    - \tilde{C}_{kk}^{(2)} = A^{(2)},
    \label{Eq:C2Diagonal}
\end{gather}

\begin{gather}
    \tilde{C}_{rk}^{(2)}
    =
    -\frac{m^2+|{\bm P}|^2}
    {2z\,m\,|{\bm P}|} A^{(2)},
    \qquad
    \tilde{C}_{kr}^{(2)}
    =
    \frac{m^2+|{\bm P}|^2}
    {2\bar z\,m\,|{\bm P}|} A^{(2)} \; .
    \label{Eq:C2OffDiagonal}
\end{gather}
These coefficients coincide with those obtained for DIS in
Ref.~\cite{Fucilla:2026mkg} in the real-photon limit, $Q^2=0$, for a
transversely polarized photon. This correspondence follows from color
coherence. Indeed, after coherently summing the contributions in which
the gluon splitting occurs before and after the interaction with the
shockwave, the non-Abelian color structure is entirely encoded in the
target correlators, while the spin dependence remains contained in the
hard splitting wave function. Up to the coupling and color factors, the
Dirac structure of the on-shell $g\to Q\bar Q$ splitting wave function is
then identical to that of $\gamma_T \to Q\bar Q$ in the photoproduction
limit. There are, however, two important differences with respect to DIS. The
first concerns the polarization content of the initial state. Since the
incoming collinear gluon has only physical transverse polarizations, the
resulting density matrix receives no longitudinal contribution. By
contrast, the experimentally measured DIS cross section generally
contains both transverse- and longitudinal-photon contributions. The second and most important difference lies in the target structure $\mathcal{G}^{ij}$. This latter can be generically decomposed into unpolarized-gluon and linearly-polarized-gluon contributions,
\begin{gather}
   \tilde{\mathcal{G}}^{ij} (z, {\bm q}) = \int \frac{d^2 {\bm v} \;  d^2 {\bm v}'}{(2 \pi)^4} \mathcal{G}^{ij}( {\bm v} , {\bm v}') e^{-i {\bm q} ( {\bm v} -{\bm v}')}  = \frac{\alpha_s }{4 C_F}  \left[\frac{\delta^{ij}}{2} {\cal F}_{g}(z, {\bm q}) +\left(\frac{q^i q^j}{ {\bm q}^2}-\frac{\delta^{ij}}{2}\right) {\cal H}_{g}(z, {\bm q} ) \right] \; ,
   \label{EqGtildeDecom}
\end{gather}
where the trace $\mathcal{F}_g \sim \tilde{\mathcal{G}}^{ij} \delta_{ij}$ defines the unpolarized
gluon distribution, whereas the traceless projection
${\cal H}_{g} \sim (q_iq_j/\boldsymbol{q}^{2}-\delta_{ij}/2) \tilde{\mathcal{G}}^{ij}$ defines its linearly polarized partner. The effective distributions $\mathcal{F}_g$ and $\mathcal{H}_g$ are combinations of the gluon TMDs~\cite{Marquet:2017xwy}
\begin{gather}
    {\cal F}_{g}(z, {\bm q} )
 =
\left(z^2+\bar z^2\right)
F_{gg}^{(1)}( {\bm q} )
+
2z\bar z\,F_{gg}^{(2)}( {\bm q} )
-
\frac{1}{N_c^2}F_{gg}^{(3)}( {\bm q} ) \; ,
\label{Eq:UnpolarDistr}
\end{gather}
\begin{gather}
    {\cal H}_{g}(z, {\bm q} ) =
\left(z^2+\bar z^2\right)
H_{gg}^{(1)}({\bm q} )
+
2z\bar z\,H_{gg}^{(2)}( {\bm q} )
-
\frac{1}{N_c^2}H_{gg}^{(3)}( {\bm q} ) \; ,
\label{Eq:LLpolarDistr}
\end{gather}
whose operator definitions and the MV-model parameterizations are given in the Appendix~\ref{CorrLim&TMDs}. \\

The final spin density matrix thus reads
\begin{gather*}
  \left[ \frac{d\sigma }{dz d^2 {\bm P}  d^2 {\bm q}} \right]_{\alpha\alpha'\beta\beta'} 
= \frac{\alpha_s^2 }{8 C_F} \frac{ p_g^+  }{(2 \pi)^4} \frac{\partial}{\partial  P^{'i} } \frac{\partial}{\partial P^j }  \psi^{  *}_{\beta\beta'}( {\bm P}' ) \psi_{\alpha\alpha'} ( {\bm P} ) \Big\arrowvert_{ {\bm P}' = {\bm P} } \nonumber \\ \times  \left[\frac{\delta_{ij}}{2} {\cal F}_{g}(z, {\bm q} ) +\left(\frac{q^i q^j}{ {\bm q}^2}-\frac{\delta_{ij}}{2}\right) {\cal H}_{g}(z, {\bm q} ) \right] \; , 
\end{gather*}
and can be finally expressed as
\begin{gather}
\left[ \frac{d\sigma }{dz d^2 {\bm P}  d^2 {\bm q}} \right]_{\alpha\alpha'\beta\beta'} = 
  \frac{\alpha_s^2 }{8 C_F} \frac{ p_g^+  }{(2 \pi)^4} \frac{1}{4} {\cal A} \; \xi^\dagger_\alpha \eta^\dagger_{\alpha'}\left(\mathbb{I}\otimes \mathbb{I}+{\cal C}_{ab}\sigma^a\otimes \sigma^b\right)\xi_\beta \eta_{\beta'} \; ,
   \label{Eq:General_Spin_Density_Matrix}  
\end{gather}
where 
\begin{gather}
  {\cal A} =   {\cal F}_{g}(z, {\bm q} )  A^{(0)}  + \cos (2 \phi_{P,q})  {\cal H}_{g}(z, {\bm q} )  A^{(2)} \; , \\  \vspace{0.2 cm}
  {\cal C}_{ab} = \frac{1}{{\cal A}}\left[ {\cal F}_{g}(z, {\bm q} ) \tilde{C}^{(0)} + \cos (2 \phi_{P,q}) {\cal H}_{g}(z, {\bm q} )  \tilde{C}^{(2)} 
  \right]_{ab} \; .
  \label{Eq:GenCoeffSpinDen}
\end{gather}
From Eqs.~(\ref{Eq:General_Spin_Density_Matrix}), (\ref{Eq:A0}) and (\ref{Eq:A2}), the total cross-section reads
\begin{gather}
    \left[ \frac{d\sigma }{dz d^2 {\bm P}  d^2 {\bm q}} \right] = \frac{\alpha_s^2 }{2 C_F} \frac{1}{\left(|{\bm P}|^2+m^2\right)^2}  \Bigg \{ \left[ \frac{\left(z^2+\bar z^2\right)}{2} +
    \frac{ 
     2 z \left(1 -  z\right) m^2|{\bm P}|^2 }
    {\left(|{\bm P}|^2+m^2\right)^2} \right] {\cal F}_{g}(z, {\bm q} ) \nonumber \\ + \cos (2 \phi_{P,q})  \frac{2 z\bar z\,m^2|{\bm P}|^2}
    {\left(|{\bm P}|^2+m^2\right)^2} {\cal H}_{g}(z, {\bm q} ) \Bigg \} \; ,
\end{gather}
in full agreement with Ref.~\cite{Marquet:2017xwy}. \\
\begin{figure}
    \centering
\includegraphics[width=0.47\linewidth]{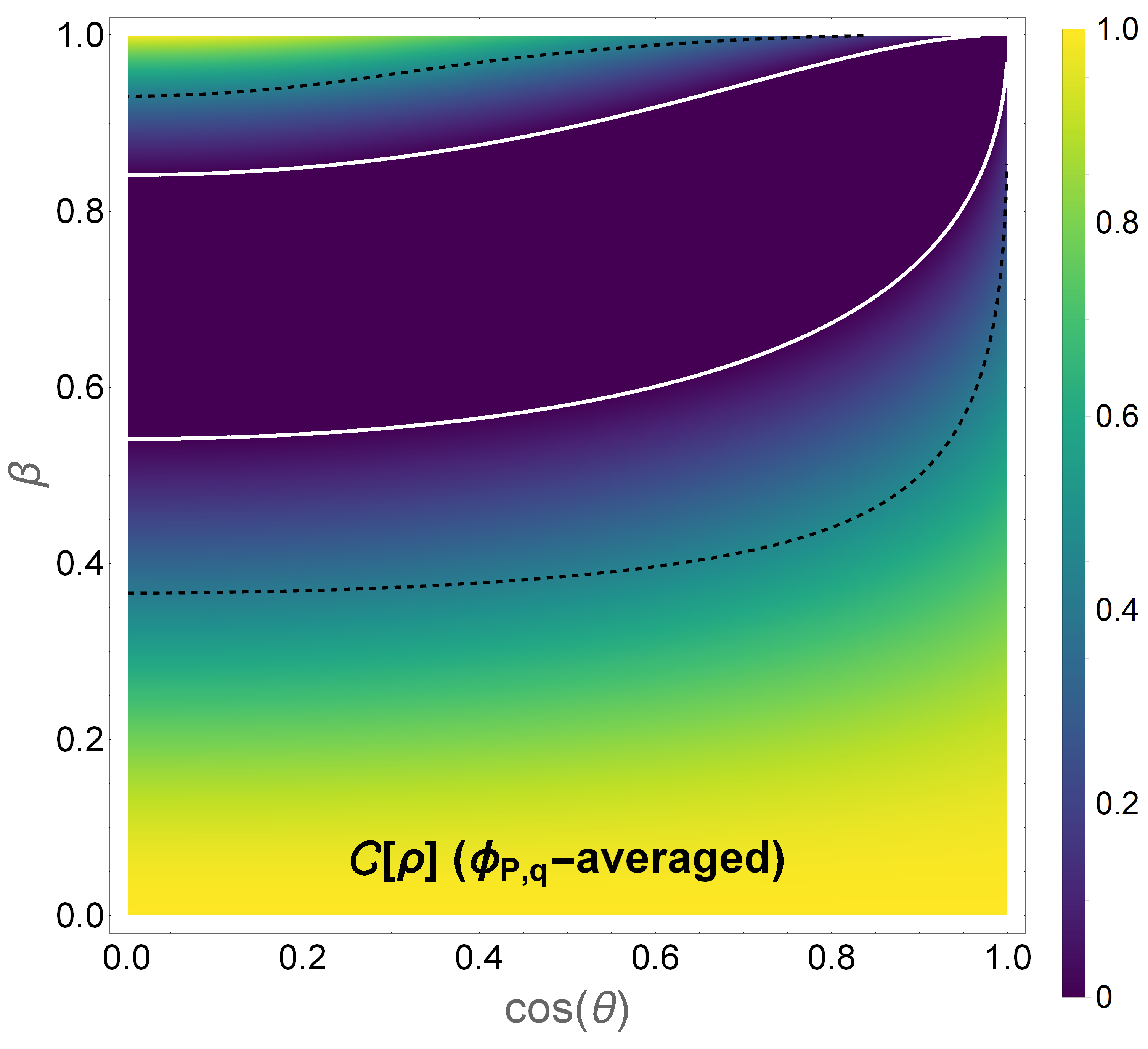} \hspace{0.5 cm}
\includegraphics[width=0.47\linewidth]{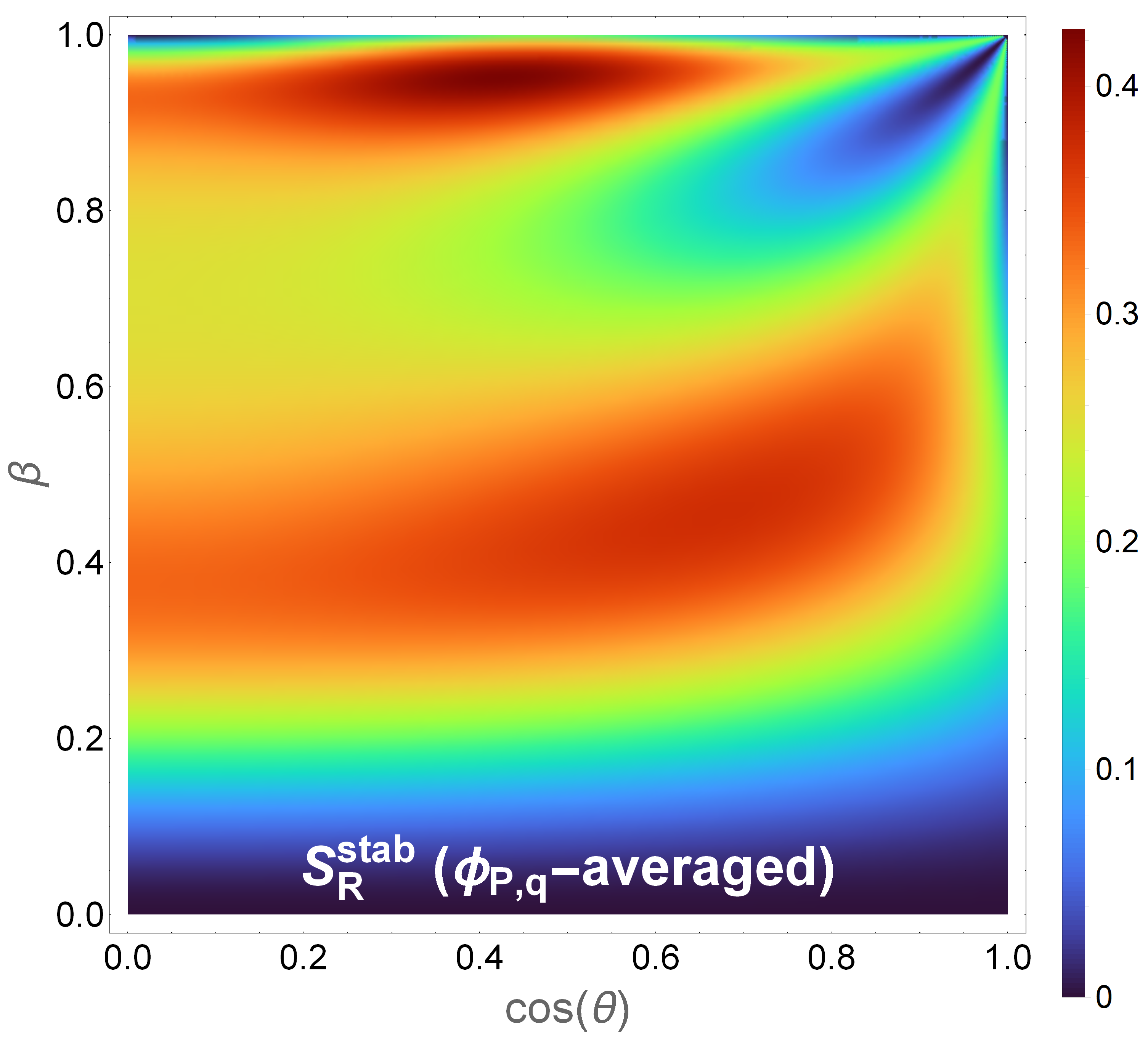}
\includegraphics[width=0.47\linewidth]{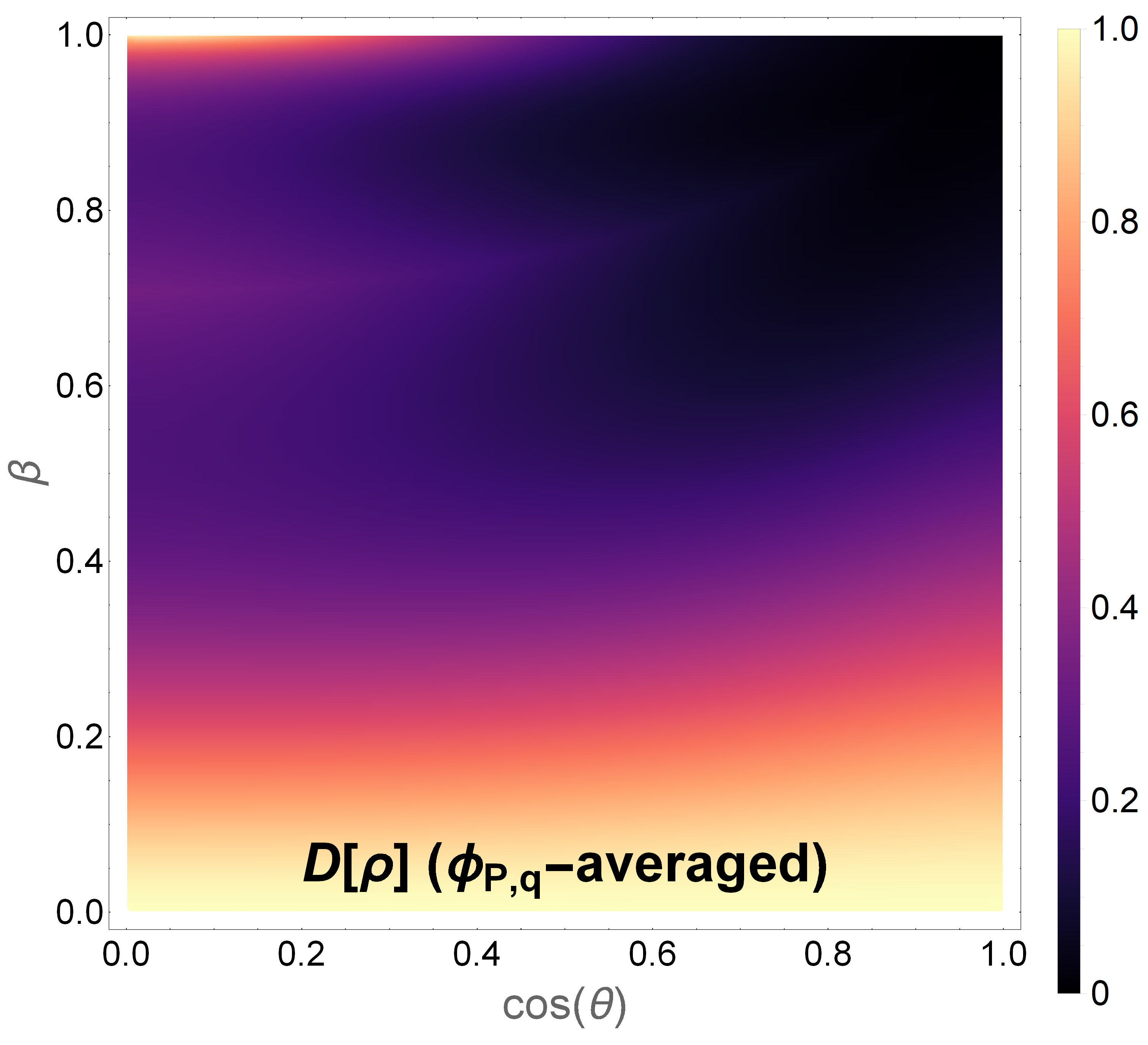}
    \caption{Concurrence (top left), stabilizer R\'enyi entropy (top right), and quantum discord (bottom) as functions of $\beta$ and $\cos\theta$, after averaging over $\phi_{P,q}$. In the concurrence panel, the solid white line separates entangled from separable states, while the dashed black line marks the boundary of the Bell-nonlocal region.}
    \label{fig:Triumvirate_Averaged}
\end{figure}

The general form of (\ref{Eq:General_Spin_Density_Matrix}) is the same as in the DIS case~\cite{Fucilla:2026mkg}. In the back-to-back limit, averaging over the relative azimuth
$\phi_{P,q}$ removes the contribution of linearly polarized gluons. The resulting normalized spin density matrix is the same in hadroproduction and photoproduction and agrees with the collinear-factorization result. At this accuracy, multiple eikonal scatterings can modify the azimuthally averaged cross section, but their dependence cancels from the normalized spin density matrix. As a benchmark of our numerical implementation, in Fig.~\ref{fig:Triumvirate_Averaged}, we reproduce the corresponding results for concurrence~\cite{Afik:2020onf,Afik:2022kwm}, stabilizer Rényi entropy~\cite{White:2024nuc}, and quantum discord~\cite{Afik:2022dgh}. These latter are studied as a function of $\beta$ and $\cos \theta$, which, in our Sudakov decomposition, reads
\begin{equation}
    \beta=\sqrt{1-\frac{4 z (1-z) m^2}{|\boldsymbol{P}|^2 + m^2 }}, \qquad \qquad \cos \theta = \frac{2z-1}{\beta} \; .
\end{equation}

For differential quantities in $\phi_{P,q}$, the density matrix is new, and the $\cos(2 \phi_{P,q})$ modulation of the various coefficients is governed by the ratio
\begin{gather}
    R_{{\cal H}_{g} / {\cal F}_{g}} (z, {\bm q}  ) \equiv \frac{ {\cal H}_{g} (z, {\bm q} )}{{\cal F}_{g} (z, {\bm q} ) } = \frac{\displaystyle \left(z^2+\bar z^2\right)
H_{gg}^{(1)}({\bm q} )
+
2z\bar z\,H_{gg}^{(2)}( {\bm q} )
-
\frac{1}{N_c^2}H_{gg}^{(3)}( {\bm q} )}{ \displaystyle \left(z^2+\bar z^2\right)
F_{gg}^{(1)}( {\bm q} )
+
2z\bar z\,F_{gg}^{(2)}( {\bm q} )
-
\frac{1}{N_c^2}F_{gg}^{(3)}( {\bm q} ) } \; .
\end{gather}

Indeed, the spin-density matrix coefficients offer a set of complementary\footnote{A standard approach to probing gluon saturation through dijet angular correlations is to study the azimuthal moment
$\langle \cos(2\phi_{P,q}) \rangle$~\cite{Marquet:2007vb,Boer:2009nc,Albacete:2010pg,Boer:2010zf,Dominguez:2011wm,Metz:2011wb,Dominguez:2011br,
Akcakaya:2012si,Dumitru:2015gaa,Hatta:2016dxp,Marquet:2016cgx,
Dumitru:2018kuw,Hatta:2020bgy,Hatta:2021jcd}.} observables to study the linearly polarized gluon distributions discussed in~\cite{Marquet:2017xwy}. The strength of the azimuthal modulation of the total cross section is
controlled by the ratio
\begin{equation}
    \Lambda(z, {\bm P} , {\bm q},m)
    =
    \frac{
        {\cal H}_{g}(z,{\bm q})A^{(2)} 
    }{
        {\cal F}_{g}(z,{\bm q})A^{(0)}
    } \; .
    \label{Eq:Lambda}
\end{equation}
In terms of this quantity, the spin-correlation coefficients are given
by
\begin{equation}
    {\cal C}_{ab}
    =
    \frac{
        C_{ab}^{(0)}
        +
        \Lambda(z, {\bm P} , {\bm q},m)\,
        C_{ab}^{(2)}
        \cos(2\phi_{P,q})
    }{
        1+
        \Lambda(z, {\bm P} , {\bm q},m)
        \cos(2\phi_{P,q})
    } \; .
    \label{Eq:SpinCorrelationLambda}
\end{equation}

As an illustrative example, let us consider the off-diagonal
correlation ${\cal C}_{kr}$ for a symmetric splitting, $z=\bar z=1/2$.
In this configuration, the azimuthally independent hard coefficient
vanishes,
\begin{equation}
    \left.C_{kr}^{(0)}\right|_{z=1/2}=0,
\end{equation}
and therefore
\begin{equation}
    \left.{\cal C}_{kr}\right|_{z=1/2}
    =
    \frac{
        \Lambda(z, {\bm P} , {\bm q},m) \,
        C_{kr}^{(2)}\big |_{z=1/2}\,
        \cos(2\phi_{P,q})
    }{
        1+
        \Lambda(z, {\bm P} , {\bm q},m)
        \cos(2\phi_{P,q})
    } \; .
    \label{Eq:CkrLambda}
\end{equation}
The remaining hard coefficient reads
\begin{equation}
    C_{kr}^{(2)}\big |_{z=1/2}
    =
    \frac{m^2+|{\bm P}|^2}
         {m|{\bm P}|} \; .
\end{equation}
In particular, for $|{\bm P}|=m$, one has
\begin{equation}
    C_{kr}^{(2)} \big |_{\substack{z=1/2\\|{\bm P}|=m}} =2,
\end{equation}
so that
\begin{equation}
    \left.{\cal C}_{kr}\right|_{\substack{z=1/2\\|{\bm P}|=m}}
    =
    \frac{
        2\Lambda\cos(2\phi_{P,q})
    }{
        1+\Lambda\cos(2\phi_{P,q})
    }
    =
    2\Lambda\cos(2\phi_{P,q})
    +{\cal O}(\Lambda^2) \; .
    \label{Eq:CkrOptimal}
\end{equation}

This off-diagonal correlation thus provides an additional channel for probing the linearly polarized gluon distributions. In particular, it has no azimuthally independent background at $z=1/2$, and its leading angular modulation is enhanced by the hard coefficient $C_{kr}^{(2)}=2$ when $|{\bm P}|=m$. \\

\subsection{BFKL vs saturation}
The difference between collinear and $k_T$ factorization (in the back-to-back limit) for these observables lies in the contribution of linearly polarized gluons~\cite{Fucilla:2026mkg}, which generate the azimuthal modulation. However, there is a profound difference in how the linear BFKL formalism~\cite{Fadin:1975cb,Kuraev:1976ge,Kuraev:1977fs,Balitsky:1978ic} and the saturation formalism~\cite{McLerran:1993ka,Balitsky:1995ub} treat linearly polarized gluons. Interestingly, in the BFKL limit, gluons are maximally linearly polarized, and one has~\cite{Altinoluk:2021ygv}
\begin{gather*}
    F_{gg}^{(1)}( {\bm q} ) = F_{gg}^{(3)}( {\bm q} ) = H_{gg}^{(1)}( {\bm q} ) = H_{gg}^{(3)}( {\bm q} ) \equiv F^{\rm BFKL} ( {\bm q} )  \; , \vspace{0.3 cm} \\
    F_{gg}^{(2)}( {\bm q} ) = H_{gg}^{(2)}( {\bm q} ) = 0 \; .
\end{gather*}
Therefore, the relevant ratio is  
\begin{gather*}
    R_{{\cal H}_{g} / {\cal F}_{g}}^{\rm BFKL} (z, {\bm q} ) = 1 \; .
\end{gather*}
\begin{figure}
    \centering    \includegraphics[width=0.75\linewidth]{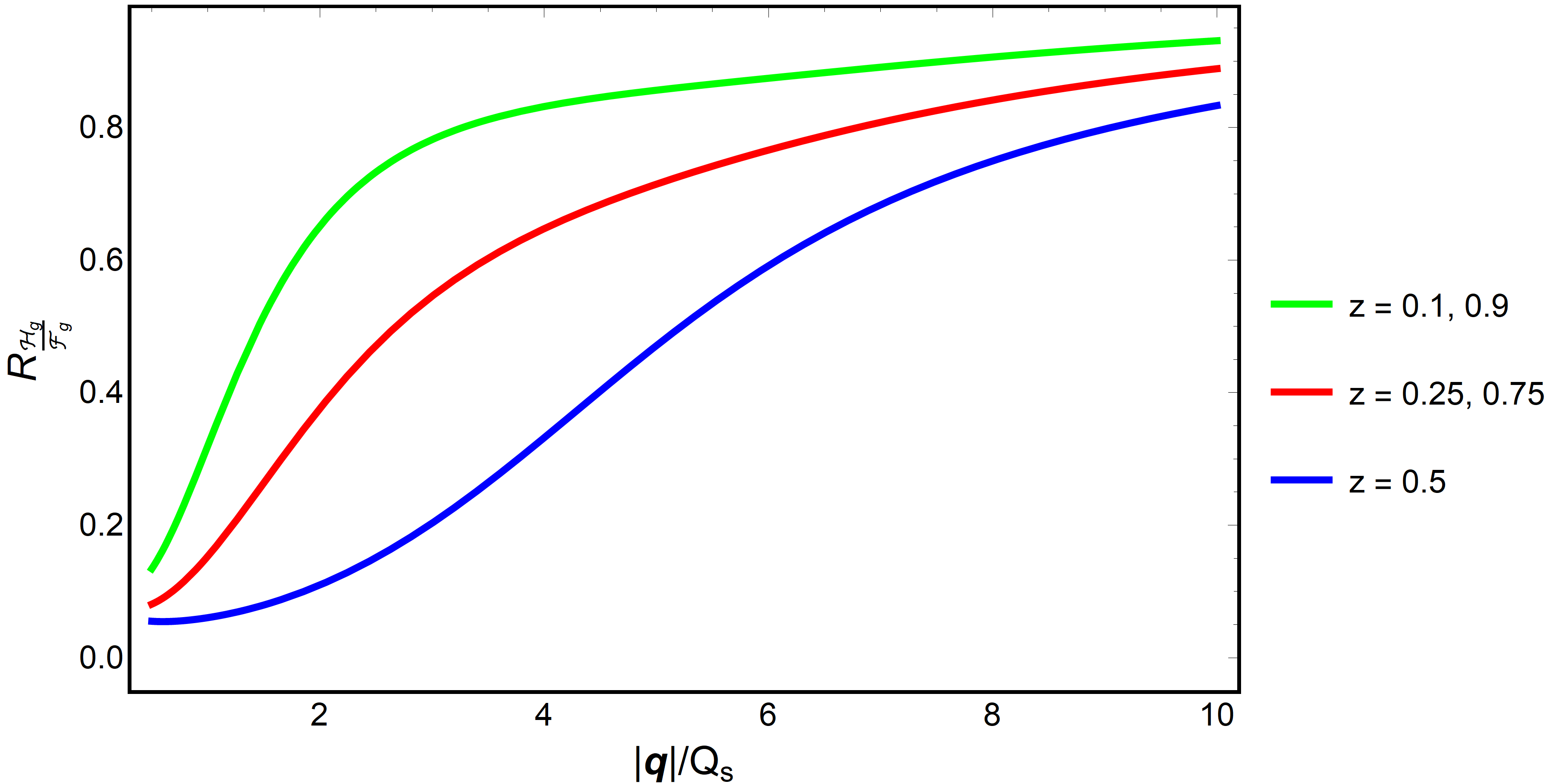}
    \caption{The ratio $R_{\mathcal H_g/\mathcal F_g}\equiv\mathcal H_g/\mathcal F_g$ as a function of $|\boldsymbol q|/Q_s$ for $z=0.1$ (green), $z=0.25$ (red), and $z=0.5$ (blue). The curves for $z$ and $1-z$ coincide.}
    \label{Fig:Relevant_Ratio}
\end{figure}
In contrast, in a more general CGC treatment, $R_{{\cal H}_{g} / {\cal F}_{g}} (z, {\bm q} )$ can be significantly smaller than 1 for ${\bm q} \sim Q_s $, with the dilute (BFKL) behaviour recovered only for $ {\bm q} \gg Q_s $. The behavior of the ratio as a function of $|\boldsymbol{q}| / Q_s$ within the McLerran-Venugopalan model, as described in Appendix~\ref{Sec:TMDs&MV_model_description}, is shown in Fig.~\ref{Fig:Relevant_Ratio}. Therefore, the coefficients in Eqs.~(\ref{Eq:GenCoeffSpinDen}) provide a new set of observables for probing gluon saturation in $pA$ collisions. We will show this explicitly in the next sections by estimating how saturation affects the entanglement, non-stabilazerness, and quantum discord patterns between the quark and the anti-quark. 
An interesting fact is that the modulation of the quantum-information quantities in DIS is governed by the ratio between the standard linearly polarized and unpolarized Weizsäcker-Williams gluon distributions~\cite{Fucilla:2026mkg}
\begin{gather*}
    R_{\rm DIS} ({\bm q}  ) = \frac{ {\cal H}_{g}^{(3)} ({\bm q} )}{{\cal F}_{g}^{(3)} ( {\bm q} ) } \; .
\end{gather*}
In inclusive $pA$ collisions, the ratio acquires a nontrivial $z$ dependence and results from the interplay among six gluon TMDs. The standard Weizs\"acker--Williams distributions, which entirely drive the corresponding DIS result, enter here with an explicit $1/N_c^2$ suppression. This color counting should, however, be interpreted with caution, since the different TMDs can have substantially different magnitudes and cancellations may occur among the leading-color contributions, making the Weizs\"acker--Williams terms numerically relevant~\cite{Marquet:2017xwy}. \\ 

\subsection{Concurrence, nonstabilizerness and quantum discord as a probe of gluon saturation}

\begin{figure}
   \includegraphics[width=0.97\linewidth]{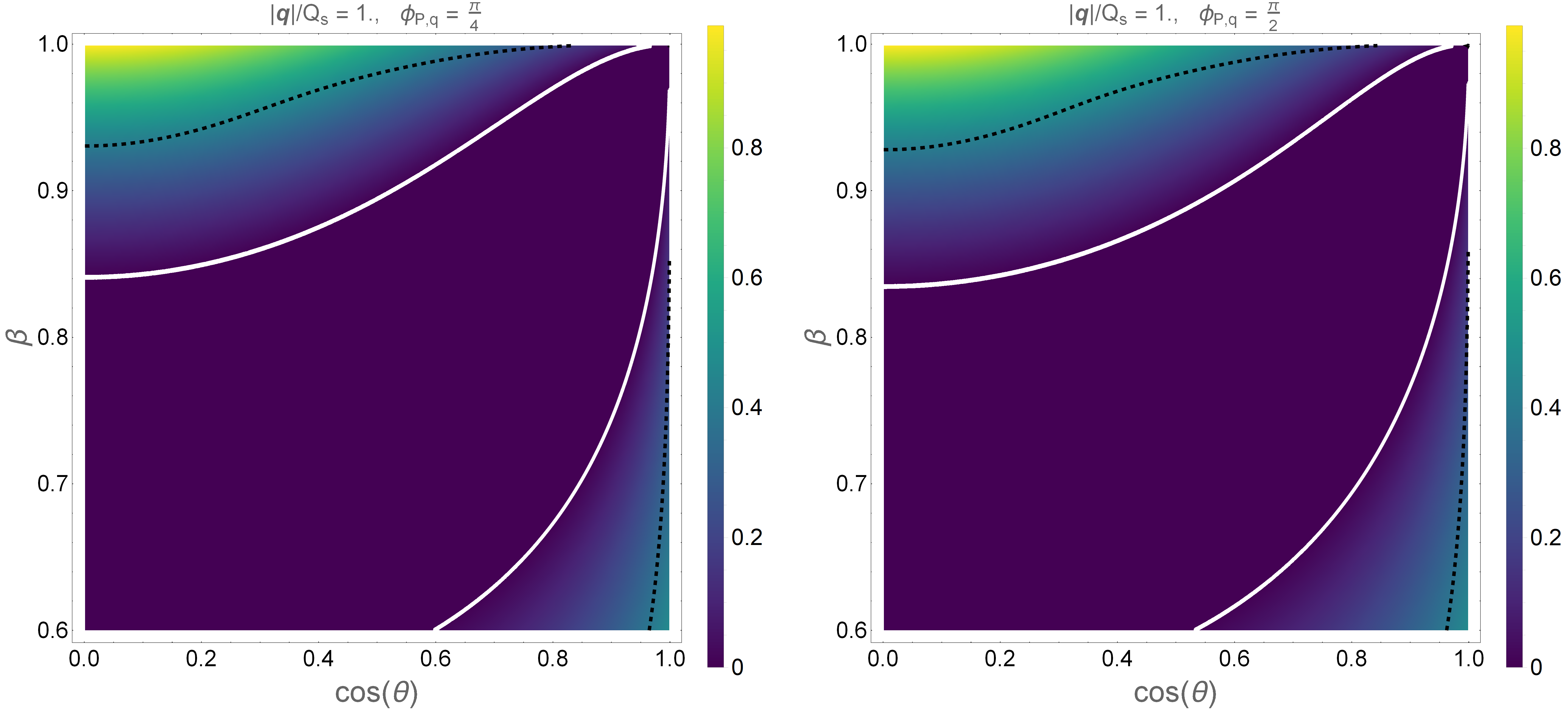} \vspace{0.5 cm} \\
   \includegraphics[width=0.475\linewidth]{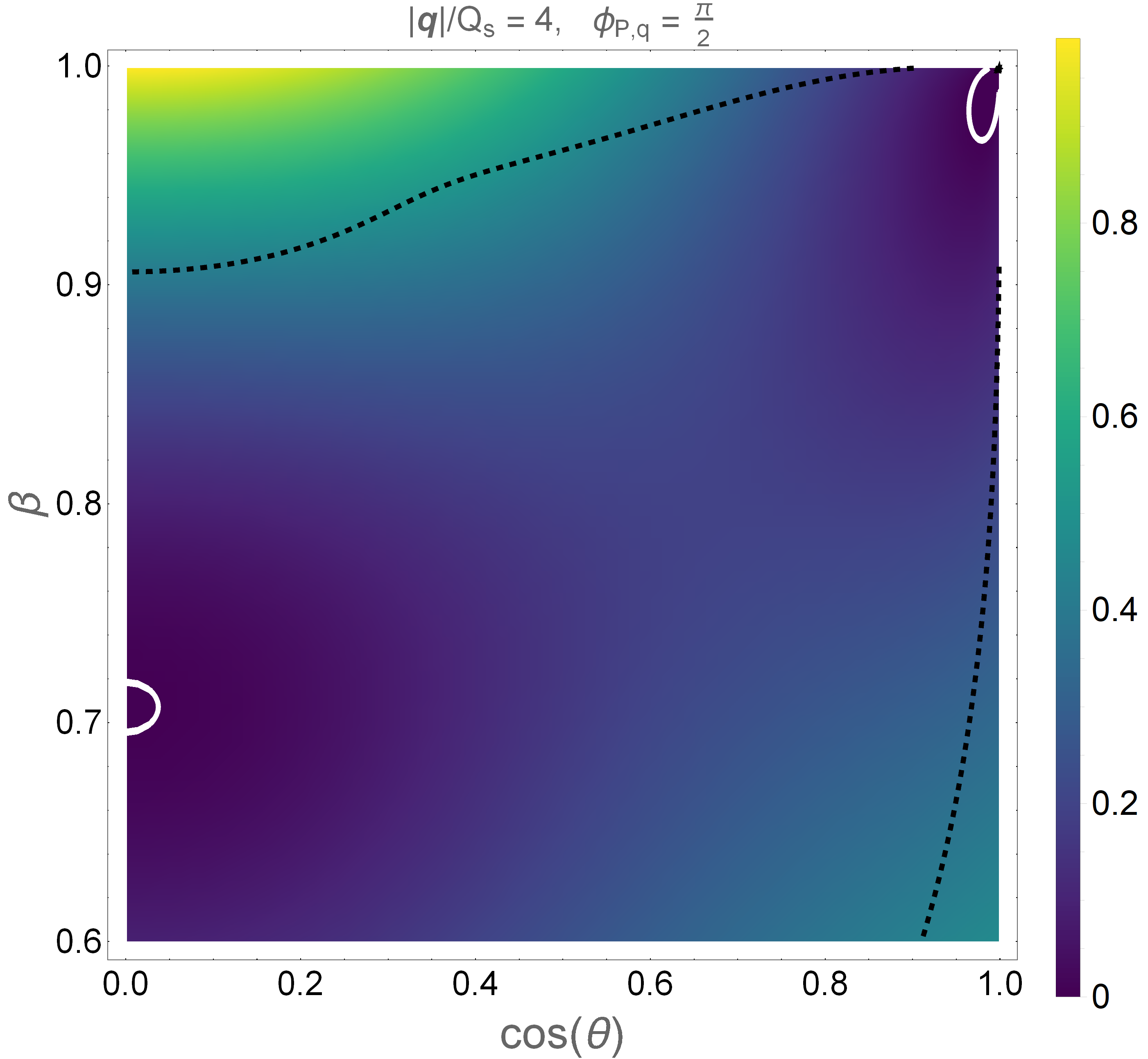} \hspace{0.06 cm}
   \includegraphics[width=0.475\linewidth]{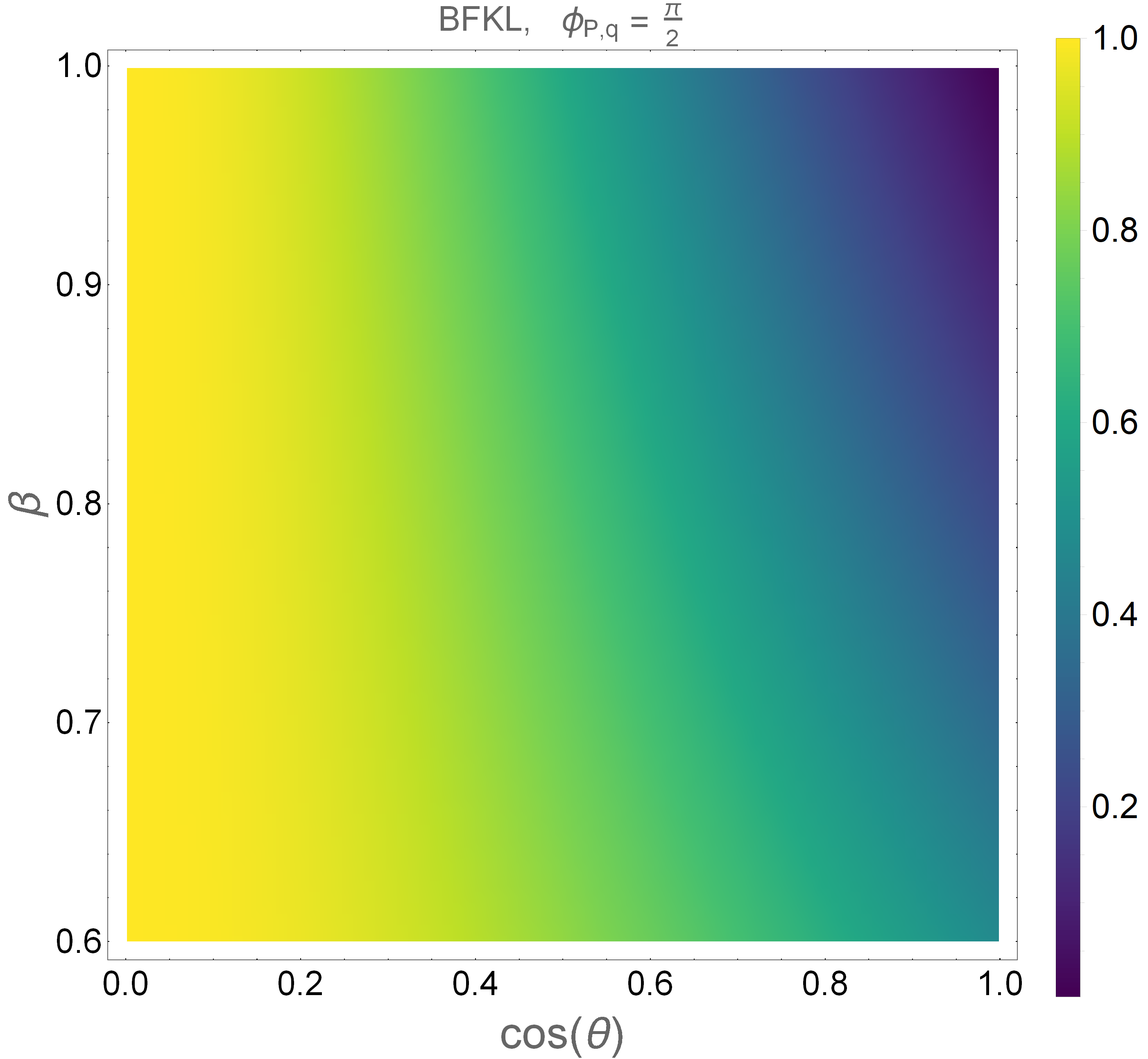}
    \caption{Concurrence for different values of $\phi_{P,q}$ and $|\boldsymbol{q}|/Q_s$.}
    \label{fig:ConcurrenceBFKLvsSat}
\end{figure}

\begin{figure}
   \includegraphics[width=0.97\linewidth]{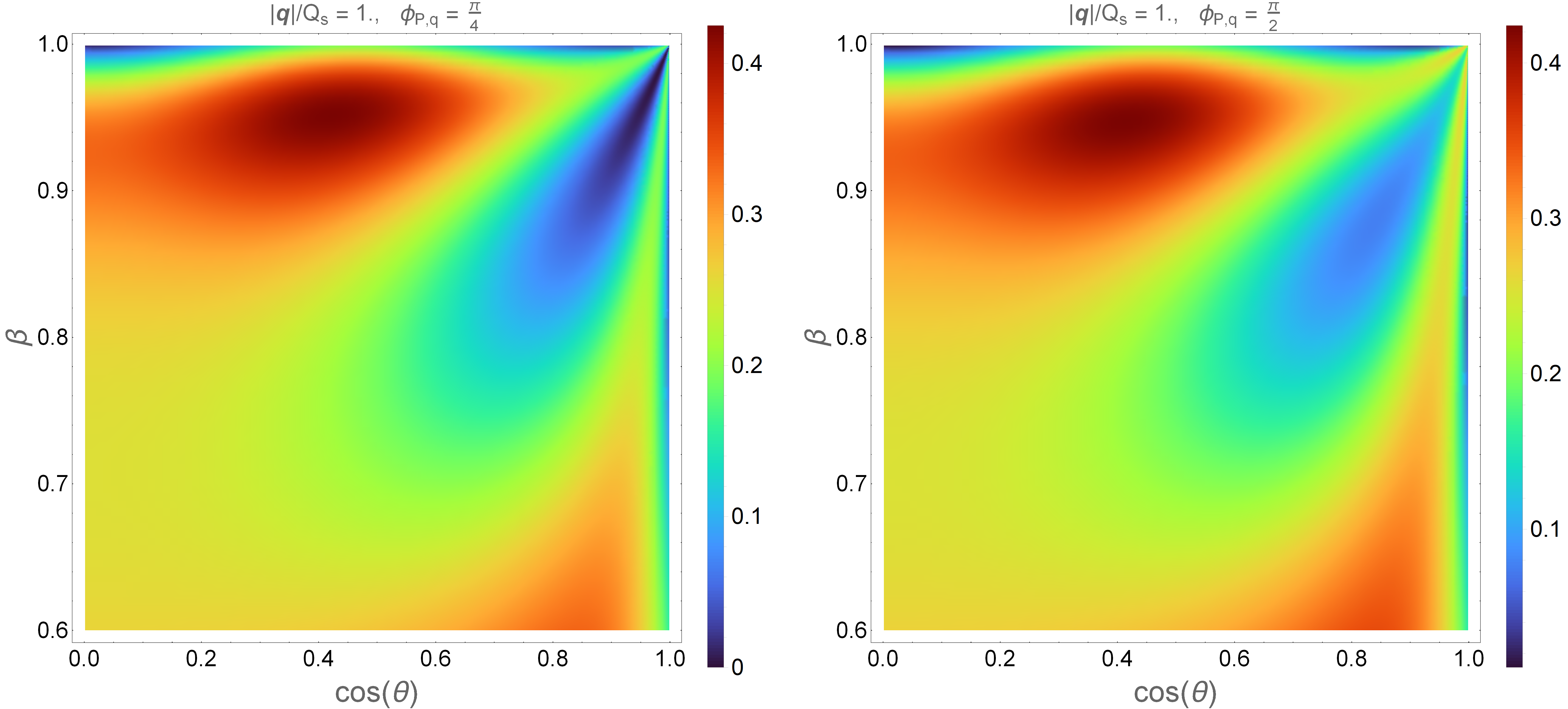} \vspace{0.5 cm}\\
   \includegraphics[width=0.475\linewidth]{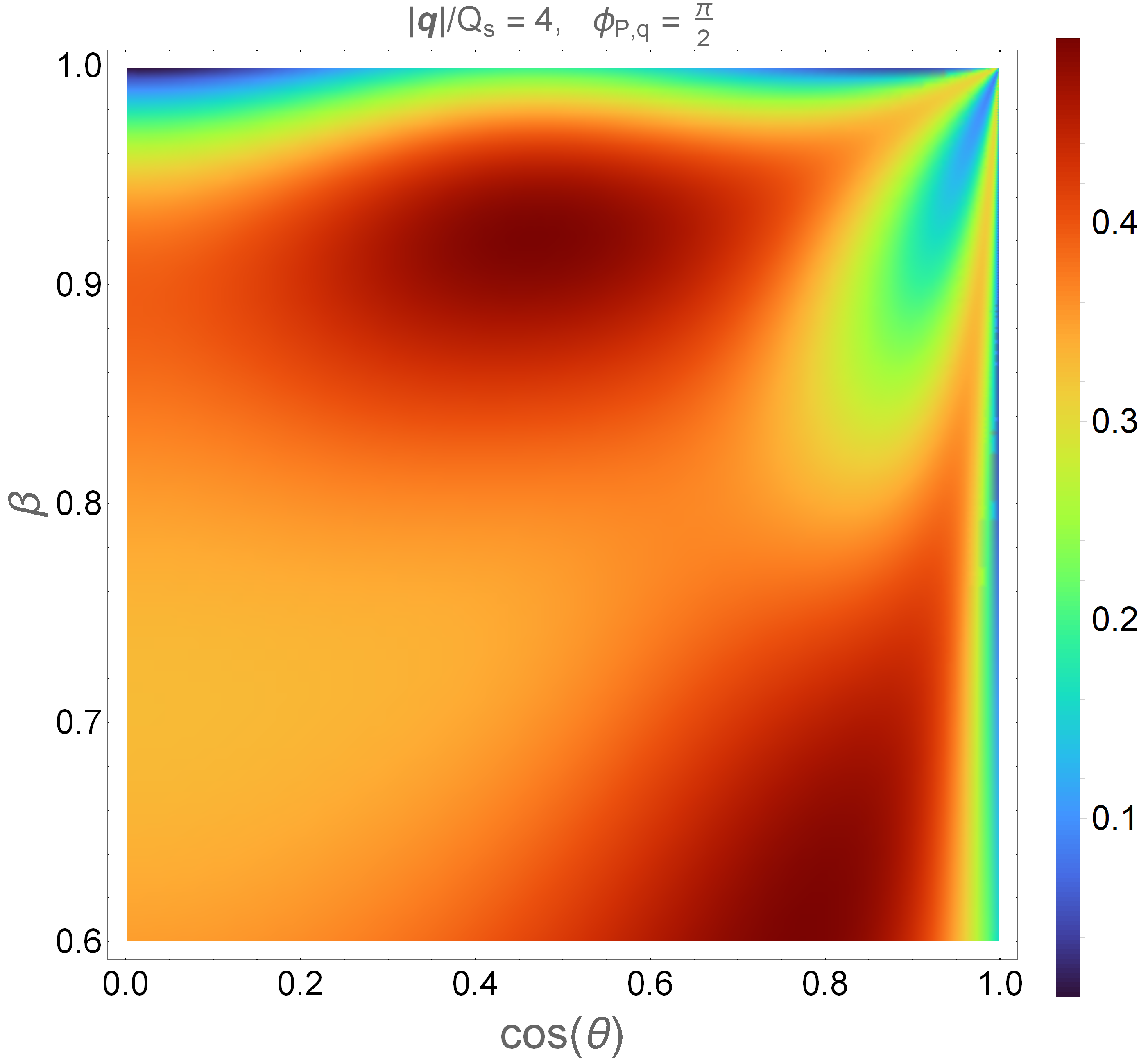} \hspace{0.06 cm}
   \includegraphics[width=0.475\linewidth]{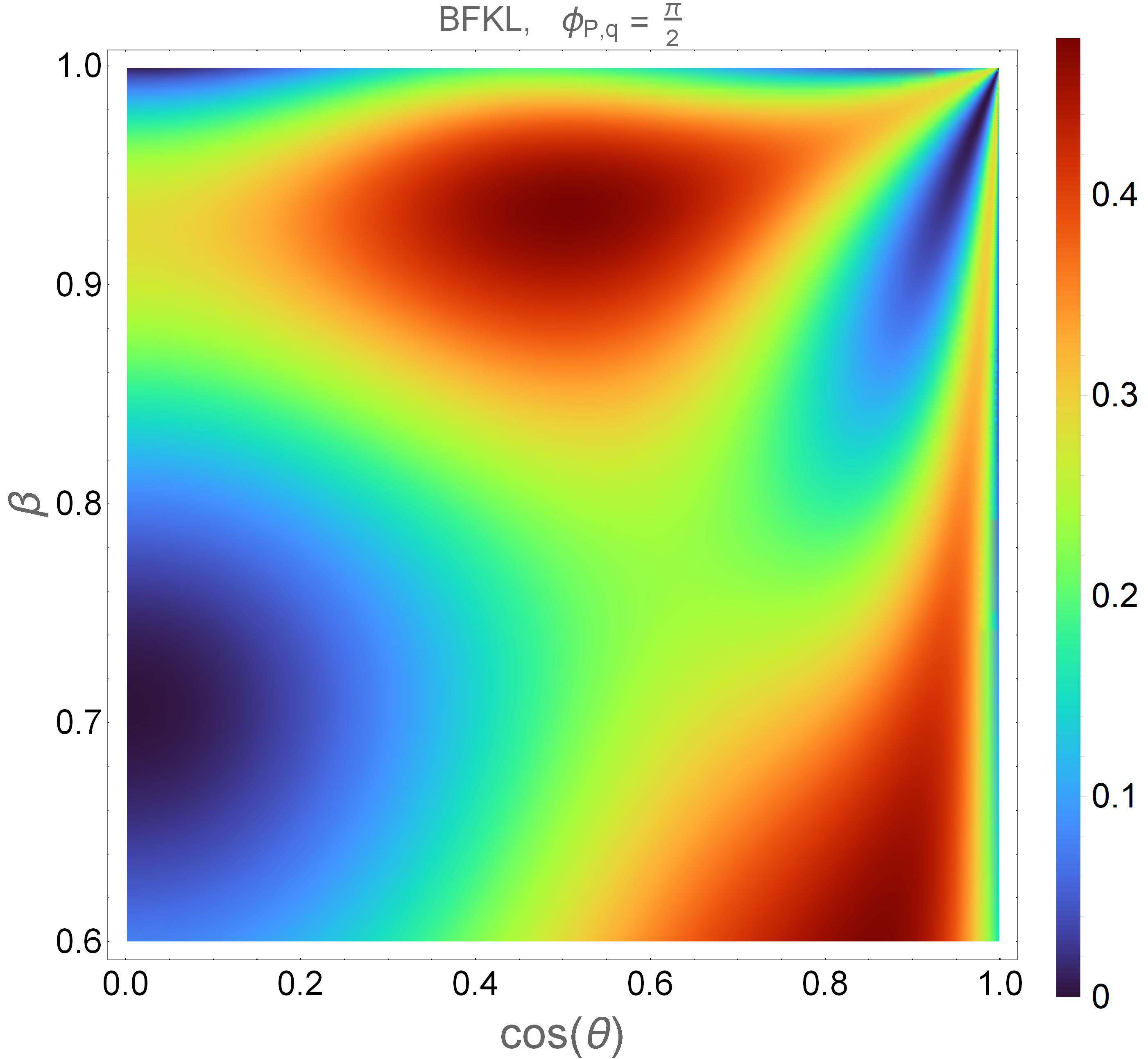}
    \caption{Magic for different values of $\phi_{P,q}$ and $|\boldsymbol{q}|/Q_s$.}
    \label{fig:MagicBFKLvsSat}
\end{figure}

\begin{figure}
   \includegraphics[width=0.97\linewidth]{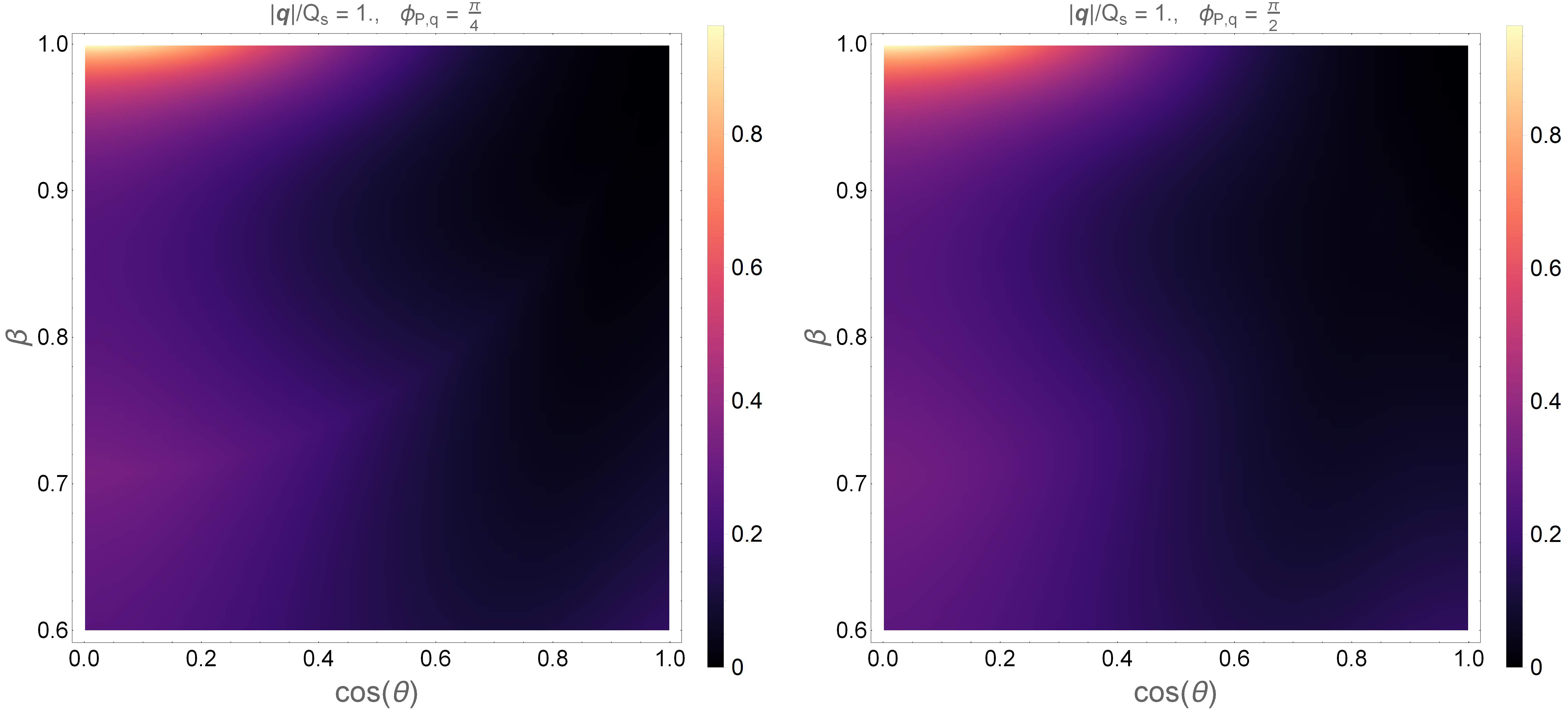} \vspace{0.5 cm}\\
   \includegraphics[width=0.475\linewidth]{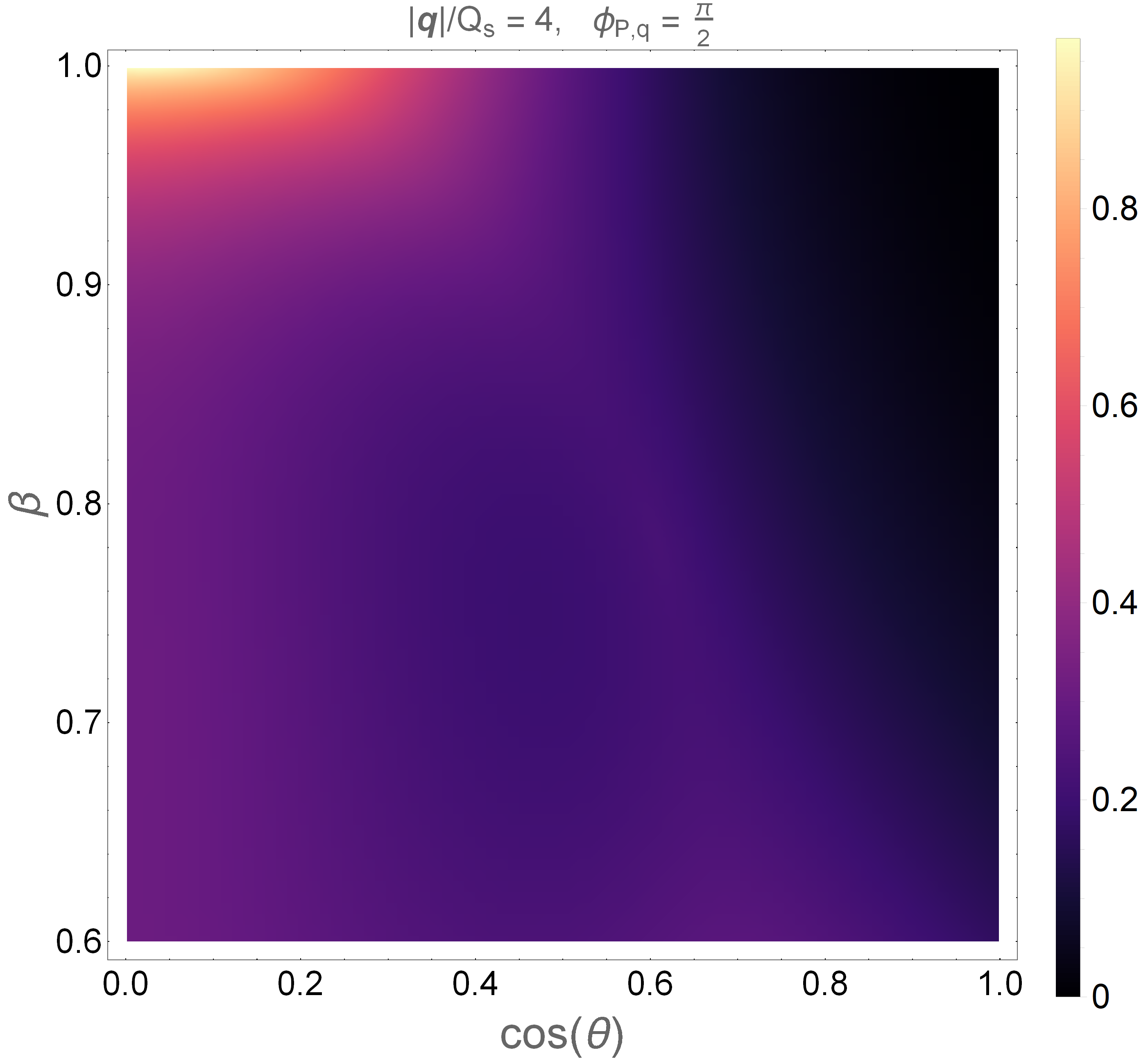} \hspace{0.1 cm}
   \includegraphics[width=0.475\linewidth]{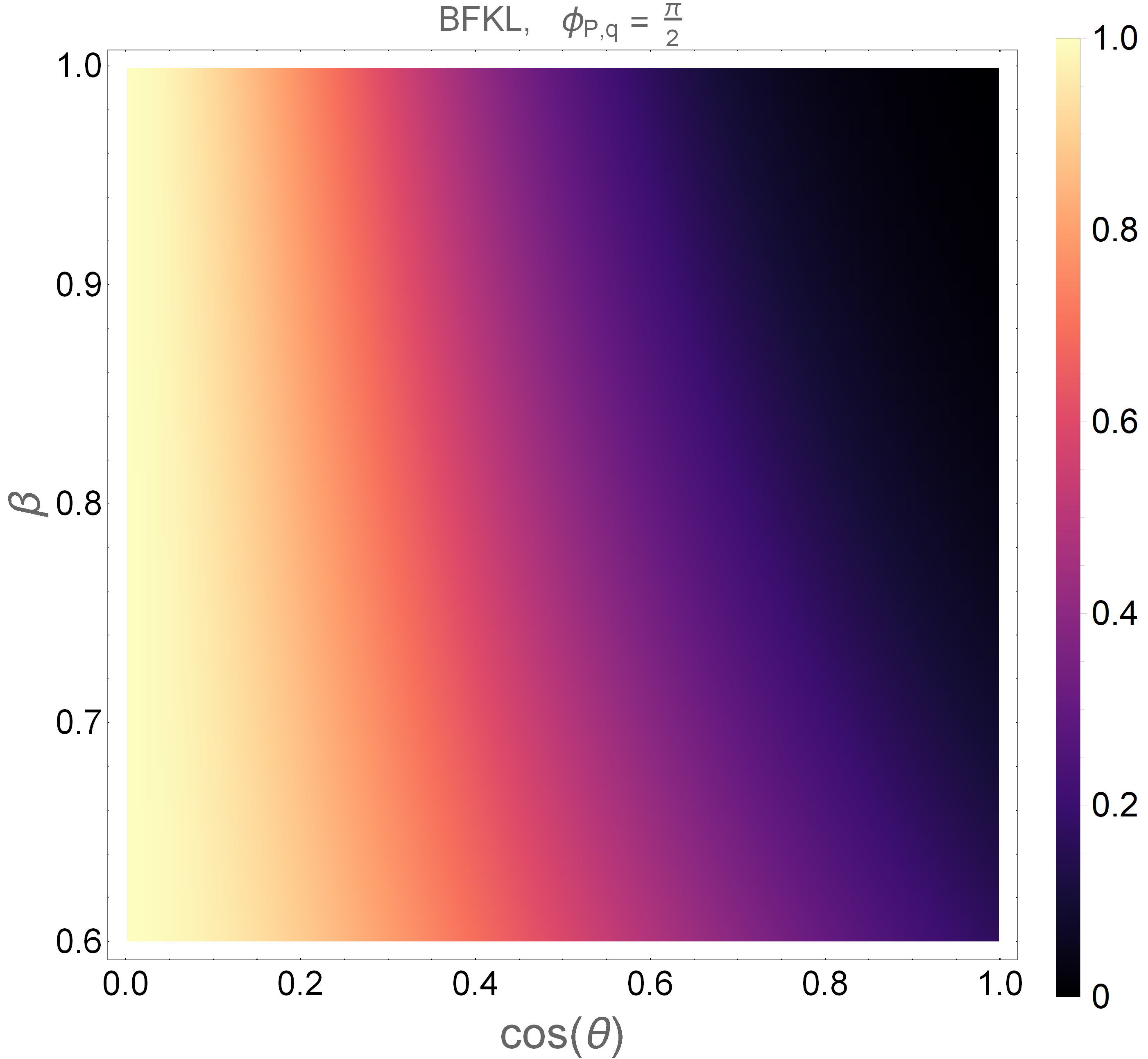}
    \caption{Quantum discord for different values of $\phi_{P,q}$ and $|\boldsymbol{q}|/Q_s$.}
    \label{fig:DiscordBFKLvsSat}
\end{figure}
In this section, we investigate the effects of linearly polarized gluons on concurrence, nonstabilizerness, and quantum discord. All plots cover the range $0.6\leq\beta<1$, focusing on azimuthal modulations analogous to those found in DIS~\cite{Fucilla:2026mkg}. From a kinematic 
standpoint, forward $b\bar b$\footnote{This is also possible for charm quarks, although the accessible range of $\beta$ is narrower.} production in proton--lead collisions\footnote{In Sec.~\ref{sec:CollinsPA}, we discuss two different strategies for extracting the spin-density matrix.} at the LHC offers a promising setting for exploring these effects over a substantial range of $\beta$. The relevant kinematics must satisfy both the back-to-back condition $|\boldsymbol q|\ll|\boldsymbol P|$ and the small-$x$ requirement\footnote{However note that in the case of a very large nucleus one can apply the present calculation for moderately small-$x$.} $x_A\simeq\mathcal M^2/(x_g s_{pA})\ll1$, where $\sqrt{s_{pA}}$ is the proton--nucleon center-of-mass energy. For top-quark pairs, the larger invariant mass at fixed $\beta$ makes these requirements more restrictive at the LHC; a future programme at the FCC would substantially extend the accessible kinematic range. \\

For each observable, we present four panels in Figs.~\ref{fig:ConcurrenceBFKLvsSat},
\ref{fig:MagicBFKLvsSat}, and \ref{fig:DiscordBFKLvsSat}, with the following kinematic configurations:
\begin{itemize}
    \item[\textbullet] The top-left panels correspond to $\phi_{P,q}=\pi/4$, for which $\cos(2\phi_{P,q})=0$, and therefore reproduce the collinear-factorization results in the kinematic window considered. One can shown that the results remain nearly unchanged throughout $\phi_{P,q}\in[0,\pi/4]$, where $\cos(2\phi_{P,q})\geq0$. Thus, the effect of linearly polarized gluons is small when the two contributions in both the numerator and the denominator of eq.~\eqref{Eq:GenCoeffSpinDen} have the same sign. The next three panels are chosen, in fact, to flip the sign of the cosine and maximize the effect of the polarized gluons by fixing $\phi_{P,q}=\pi/2$.
    \item[\textbullet] The down-left panels correspond to $\phi_{P,q}=\pi/2$ and $Q_s = 4 |\boldsymbol{q}|$.
    In this way, we study the effect of linearly polarized gluons in a transition region between the non-linear and linear regimes. 
    \item[\textbullet] The top-right panels correspond to $\phi_{P,q}=\pi/2$ and $Q_s = |\boldsymbol{q}|$, i.e. a deeply saturated regime.
    \item[\textbullet] Finally, the down-right panels correspond to $\phi_{P,q}=\pi/2$ within the BFKL approximation.
\end{itemize}
We first comment on the general pattern of the three quantities in Figs.~\ref{fig:ConcurrenceBFKLvsSat},
\ref{fig:MagicBFKLvsSat}, and \ref{fig:DiscordBFKLvsSat}, which, for the concurrence and nonstabilizerness, follows that already observed in DIS~\cite{Fucilla:2026mkg}. However, the effect of linearly polarized gluons is stronger for a given $| \boldsymbol{q}|/Q_s$ value and tends to alter the plots to a greater extent when $| \boldsymbol{q}| \geq 2$. A notable fact is that BFKL predicts a radical change in the predictions at $\phi_{P,q}=\pi/2$, which is, much like in the case of collinear factorization, completely independent of the target and the kinematic condition $|\boldsymbol{q}|/Q_s$. A saturation model, instead, in addition to removing this effect when $|\boldsymbol{q}|/Q_s \sim 1$, predicts that it remains milder even in a region where $ |\boldsymbol{q}|$ is already quite larger than $Q_s$.  Although the prediction of saturation depends on the target and is therefore influenced by our model (Appendix~\ref{Sec:TMDs&MV_model_description})—which is strictly valid only for a heavy nucleus—a precise determination of the ratio shown in Fig.~\ref{Fig:Relevant_Ratio} for a proton is expected to exhibit qualitatively similar behavior. The approach to the perturbative regime is expected to be faster for a proton. However, accessing proton saturation requires quite small $x$, making JIMWLK evolution of the gluon TMDs~\cite{Dominguez:2011gc,Dominguez:2011br,Dumitru:2015gaa,Marquet:2016cgx,Marquet:2017xwy,Cougoulic:2026drr,Cougoulic:2026sww} essential for reliable predictions.\footnote{For a large nucleus, saturation effects can already be significant at moderately small $x$. Moreover, since the observable is a ratio, evolution effects may partially cancel, reducing its sensitivity to small-$x$ evolution.} This development is left for future work. 

We now comment on the behavior of the three distinct quantities more specifically. The concurrence shown in the top-right of Fig.~\ref{fig:ConcurrenceBFKLvsSat} is characterized by a broad region of separability, interrupted by regions of entanglement (and Bell nonlocality) in the left-top and right-down corners. The concurrence tends to increase when $\{\beta, \cos(\theta)\} \rightarrow \{1, 0\}$ or $\{\beta, \cos(\theta)\} \rightarrow \{0.6, 1\}$. BFKL predicts (in the bottom-right panel of Fig.~\ref{fig:ConcurrenceBFKLvsSat}), at $\phi_{P,q}=\pi/2$, a complete transformation of the behavior, with the entire space characterized by positive concurrence, and thus entanglement. Saturation also predicts an increase in entanglement toward the dilute regime, which becomes numerically significant for $|\boldsymbol{q}| \geq 2 Q_s$ but, as already stated, is much milder than BFKL even at $|\boldsymbol{q}| \geq 4 Q_s$. The modulation is washed out for $|\boldsymbol{q}| < 2 Q_s$. 

The behavior of nonstabilizerness in Fig.~\ref{fig:MagicBFKLvsSat} exhibits a pattern similar to concurrence, though with some differences. Saturation implies that, as one moves toward the linear regime, the magic increases (most notably in the small-$\beta$ regions of the plot). However, this phenomenon is not constant; as saturation effects are more and more washed away, bubbles of zero magic appear. This is clearly visible within the BFKL approximation of the down-left panel of Fig.~\ref{fig:MagicBFKLvsSat}. Also, for the nonstabilizerness, the effect is washed out for $|\boldsymbol{q}| < 2 Q_s$. Interestingly, when $ 0< \cos \theta < 0.2 $, BFKL predicts maximally entangled states that can be either stabilizer states or states with a substantial amount of magic. 

We conclude this section by analyzing the behavior of quantum discord in Fig.~\ref{fig:DiscordBFKLvsSat}. In general, its behavior shows that separable regions are characterized by non-zero discord. As suggested by the behavior of the concurrence, the effect of linearly polarized gluons tends to increase the discord. An important point, which is not clearly visible from the density plots, is that the discord never vanishes (even in the black regions, it is numerically small but not zero). This is true also when the effect of linearly polarized gluons is taken into account within the saturation formalism\footnote{For BFKL ($ R_{{\cal H}_{g} / {\cal F}_{g}}^{\rm BFKL} (z, {\bm q} )$ strictly equal 1), at $\phi_{P,q}=0$ and $|\boldsymbol{P}| =m$, there is phase space in which the discord is strictly zero.}.

\section{Diffraction}
\label{Sec:DiffractiveChannel}
In this section, we study diffractive gluon dissociation into a heavy-quark pair, the hadroproduction counterpart of diffractive quark-pair production in DIS. We show that the two processes have closely related quantum-information properties, then revisit the peculiar entanglement and Bell-nonlocality patterns found in Ref.~\cite{Fucilla:2025kit} and clarify their origin.

\subsection{Spin density matrix for diffractive gluon dissociation}

The amplitude in eq.~(\ref{Eq:FullAmpInclusive}) can also be used to describe coherent diffractive production,
\begin{equation}
    g+A\longrightarrow Q+\bar Q+A,
\end{equation}
where the target remains intact. The complete hadronic process still involves an inclusive remnant, $X_p$, and reads
\begin{gather}
    p + A \longrightarrow Q+\bar Q + X_p  + A \; .
\end{gather}
However, the crucial difference is that $X_p$ is associated with the breaking of the projectile proton only, while the dense target $A$ stays intact and has rapidity gap separation from the whole diffractive state $Q + \bar Q + X_p$. This process has previously been studied at the unpolarized level in the two-gluon-exchange approximation~\cite{Luszczak:2013diffractive}. Its experimental realization in hadronic collisions is, however, more challenging than the standard diffractive photoproduction. Soft interactions involving the proton spectators can destroy the rapidity gap or break up the target, so that hard-diffractive factorization is not expected to hold without an additional gap-survival correction~\cite{Luszczak:2013cba,Khoze:2000wk}. Schematically,
\begin{equation}
d\sigma^{pA,{\rm diff}}
\sim
\int dx_g\,
x_g g(x_g,\mu^2)\,
S_{\rm gap}^2\,
d\hat{\sigma}^{gA,{\rm diff}} .
\label{Eq:DiffHadronicCrossSection}
\end{equation}
The uncertain factor \(S_{\rm gap}^2\) mainly affects the absolute normalization and should cancel from normalized spin correlations if it is spin-independent. Nevertheless, the coherent-target requirement, the reduced event rate and the reconstruction of the heavy-quark spin analyzers make this channel experimentally demanding.  

However, from a theory perspective, this process is particularly intriguing because it provides the hadroproduction analogue of diffractive DIS\footnote{Besides diffractive DIS, we have proposed accessing this spin-density matrix through photon-initiated diffraction in ultra-peripheral collisions (UPCs) at the LHC~\cite{Fucilla:2025kit,Fucilla:2026zhy}. The UPC channel largely avoids the hadronic rapidity-gap survival suppression affecting diffractive hadroproduction. The latter may nevertheless offer higher event rates because of its strong-interaction coupling and gluon luminosity. An analogous comparison of coupling factors, gluon PDFs, and equivalent-photon fluxes is given in Ref.~\cite{Bolognino:2019yls}. A quantitative comparison for the present diffractive channels must also account for the nuclear photon flux and gap survival.}, whose spin-density matrix was constructed in Ref.~\cite{Fucilla:2025kit}. In that case, the color-singlet exchange was found to produce a remarkably distinctive spin state. For a transversely polarized photon, the state is entangled throughout the entire phase space, with entanglement implying Bell nonlocality and vice versa\footnote{Such an equivalence generally holds for pure states as a consequence of Gisin's theorem~\cite{Gisin:1991vpb}. Remarkably, Ref.~\cite{Fucilla:2025kit} established it for the entire family of generally mixed states generated by this process.}. Moreover, for symmetric splittings and transverse momenta of the order of the heavy-quark mass, the state becomes maximally entangled and maximally violates the Bell--CHSH inequality. We would like therefore to compare the situation to the present case, and stress a few key features.

For coherent diffraction, the average over the target fields must be performed at the amplitude level, $\left\langle\mathcal{M}_{\alpha\alpha'}^a\right\rangle \langle\mathcal{M}_{\beta\beta'}^a \rangle^* $, rather than the inclusive average $\langle\mathcal{M}_{\alpha\alpha'}^a
\mathcal{M}_{\beta\beta'}^{a*}\rangle$. To enforce the exchange of a color singlet in the $t$-channel, we have to project the final state quark pair onto a color octet. We can thus define the target structure
\begin{equation}
{\cal T}_g(\boldsymbol{u},\boldsymbol{v})
=
\frac{ 2  }{N_c^2 -1}
\left\langle
{\rm Tr}_F\!\left[
t^a \mathcal{U}_F( {\bm v} + {\bar z} {\bm u} )t^a
\mathcal{U}_F^\dagger( {\bm v} - z {\bm u} )
-
t^a \mathcal{U}_F( {\bm v}   )t^a
\mathcal{U}_F^\dagger( {\bm v}   )
\right]
\right\rangle  \; .
\label{Eq:DiffTargetProjection}
\end{equation}
At zero momentum transfer, we define
\begin{equation}
    {\cal T}_g(\boldsymbol{u})
    =
    \int d^2\boldsymbol{v}\,
    {\cal T}_g(\boldsymbol{u},\boldsymbol{v}),
    \qquad
    T_g(\boldsymbol{p})
    =
    \int\frac{d^2\boldsymbol{u}}{(2\pi)^2}\,
    e^{-i\boldsymbol{p}\cdot\boldsymbol{u}}\,
    {\cal T}_g(\boldsymbol{u}).
\label{Eq:DiffTargetFormFactor}
\end{equation}
Note that color coherences require ${\cal T}_g(\boldsymbol{u}=0,\boldsymbol{v})=0$, which immediately implies
\begin{equation}
    \int d^2\boldsymbol{p}\,T_g(\boldsymbol{p})=0.
\label{Eq:DiffColorCoherence}
\end{equation}
The color-projected amplitude becomes
\begin{equation}
\left.
\left\langle
\mathcal{M}_{g\rightarrow Q\bar Q}
\right\rangle  
\right|_{\boldsymbol{\Delta}=0}
=
-ig 
\int d^2\boldsymbol{u}\,
e^{-i\boldsymbol{P}\cdot\boldsymbol{u}}\,
\Phi_{\alpha\alpha'\perp}(\boldsymbol{u})\,
{\cal T}_g(\boldsymbol{u}) .
\label{Eq:DiffAveragedAmplitude}
\end{equation}
The spinor structure of eq.~(\ref{Eq:DiffAveragedAmplitude}) is the same as that of a transversely polarized real photon. All differences between the two processes can therefore be encoded in the replacement of the dipole form factor \(T(\boldsymbol{p})\) of Ref.~\cite{Fucilla:2025kit} by the coherent gluon-induced form factor \(T_g(\boldsymbol{p})\). In analogy with Ref.~\cite{Fucilla:2025kit}, we introduce
\begin{align}
T_{g,1}(\boldsymbol{P})
=
\int d^2\boldsymbol{p}\,
\frac{T_g(\boldsymbol{p})}
     {(\boldsymbol{P}-\boldsymbol{p})^2+m^2}, \hspace{1.5 cm}
T_{g,2}(\boldsymbol{P})
=
-\frac{1}{\boldsymbol{P}^2}
\int d^2\boldsymbol{p}\,
\frac{\boldsymbol{P}\cdot\boldsymbol{p}\,
      T_g(\boldsymbol{p})}
     {(\boldsymbol{P}-\boldsymbol{p})^2+m^2}.
\label{Eq:Tg1Tg2}
\end{align}
The spin-summed hard factor consequently has the form
\begin{equation}
{\cal A}^{\rm diff}_g
=
{\cal N}^{\rm diff}_g
\left[
\left(z^2+\bar z^2\right)
\boldsymbol{P}^2
\left|T_{g,1}+T_{g,2}\right|^2
+
m^2\left|T_{g,1}\right|^2
\right],
\label{Eq:DiffUnpolarizedFactor}
\end{equation}
where \({\cal N}^{\rm diff}_g\) contains the overall coupling, color and phase-space factors. The normalized spin-correlation coefficients follow directly from the transverse-photon result:
\begin{equation}
C_{ab}^{g {\mathbb P}}
=
\left.
C_{Tab}^{\gamma^*{\mathbb P}}
\right|_{
Q^2=0,\,
T_i\rightarrow T_{g,i},\,
} \; ,
\label{Eq:DiffSpinMatrixReplacement}
\end{equation}
where the two form factors, $T_1 (\boldsymbol{P})$ and $T_2 (\boldsymbol{P})$, of Ref.~\cite{Fucilla:2025kit}, are obtained from from the target structure\footnote{They coincide with the form factors of
Ref.~\cite{Fucilla:2025kit} up to the overall normalization convention
adopted there, which is irrelevant for the normalized spin-density
matrix.} 
\begin{gather*}
    {\cal T}_{\gamma} (\boldsymbol{u},\boldsymbol{v})
= 1- \frac{1}{N_c} \left\langle
{\rm Tr}_F\!\left[ \mathcal{U}_F( {\bm v} + {\bar z} {\bm u} ) 
\mathcal{U}_F^\dagger( {\bm v} - z {\bm u} )
\right]
\right\rangle \; ,
\end{gather*}
It is instructive to consider the dilute BFKL limit again. To this aim, we perform the standard expansion of the Wilson lines in Reggeized-gluon fields~\cite{Caron-Huot:2013fea} 
\begin{gather}
\mathcal{U}_F(\boldsymbol{z})
=
\exp \left \{ -ig t^a R^a( \boldsymbol{z}) \right \} = 1
    - i g t^a R^a (\boldsymbol{z})
    - \frac{g^2}{2} t^a t^b
    R^a ( \boldsymbol{z})
    R^b ( \boldsymbol{z}) + \mathcal{O} (g^3 R^3) \; ,
\label{Eq:UniversExp}
\end{gather}
neglecting corrections beyond the two-Reggeon approximation. Then, in the forward limit $\bm \Delta = {\bm p}_q + {\bm p}_{\bar q} =0$, it is not difficult to show that, thanks to the absence of the phase,
\begin{gather*}
    \int d^2 {\bm v} {\cal T}_g^{\rm dilute} (\boldsymbol{u},\boldsymbol{v}) = \frac{1}{N_c^2-1} \int d^2 {\bm v} {\cal T}_{\gamma}^{\rm dilute} (\boldsymbol{u},\boldsymbol{v}) + \mathcal{O} (g^3 R^3)
\end{gather*}
Consequently, the gluon- and photon-induced target form factors are proportional,
\begin{equation}
T_{g,i}^{\rm dilute}(\boldsymbol{P})
=
\frac{T_i^{\rm dilute}  (\boldsymbol{P})}{2N_c C_F} ,
\label{Eq:DiffDiluteRelation}
\end{equation}
At $\boldsymbol{\Delta}=0$, the constant color factor cancels from the normalized density matrix, yielding
\begin{equation}
\left.
C_{ab}^{g {\mathbb P}}
\right|_{\rm dilute}
=
\left.
C_{Tab}^{\gamma^*{\mathbb P}}
\right|_{Q^2=0,\,{\rm dilute}} .
\label{Eq:DiffBFKLSpinEquality}
\end{equation}
This equality concerns the normalized spin state; the absolute cross sections remain different because of the coupling, color, and projectile-flux factors. The proportionality in eq.~(\ref{Eq:DiffDiluteRelation}) is not expected to survive unchanged in the saturation regime and, in particular, beyond the two Reggeized gluon approximation. The photon probes a color-singlet fundamental dipole, whereas eq.~(\ref{Eq:DiffTargetProjection}) describes the coherent propagation of a color-octet \(Q\bar Q\) configuration relative to its parent gluon. Multiple scattering therefore gives, in general,
\begin{equation}
T_{g,i}^{\rm sat}(\boldsymbol{P})
\not\propto
T_{\gamma,i}^{\rm sat}(\boldsymbol{P}) \; .
\label{Eq:DiffSaturationBreaking}
\end{equation}
Saturation effects can consequently modify the numerical spin-density matrix and, for instance, shift the kinematic region in which maximal entanglement is approached. Nevertheless, as long as the amplitude retains the two-form-factor structure in eq.~(\ref{Eq:Tg1Tg2}), the algebraic relations between entanglement and Bell nonlocality derived in Ref.~\cite{Fucilla:2025kit} remain valid after the replacement in eq.~(\ref{Eq:DiffSpinMatrixReplacement}). \\

\subsection{Why does the Pomeron give so much entanglement?}
\label{sec:Coherent_Diffraction}

The calculation in Ref.~\cite{Fucilla:2025kit}, and consequently the present one, shows that transverse diffraction leads to an extremely complex density matrix that depends on the non-perturbative physics of QCD in a highly intricate way. Nevertheless, the non-trivial relations (14) and (15) from~\cite{Fucilla:2025kit}, involving the coefficients $C_{ab}$, ensure that the pair remains entangled throughout the entire kinematic space, except at the endpoints $z=0, 1$. It is therefore interesting to discuss why this happens. Since neither the inclusive counterpart~\cite{Qi:2025onf,Fucilla:2026mkg} nor the more general exclusive calculation~\cite{Hatta:2025obw} outside the Regge limit produces this situation, it is clear that the peculiarity must arise from the combination of exclusivity and eikonality of the Pomeron calculation. \\

\paragraph{Exclusivity and eikonality as constraints on the density matrix.} First, we want to see that exclusivity and eikonality limit the rank of the density matrix to be at most two. If one of the four states of the pair is identified by $s$ and the state of the outgoing proton by $X$, then, in general, the state of the system for a given photon polarization, $\lambda$, must be\footnote{In the leading-eikonal, near-forward approximation used in \cite{Fucilla:2025kit} and here, the interaction does not couple the spins of the projectile proton and the target. Their spin degrees of freedom therefore introduce no additional structure in the $q\bar q$ spin-density matrix.}
\begin{gather}
    | \Psi_{\lambda} \rangle = \sum_{s,X} \mathcal{C}_{\lambda,sX} |s \rangle_{q \bar q} | X\rangle_A \; ,
\end{gather}
where $|s \rangle_{q \bar q} = \{ | \uparrow \uparrow \rangle , | \downarrow \downarrow \rangle , | \uparrow \downarrow \rangle, | \downarrow \uparrow \rangle \} $. In the inclusive case, there is no restriction on the state $|X \rangle$, but in coherent diffraction, the state of the target is completely fixed. We thus have
\begin{equation}
    | \Psi_{\lambda} \rangle = \sum_{s} \mathcal{C}_{\lambda,s} |s \rangle_{q \bar q}  \; .
\end{equation}
For unpolarized photon (or gluon) production, the density matrix must be a perfect statistical mixture of the two resulting states. We thus have
\begin{equation}
    \rho_T = \frac{1}{2 N} \bigg( | \Psi_{+} \rangle \langle \Psi_{+}| + |\Psi_{-} \rangle \langle \Psi_{-}|  \bigg) \; .
    \label{Eq:DensityAlaLightFront}
\end{equation}
This matrix evidently has rank at most two\footnote{Please note that this is in strong contrast with the inclusive case, in which the final density matrix must include a sum over all final states of the proton and can have arbitrary rank.}. 

Expressing the spin amplitudes of Ref.~\cite{Fucilla:2025kit} in the light-front helicity basis (see, e.g., Ref.~\cite{Lappi:2020ufv}) yields the density matrix in eq.~\eqref{Eq:DensityAlaLightFront}. We show this change of basis explicitly in the Appendix \ref{HelicityBasis}. In the notation of Ref.~\cite{Fucilla:2025kit}, we obtain
\begin{gather}
    | \Psi_+ \rangle = m T_1 (\boldsymbol{P}) | \uparrow \uparrow \rangle + |\boldsymbol{P}| (T_1 (\boldsymbol{P}) + T_2 (\boldsymbol{P}) ) \big( z | \uparrow \downarrow \rangle - \bar z | \downarrow \uparrow \rangle \big) \nonumber \\
    | \Psi_- \rangle = m T_1 (\boldsymbol{P}) | \downarrow \downarrow \rangle + |\boldsymbol{P}| (T_1 (\boldsymbol{P}) + T_2 (\boldsymbol{P}) ) \big( \bar z | \uparrow \downarrow \rangle -  z | \downarrow \uparrow \rangle \big) \; .
    \label{Eq:StatesInLightFront}
\end{gather}
The normalization in eq.~\eqref{Eq:DensityAlaLightFront} is
\begin{gather}
    N =   \langle \Psi_+  | \Psi_+ \rangle  =  \langle \Psi_- | \Psi_- \rangle =  m^2 T_1^2 (\boldsymbol{P}) +  (z^2 + \bar z^2) |\boldsymbol{P}|^2 (T_1 (\boldsymbol{P}) + T_2 (\boldsymbol{P}) )^2 \; ,
    \label{Eq:NormalizationDensPsi}
\end{gather}
which is just proportional to the unpolarized cross-section~\cite{Fucilla:2025kit}. Moreover, the two states overlap,
\begin{gather}
  {\rm O} = \langle \Psi_+  | \Psi_- \rangle = 2 z \bar z |\boldsymbol{P}|^2 (T_1 (\boldsymbol{P}) + T_2 (\boldsymbol{P}) )^2 \; .
  \label{Eq:OverlapDensPsi}
\end{gather}

\paragraph{Maximally mixed marginals and local rotations.}  The fact that the matrix $\rho_T$ has at most rank two is not sufficient to guarantee entanglement. However, a fundamental difference between the high-energy result and that of collinear factorization is related to the individual polarizations of the quarks. Indeed, in Ref.~\cite{Hatta:2025obw}, it was shown that in the collinear calculation the quark and antiquark acquire an induced polarization even in unpolarized collisions. In the Regge limit, these latter are vanishing, and the density matrix has maximally mixed marginals (see eq.~(\ref{Eq:MaximallyMixedMarginals})). As explained when we discussed the quantum discord, a density matrix of this type can be brought into a Bell-diagonal form by local spin rotations~\cite{Lang:2010xtw}. This transformation preserves the concurrence. Since the maximum permitted rank is two, the resulting matrix must be, at most, a combination of two Bell states. In the Appendix~\ref{HelicityBasis}, it is demonstrated that a local spin rotations~\cite{Lang:2010xtw} brings the matrix in the form
\begin{equation}
    \rho_T^{\rm rot} = \frac{1+t}{2} | \Phi^{+} \rangle \langle \Phi^{+} | + \frac{1-t}{2} | \Phi^{-} \rangle \langle \Phi^{-} | \; ,
    \label{Eq:RotatedMatrix}
\end{equation}
where the two Bell states are
\begin{equation}
    | \Phi^{\pm} \rangle = \frac{| \uparrow \uparrow \rangle \pm | \downarrow \downarrow \rangle  }{ \sqrt{2}} \; ,
\end{equation}
and 
\begin{equation}
    t = \frac{2 z \bar z \boldsymbol{P}^2 (T_1 (\boldsymbol{P}) + T_2 (\boldsymbol{P}) )^2 }{ (z^2 + \bar z^2) \boldsymbol{P}^2 (T_1 (\boldsymbol{P}) +T_2 (\boldsymbol{P}))^2 + m^2 T_1^2 (\boldsymbol{P}) }  \; .
    \label{Eq:tDef}
\end{equation}
We can therefore write
\begin{equation}
    \rho_T^{\rm rot} = \frac{1}{2} \bigg( | \uparrow \uparrow \rangle \langle \uparrow \uparrow | + | \downarrow \downarrow \rangle \langle \downarrow \downarrow |  + t ( | \uparrow \uparrow \rangle \langle \downarrow \downarrow | + | \downarrow \downarrow \rangle \langle \uparrow \uparrow | ) \bigg)  \; ,
    \label{Eq:BellRotatBasis}
\end{equation}
The state described by $\rho_T^{\rm rot}$ is always entangled and Bell nonlocal except when $t=0$ and the matrix becomes separable. As already stated, the concurrence is preserved by the rotation and reads  
\begin{equation}
 {\cal C}(\rho_T^{\rm rot}) = {\cal C}(\rho_T) = t \; .
\end{equation} 
Indeed, $t$ is nothing but $-C_{nn}^T$ of Ref.~\cite{Fucilla:2025kit} and the separability is only obtained at $z=0,1$, i.e. when $t=-C_{nn}^T=0$. A maximally entangled state, which in the rotated frame is the Bell state $| \Phi^+ \rangle$ and maximizes $t$, is obtained at $z=1/2$ and $k_{\perp} \approx m$ as first noted in Ref.~\cite{Fucilla:2025kit}. \\

Although we have derived the result explicitly, we would like to note that it follows from very general properties of the Pomeron interaction. Indeed, coherent diffraction fixes the final state of the target; averaging over the two transverse polarizations gives a spin-density matrix of rank at most two. In the eikonal interaction, single-spin polarizations cannot be induced, so both quark marginals are maximally mixed. A two-qubit state with these properties must be equivalent to a mixture of at most two Bell states~\cite{Lang:2010xtw}, as in eq.~\eqref{Eq:RotatedMatrix}. A mixture of two Bell states is separable only for equal weights. However, the inequality of the two Bell-state weights everywhere except $z=0,1$ seems to be just dictated by the light-front form of the two amplitudes in (\ref{Eq:StatesInLightFront}). 

\subsection{The special relations in Ref.~[28]}
\label{sec:The_Magic_Relation}

To close the circle, we would like to reinterpret the results found in Ref.~\cite{Fucilla:2025kit}, in light of our previous discussion. Indeed, in Ref.~\cite{Fucilla:2025kit}, the entanglement through the whole physical phase space and the equivalence between entanglement and Bell nonlocality even for mixed spin state, follow from the following relations among the spin-correlation coefficients:\footnote{See eqs.~(14) and (15) of~\cite{Fucilla:2025kit}.} 
\beq
(C_{rr} )^2+(C_{rk} )^2+(C_{kr} )^2+(C_{kk} )^2 -(C_{nn} )^2 = 1 \; , \label{id1}
\eeq
\beq
C_{nn}  = -C_{rr}  C_{kk} +C_{rk} C_{kr}  \; , \label{id2}
\eeq
obtained in the $\{ \hat{n}, \hat{r}, \hat{k} \}$. Based on what was described in the previous section, it is natural to expect that eqs.~(\ref{id1}) and~(\ref{id2}) are a consequence of the restrictions on the rank and the marginals of the density matrix. In fact, by analyzing various processes, one can see the pattern described in the table \ref{tab:VariousProcess}\footnote{At leading order, $e^+e^-\to\gamma^*\to q\bar q$ with massless electrons involves only the two transverse helicities of the virtual photon. With no hadronic target states to trace over, the resulting spin-density matrix has rank at most two and maximally mixed marginals.}.
\begin{table}[t]
    \centering
    \begin{tabular}{|c|c|c|c|}
        \hline
         & rank $\rho_T \leq2 $  & $\rho_q  = \rho_{\bar q} =\mathbb{I}/2 $ & (\ref{id1}) and (\ref{id2})  \\ \hline
        $\gamma_T (Q^2=0) \; P \rightarrow q \bar q P$ (Regge) & Yes & Yes & Yes \\ \hline
        $g  P \rightarrow q \bar q P$ (Regge) & Yes & Yes & Yes \\ \hline
        $e^+ e^- \rightarrow\gamma^{*} \rightarrow q \bar q$ ($m_e=0$) & Yes & Yes & Yes \\ \hline
        $\gamma_T (Q^2=0) \; p \rightarrow q \bar q X$ & No & Yes & No \\ \hline
        $\gamma_T (Q^2=0) \; P \rightarrow q \bar q P$ (collinear) & No & No & No \\ \hline
    \end{tabular}
    \caption{Validity of the relationships depending on the process and based on the constraints. The statement ``No" does not mean that the condition cannot occur, but rather that it is not guaranteed across the entire phase space and that there does not appear to be any \textit{a priori} principle to ensure it.}
    \label{tab:VariousProcess}
\end{table}
To prove that this is indeed the case, let us recall that with vanishing polarization and parity-conserving processes we have
\begin{equation}
    \rho_T
=
\frac{1}{4}
\left[
\mathbb{I}\otimes\mathbb{I}
+C^{ab}\sigma^a\otimes\sigma^b
\right] \; ,
    \label{eq:rho_rank_argument}
\end{equation}
and
\begin{equation}
    C=
    \begin{pmatrix}
        C_{nn} & 0 & 0\\
        0 & C_{rr} & C_{rk}\\
        0 & C_{kr} & C_{kk}
    \end{pmatrix}.
    \label{eq:C_rank_argument}
\end{equation}
It is convenient to choose the following basis of Pauli matrices
\begin{equation}
    \sigma^n=\sigma^3,\qquad
    \sigma^r=\sigma^1,\qquad
    \sigma^k=\sigma^2.
\end{equation}
Now, we can represent $\rho_T$ in the basis of spin eigenstates with respect to the direction $n$, choosing the ordering $\{|\uparrow\uparrow\rangle_n,|\downarrow\downarrow\rangle_n,
|\uparrow\downarrow\rangle_n,|\downarrow\uparrow\rangle_n\}$. Please note that this means, for instance, 
\begin{equation}
    \sigma^n \otimes \sigma^n = \sigma^3 \otimes \sigma^3 = \begin{pmatrix}
        1 & 0   & 0   & 0\\
        0 & -1 & 0   & 0\\
        0   & 0   & -1 & 0\\
        0   & 0   & 0 & 1
    \end{pmatrix}_{\rm standard \; order}  \hspace{1 cm} \xrightarrow[]{\rm re-ordering} \hspace{1 cm} \begin{pmatrix}
        1 & 0   & 0   & 0\\
        0 & 1 & 0   & 0\\
        0   & 0   & -1 & 0\\
        0   & 0   & 0 & -1
    \end{pmatrix} \; .
\end{equation}
In this representation, $\rho_T$ is block diagonal
\begin{equation}
  \rho_T=\frac{1}{4}
    \begin{pmatrix}
        1+C_{nn} & U   & 0   & 0\\
        U^* & 1+C_{nn} & 0   & 0\\
        0   & 0   & 1-C_{nn} & V\\
        0   & 0   & V^* & 1-C_{nn}
    \end{pmatrix}
    \equiv \rho_+\oplus\rho_- \; , 
\end{equation}
where we defined 
\begin{align}
    U&=C_{rr}-C_{kk}
       -i(C_{rk}+C_{kr}) \; ,
    \nonumber\\
    V&=C_{rr}+C_{kk}
       +i(C_{rk}-C_{kr}) \; .
    \label{eq:UV_rank_argument}
\end{align}
The $2 \times 2$ matrix $\rho_+$ acts on $\{|\uparrow\uparrow\rangle,|\downarrow\downarrow\rangle\}$
while the $\rho_-$ acts on
$\{|\uparrow\downarrow\rangle,|\downarrow\uparrow\rangle\}$. Let us consider $|C_{nn}| \neq 1$\footnote{In this case, we know from the Bell basis, eq.~(\ref{Eq:BellRotatBasis}), that for $C_{nn} = \pm 1$ we obtain a Bell state.}, we then have 
\begin{equation}    \operatorname{Tr}\rho_+=\frac{1+C_{nn}}{2}>0,
    \qquad
    \operatorname{Tr}\rho_-=\frac{1-C_{nn}}{2}>0 \; ,
    \label{eq:positive_block_traces}
\end{equation}
implying that each block has rank at least one. On the other hand, the restriction on the total rank of $\rho_T$ imposes
\begin{equation}
    \operatorname{rank}\rho_T
    =\operatorname{rank}\rho_+
     +\operatorname{rank}\rho_-
    \leq 2 \; .
\end{equation}
The only possibility is therefore
\begin{equation}
    \operatorname{rank}\rho_+
    =\operatorname{rank}\rho_-=1,
    \implies 
    \det\rho_+=\det\rho_-=0.
    \label{eq:both_blocks_rank_one}
\end{equation}
Computing the two determinants explicitly
\begin{equation}
  1-C_{kk}^2 - C_{kr}^2 + 2 C_{nn} + C_{nn}^2   - 2 C_{kr} C_{rk}  - C_{rk}^2 +2 C_{kk} C_{rr}- C_{rr}^2 =0 \; ,
  \label{Eq:PlusDet}
\end{equation}
and
  \begin{equation}
  1-C_{kk}^2 - C_{kr}^2 - 2 C_{nn} + C_{nn}^2   + 2 C_{kr} C_{rk}  - C_{rk}^2 -2 C_{kk} C_{rr}- C_{rr}^2 =0 \; .
  \label{Eq:MinusDet}
\end{equation}
Taking (\ref{Eq:PlusDet}) plus (\ref{Eq:MinusDet}) one obtains (\ref{id1}), while (\ref{Eq:PlusDet}) minus (\ref{Eq:MinusDet}) gives (\ref{id2}). We have thus recovered the special relations observed in Ref.~\cite{Fucilla:2025kit} for diffractive photoproduction and in Ref.~\cite{Fucilla:2026zhy} for massive heavy-quark pair production in $e^+ e^-$-annihilation, and shown that they are a direct consequence of the rank $\rho_T \leq 2$ and the maximally-mixed marginals. In summary, we explained why the pair is generically entangled and why, in this class of channels, entanglement coincides with Bell nonlocality even beyond the pure states, for which the Gisin's theorem holds~\cite{Gisin:1991vpb}. \\

This line of reasoning could also be apply to the longitudinal case, where the system is always in a maximally entangled pure state\footnote{In this case the relations (\ref{id1}) and (\ref{id2}) hold more trivially.}, corresponding to a Bell state. However, this does not occur solely in the exclusive case and the Regge limit but in all processes of table \ref{tab:VariousProcess}~\cite{Qi:2025onf,Hatta:2025obw,Fucilla:2026mkg}. This universality is therefore expected to be more general.

\section{Accessing spin correlations through TMD and dihadron FFs}
\label{sec:CollinsPA}
In this section, we provide some proposals on how to reconstruct this density matrix, distinguishing between the case of heavy quarks and light quarks. Please note that the strategies are intended to apply beyond the specific production channels and kinematic regime studied in the previous sections. \\

\paragraph{Massive quarks.} For charm and bottom quarks, a possibility is to use the polar angle distribution of particles produced after a decay chain~\cite{Afik:2025grr}. For bottom and charm quark, in Ref.~\cite{Kats:2023zxb}, the following channels have been identified
\begin{gather}
    b\xrightarrow{7.0\%}  \Lambda_b\xrightarrow{11\%} X^+_c+\mu^-+\bar{\nu}_\mu, \\
    c\xrightarrow{6.4\%}\Lambda_c^+ \xrightarrow{3.5\%} \Lambda+\mu^++\nu_\mu,
    \label{Eq:DecayChannels}
\end{gather}
and their charge-conjugate counterparts for $\bar{b},\bar{c}$. The numbers above the first arrow are the fragmentation probabilities. The spin of the  (anti)quark is largely inherited by the daughter (anti)baryon \cite{Galanti:2015pqa}. The numbers above the second arrow are the branching ratios. Thanks to the large cross-section, following this strategy is feasible in inclusive $pp$ or $pA$ collisions~\cite{Afik:2025grr}. \\ 

However, as already demonstrated for DIS in Ref.~\cite{Fucilla:2026zhy}, an alternative strategy is to use the Collins effect~\cite{Collins:1992kk,Boer:1997mf} to access the transverse correlations, and dihadron fragmentation functions~\cite{Artru:1995zu,Bianconi:1999cd,Metz:2016swz} for longitudinal correlations~\cite{Cheng:2025cuv}. Extraction and phenomenological applications of the Collins and dihadron fragmentation functions represents a huge and highly-developed field of research~\cite{Boer:2003ya,Anselmino:2007fs,Bacchetta:2008wb,Yuan:2008yv,DAlesio:2010sag,Anselmino:2013vqa,Pitonyak:2013dsu,Anselmino:2015sxa,Kang:2015msa,Metz:2016swz,Anselmino:2018psi,DAlesio:2017bvu,Kang:2017btw,Arratia:2020nxw,Gamberg:2021lgx,DAlesio:2021dcx,Bacchetta:2023njc,Boglione:2024dal,DAlesio:2025jmr,Mahaut:2025hie,Anedda:2026pcn}. Here, our intention is to stress the potential for the scope of accessing entanglement at hadron-hadron colliders, since it has the considerable advantage of applying to both light and heavy quarks. Furthermore, the spin-dependent FFs (for the transverse polarization) are not strongly suppressed compared to the spin-independent FFs, and therefore the suppression is less severe than in eq.~(\ref{Eq:DecayChannels}). To illustrate this strategy, consider a hadron $h_1$ inside the quark jet and a hadron $h_2$ inside the antiquark jet. Their momentum fractions relative to the respective jets are denoted by $z_1$ and $z_2$, to distinguish them from the splitting fraction $z$. We work in the $q\bar q$ center-of-mass frame and define $\phi_1$ and $\phi_2$ as the azimuthal angles of the hadron momenta around the quark and antiquark jet axes, measured with respect to the production plane. The common transverse spin axes are $\boldsymbol r$ and $\boldsymbol n$, with $\epsilon^{rn}=1$. Our definitions and conventions follow~\cite{Fucilla:2026zhy} (see sections III and IV there). The following expressions apply at leading power in the hadron transverse momenta relative to the jet momenta. For an isolated $f\bar f$ channel, the angular distribution, normalized to its azimuthal average, becomes~\cite{Fucilla:2026zhy}
\begin{align}
{\cal R}_f^{\rm Collins}(\phi_1,\phi_2) = 1+a_f\bar a_f \left[ \frac{{\cal C}_{rr}^{f}-{\cal C}_{nn}^{f}}{2} \cos(\phi_1+\phi_2) -\frac{{\cal C}_{rr}^{f}+{\cal C}_{nn}^{f}}{2} \cos(\phi_1-\phi_2) \right].
\label{Eq:pACollinsAngular}
\end{align}
where
\begin{equation}
a_f(z_1)=
\frac{H_1^{\perp(1)\,h_1/f}(z_1)}
     {z_1D_1^{h_1/f}(z_1)},
\qquad
\bar a_f(z_2)=
\frac{H_1^{\perp(1)\,h_2/\bar f}(z_2)}
     {z_2D_1^{h_2/\bar f}(z_2)} .
\label{Eq:pACollinsAnalyzingPowers}
\end{equation}
with the integrated Collins functions being
\begin{equation}
H_1^{\perp(1)\,h_i/f}(z_i)
=
\int d^2\boldsymbol p_{h\perp}\,
\frac{|\boldsymbol p_{h\perp}|}{M_h}\,
H_1^{\perp\,h_i/f}
   (z_i,\boldsymbol p_{h\perp}^{\,2}) .
\label{Eq:pACollinsMoment}
\end{equation}
and the $D_1^{h_i/f}(z_i)$ the standard unpolarized FFs. For massive quarks, the coefficients ${\cal C}_{rr}$ and ${\cal C}_{nn}$ can be extracted from the two harmonics in eq.~(\ref{Eq:pACollinsAngular}). However, the Collins fragmentation functions alone do not analyze spin along $\boldsymbol k$: they do not determine ${\cal C}_{kk}$, ${\cal C}_{rk}$, or ${\cal C}_{kr}$. Access to these components requires an additional longitudinal-spin analyzer, such as the dihadron fragmentation function. This strategy is discussed in
Ref.~\cite{Cheng:2025cuv,Fucilla:2026zhy} and remains identical in the present case. For example, identifying an unpolarized hadron pair in each jet
allows one to probe ${\cal C}_{kk}$ through the helicity-dependent
dihadron FF $G_1^\perp$. Denoting by
$\varphi_i=\phi_{R_i}-\phi_{P_{hi}}$ the relative azimuthal angle
between the relative and total transverse momenta of the hadron
pair in jet $i$, the normalized angular distribution takes the form
\begin{equation}
{\cal R}_f^{\rm dihadron}(\varphi_1,\varphi_2)
=
1+
{\cal C}_{kk}^{f}\,
\frac{
G_1^{\perp(1)\,h_1h_2/f}(z_1)\,
G_1^{\perp(1)\,h_3h_4/\bar f}(z_2)
}{
z_1z_2\,
D_1^{h_1h_2/f}(z_1)\,
D_1^{h_3h_4/\bar f}(z_2)
}
\sin\varphi_1\sin\varphi_2 .
\label{Eq:pADihadronLongitudinal}
\end{equation}
where $D_1^{h_1h_2/f}$ denotes the integrated unpolarized
dihadron FF, and $G_1^{\perp(1)\,h_1h_2/f}$ is the
helicity-dependent moment defined in
Ref.~\cite{Fucilla:2026zhy} and $z_i$ denotes the momentum fraction carried by the hadron
pair. 

\paragraph{Massless quark.} The massless limit is very interesting. For light quarks, $A^{(2)}=0$ and $\tilde{C}^{(2)}=0$. Consequently, the $\cos(2\phi_{P,q})$ terms in eq.~(\ref{Eq:GenCoeffSpinDen}) vanish, and one immediately obtains
\begin{gather}
    {\cal C}_{ab} = \frac{{\cal F}_{g}(z, {\bm q}) \tilde{C}_{ab}^{(0)}}{{\cal A}} = C_{ab}^{(0)} \bigg|_{m=0} = \begin{pmatrix} -\frac{2 z \bar{z}}{z^2 + \bar{z}^2} & 0 & 0 \\
    0 & \frac{2 z \bar{z}}{z^2 + \bar{z}^2} & 0 \\
    0 & 0 & 1 \end{pmatrix}_{ab} \;. \label{masslessc0}
\end{gather}
For this density matrix,
\begin{gather}
    \Delta_2 = - \Delta_1 = \frac{4 z \bar{z}}{z^2 + \bar{z}^2} \; ,
\end{gather}
and therefore the concurrence of the bipartite system reads
\begin{gather}
    {\cal C}[\rho]  \big|_{m=0} = C_{rr}^{(0)} \big|_{m=0} = - C_{nn}^{(0)} \big|_{m=0} \; .
\end{gather}
The concurrence is positive for $0<z<1$ and reaches unity at $z=1/2$: the pair is therefore entangled throughout the interior
of phase space and maximally entangled for a symmetric splitting\footnote{At this point the matrix describes the Bell state $| \Phi^+ \rangle$~\cite{Qi:2025onf}.}. Unlike DIS at finite $Q^2$, this process receives no contribution from longitudinal polarizations, making the prediction for ${\cal C}[\rho]$ cleaner in this respect. Within the isolated leading-order $g\to q\bar q$ channel, the relations $C_{kk}^{(0)} \big|_{m=0}=1$ and $C_{nn}^{(0)} \big|_{m=0}=-C_{rr}^{(0)} \big|_{m=0}$ imply that a nonzero transverse spin correlation entails entanglement. At the hadron level, the final dijet pair can also be produced by $qg$ and $gg$ final states (see~\cite{Yang:2026rye} for a recent study). However, by selecting an unlike-sign minus like-sign cross-section difference in conjugated charged hadrons ($h$ and $\bar h$) production, i.e.
\begin{equation}
    d \sigma^{\rm U-L} = (d \sigma^{h \bar h} + d \sigma^{\bar h h}) - (d \sigma^{hh} + d \sigma^{\bar h \bar h}) \; , 
\label{Eq:DoubleCollinsAsymNice}
\end{equation} 
a given partonic cross-section takes the schematic form
\begin{equation}
    d \sigma_{ij} \left[ F_i^h F_j^{\bar h} +  F_i^{\bar h} F_j^{h} - F_i^{h} F_j^{h} - F_i^{\bar h} F_j^{ \bar h} \right] =  d \sigma_{ij} \left[ (F_i^h- F_i^{\bar h}) (F_j^{\bar h} - F_j^{ h}) \right] \; ,
\end{equation}
where $F_i^h$ is the fragmentation function of a parton $i$ in a hadron $j$. Charge conjugation implies
\begin{equation}
    F_g^h=F_g^{\bar h},
    \qquad
    F_{ q}^{h}=F_{\bar q}^{\bar h},
\end{equation}
whereas $F_q^h-F_q^{\bar h}$ is, in general, nonzero. Thus,
eq.~\eqref{Eq:DoubleCollinsAsymNice} cancels contributions with a gluon
in either final-state jet, isolating the $q \bar q$ channel. Then, the relation $C_{rr}^{(0)} \big|_{m=0}=-C_{nn}^{(0)} \big|_{m=0}$ gives the transverse-spin structure familiar from the double-Collins effect, with eq.~(\ref{Eq:pACollinsAngular}) then reducing to the standard double-Collins form
\begin{equation}
{\cal R}_f(\phi_1,\phi_2)
=
1+a_f(z_1)\bar a_f(z_2) {\cal C}[\rho] \big|_{m=0}
 \cos(\phi_1+\phi_2) 
\label{Eq:pAMasslessDoubleCollins}
\end{equation}
The double-Collins observable in unpolarized $pp$ collisions is studied in Ref.~\cite{Kang:2010vd}. As a consistency check, one can see that, in the massless limit, the hard factor of the $gg \rightarrow q \bar q$-channel is nothing but ${\cal C}[\rho] \big|_{m=0}$. One final point concerns the quark-quark scattering background. The unlike-sign
minus like-sign combination removes channels with a gluon in either
final-state jet, but does not isolate the $gg\to q\bar q$ production
mechanism. At leading order, the double-Collins hard factor vanishes
for $q_iq_j\to q_iq_j$ with $i\neq j$, whereas identical-flavor
$qq\to qq$ and quark--antiquark initiated channels can contribute
~\cite{Kang:2010vd}. In $pp$ collisions, one can seek kinematics in
which the gluon--gluon luminosity dominates over the relevant quark
luminosities; the $q \bar q$ channels are additionally limited by the antiquark PDFs, e.g. in a proton-proton collision. In forward scattering the quark content of the ``target" hadron is eikonally suppressed and thus discarded. In this sense, at small-$x$, the asymmetry (\ref{Eq:DoubleCollinsAsymNice}) is an ideal entanglement witness, whose maximal value signals a maximally entangled Bell state. However, even in the moderate-$x$ (collinear) regime, suitable identified-hadron selections may further improve the relative weight of $gg\to q\bar q$\footnote{For instance, hadrons containing $s$ or $c$ quarks can strongly suppress the $qq \longrightarrow qq$ channel in proton-proton collisions.}. Therefore, this peculiar double-Collins asymmetry can provide a promising entanglement witness, although quantitative studies are necessary to establish full feasibility. We also observe that, in this case, the asymmetry is identical to that of $e^+e^-$ annihilation in the massless limit; indeed, the two partonic density matrices coincide after identifying the corresponding kinematic variables. Indeed, for light quarks, these correlations could be alternatively accessed through the dihadron fragmentation~\cite{Cheng:2025cuv}, although we believe that the Collins effect represents the best strategy. \\

We close by noting that the $qq$, $q\bar q$, and $qg$ channels can be included in a complete treatment of the hadronic observable. We leave this extension for future work. Our focus on $gg\to q\bar q$ has been motivated by a special partonic relation: for massless quarks, the ratio of its double-Collins and unpolarized hard factors equals the concurrence of the produced pair. We also want to stress that Ultra-peripheral collisions could offer another promising setting for this measurement. The photons are quasi-real, so longitudinal contributions are negligible and vanish in the photoproduction limit. In this case, the process is fully dominated by photon--gluon fusion, $\gamma g\to q\bar q$, and the asymmetry (\ref{Eq:pAMasslessDoubleCollins}) is a perfect entanglement witness, even without forming the unlike-sign minus like-sign difference.

\section{Conclusion and outlook}
We studied the spin-density matrix of a $q\bar q$ pair produced in inclusive and coherent diffractive $pp$ ($pA$) collisions in the back-to-back limit. We showed how linearly polarized gluons and saturation affect its quantum-information properties, and gave a simple explanation of why, in the diffractive channel, the pair is entangled throughout almost the entire physical phase space, with Bell nonlocality arising simultaneously with the entanglement. We trace this behavior to a spin-density matrix of rank at most two with maximally mixed marginals. For this class of states, entanglement and Bell nonlocality coincide even when the state is mixed, while the explicit form of the light-front helicity amplitudes ensures that both persist in the whole phase space. We proposed using Collins and dihadron fragmentation functions to access quark spin correlations in unpolarized hadron-hadron collisions, including for light quarks. For massless quarks at leading order, the spin coefficient of a suitably chosen double-Collins asymmetry provides a promising entanglement witness. We close by discussing promising directions for extending this analysis. \\

\paragraph*{Beyond the back-to-back limit and/or the leading order.} In section~\ref{sec:IncluCollis}, we computed the spin-density matrix starting from the shockwave result and taking the back-to-back (correlation) limit. However, an all-twist calculation in the shockwave will provide better control over the higher-power corrections and could allow us to study the sensitivity of the coefficients to high-energy dynamics without advocating for a $\cos (2 \varphi)$ measurement\footnote{As this work was being finalized, the first study of next-to-leading-power corrections to spin correlations in small-$x$ DIS appeared~\cite{Gao:2026ipb}.}. Next-to-leading order corrections to the spin-density matrix in the shockwave formalism are also an intriguing direction. Going beyond the leading order\footnote{See Ref.~\cite{Agrawal:2026zwa} for a next-to-leading order (collinear factorization) calculation in the soft gluon limit.} in our calculation would be useful to investigate the interplay between Sudakov and small-$x$ effects~\cite{Gao:2026azd}, which is particularly relevant for what we discussed in section~\ref{sec:IncluCollis}\footnote{Another potential issue to analyze is related to decoherence~\cite{Aoude:2025ovu,Gu:2025ijz,Cheng:2026zfb,Liu:2026ees}.}.

\paragraph*{Higher-order corrections beyond the top sector.} In section \ref{sec:CollinsPA}, we propose to use the Collins and dihadron fragmentation functions to directly probe the entanglement, following the recent proposal for the EIC~\cite{Fucilla:2026zhy}. Developing such a set-up beyond the leading order is a goal for the future. Spin correlations in top-quark pair production and decay have been computed up to NNLO in QCD within the narrow-width approximation~\cite{Bernreuther:2001rq,Bernreuther:2004jv,Melnikov:2009dn,Behring:2019iiv,Czakon:2020qbd}. The top provides a particularly clean spin probe: its mass regulates collinear singularities associated with radiation from the heavy-quark lines, while its decay before hadronization allows its spin to be inferred from the angular distributions of its decay products. For light-quark spin measurements based on identified hadrons, NLO predictions in the massless approximation contain final-state collinear singularities that must be absorbed into the appropriate fragmentation correlators.\footnote{For charm- and bottom-flavoured hadrons produced at transverse momenta $p_{hT}^2\gg m_Q^2$, the finite quark mass regulates these singularities but leaves potentially large logarithms of $p_{hT}^2/m_Q^2$, which require resummation.}
The required factorization and evolution depend on the observable used to analyse the quark spin. Extending the present analysis to NLO therefore requires matching the spin-dependent hard coefficients to the fragmentation correlators appropriate to each measurement. 
\paragraph*{Universal properties of the spin-density matrix in QCD.} In Sec.~\ref{sec:CollinsPA}, we revisited spin entanglement in coherent diffraction~\cite{Fucilla:2025kit,Hatta:2025obw}. We showed that color coherence leads to identical normalized spin-density matrices for transverse-photon- and gluon-initiated production in the dilute regime, while saturation effects break this equivalence. We then traced the peculiar diffractive entanglement pattern to the constraints imposed by exclusivity and the spin structure of eikonal Pomeron exchange. This interpretation motivates investigating whether analogous constraints underlie other recurring features of recent leading-order calculations. A clear example is the maximally entangled \(q\bar q\) spin state produced by a longitudinally polarized photon. This result holds in both inclusive and exclusive production, within both high-energy and collinear descriptions~\cite{Qi:2025onf,Fucilla:2025kit,Hatta:2025obw,Fucilla:2026mkg}.

\section*{Acknowledgments}
Many of the ideas developed in this paper grew out of discussions with Yoshitaka Hatta during our collaborations on Refs.~\cite{Fucilla:2025kit,Fucilla:2026mkg,Fucilla:2026zhy}. I am grateful to him for these discussions, for carefully reading the manuscript, and for several constructive comments on the content. I also thank Yuxun Guo, Bo-wen Xiao, Tolga Altinoluk, Guillaume Beuf, Florian Cougoulic, Edmond Iancu, Cyrille Marquet, Alessandro Papa, Lech Szymanowski, and Samuel Wallon for discussions.

M.~F.\ acknowledges support from the Polish National Agency for Academic Exchange (NAWA) through the Ulam Fellowship Programme under grant No.~BNI/ULM/2024/1/00065, ``Color glass condensate effective theory beyond the eikonal approximation''.

\appendix

\section{Correlation limit and gluon TMDs}
\label{CorrLim&TMDs}

\subsection{Derivation of eq.~(\ref{Eq:GijFinal_MainTex})}

Here we give some useful details on the derivation of eq.~(\ref{Eq:GijFinal_MainTex}). To this end, it is necessary to use the Fierz identity
\begin{equation}
    t^a_{ij}t^a_{kl}
    =
    \frac{1}{2}
    \left(
    \delta_{il}\delta_{jk}
    -
    \frac{1}{N_c}\delta_{ij}\delta_{kl}
    \right),
\label{Eq:FierzIdentity}
\end{equation}
and the relation
\begin{equation}
    \mathcal{U}_A^{ba} ({\bm x}) t^b =  \mathcal{U}_F ({\bm x}) t^a \mathcal{U}_F^{\dagger} ({\bm x}) \; .
\label{Eq:AdgToFund}
\end{equation}
Applying eq.~(\ref{Eq:FierzIdentity}) to the four-, three- and two-point
correlators, we can write
\begin{align}
S^{(4)}_{q\bar q\bar q q}
(\boldsymbol{x}_1,\boldsymbol{x}_2,
 \boldsymbol{x}_{1'},\boldsymbol{x}_{2'})
 =
\frac{N_c}{2C_F}
\left\langle
D(\boldsymbol{x}_1,\boldsymbol{x}_{1'})
D(\boldsymbol{x}_{2'},\boldsymbol{x}_2)
-
\frac{1}{N_c^2}
Q(\boldsymbol{x}_1,\boldsymbol{x}_2,
  \boldsymbol{x}_{2'},\boldsymbol{x}_{1'})
\right\rangle ,
\label{Eq:S4DipoleQuadrupole}
\end{align}
\begin{align}
S^{(3)}_{qg\bar q}
(\boldsymbol{v},\boldsymbol{x}_{1'},\boldsymbol{x}_{2'}) =
\frac{N_c}{2C_F}
\left\langle
D(\boldsymbol{v},\boldsymbol{x}_{1'})
D(\boldsymbol{x}_{2'},\boldsymbol{v})
-
\frac{1}{N_c^2}
D(\boldsymbol{x}_{2'},\boldsymbol{x}_{1'})
\right\rangle ,
\label{Eq:S3FirstDipole}
\end{align}
\begin{align}
S^{(3)}_{qg\bar q}
(\boldsymbol{v}',\boldsymbol{x}_2,\boldsymbol{x}_1)  =
\frac{N_c}{2C_F}
\left\langle
D(\boldsymbol{x}_1,\boldsymbol{v}')
D(\boldsymbol{v}',\boldsymbol{x}_2)
-
\frac{1}{N_c^2}
D(\boldsymbol{x}_1,\boldsymbol{x}_2)
\right\rangle ,
\label{Eq:S3SecondDipole}
\end{align}
\begin{equation}
    S^{(2)}_{gg}(\boldsymbol{v},\boldsymbol{v}')
    =
    \frac{N_c}{2C_F}
    \left\langle
    D(\boldsymbol{v},\boldsymbol{v}')
    D(\boldsymbol{v}',\boldsymbol{v})
    -
    \frac{1}{N_c^2}
    \right\rangle ,
\label{Eq:S2AdjointDipole}
\end{equation}
where we defined the dipole, eq.~(\ref{Eq:DipoleDefinition}), and quadrupole, 
\begin{align}
    Q(\boldsymbol{x},\boldsymbol{y},
      \boldsymbol{z},\boldsymbol{w})
    &=
    \frac{1}{N_c}
    {\rm Tr}_F\left[
    \mathcal{U}_F(\boldsymbol{x})
    \mathcal{U}_F^\dagger(\boldsymbol{y})
    \mathcal{U}_F(\boldsymbol{z})
    \mathcal{U}_F^\dagger(\boldsymbol{w})
    \right] \; ,
    \label{Eq:QuadrupoleDefinition}
\end{align}
operators. Combining Eqs.~(\ref{Eq:S4DipoleQuadrupole})--%
(\ref{Eq:S2AdjointDipole}), the exact quantity
$S(\boldsymbol{v},\boldsymbol{v}',\boldsymbol{u},\boldsymbol{u}')$
can be written as
\begin{gather}
S={}
\frac{N_c}{2C_F}
\Bigg\langle
D(\boldsymbol{x}_1,\boldsymbol{x}_{1'})
D(\boldsymbol{x}_{2'},\boldsymbol{x}_2)
+
D(\boldsymbol{v},\boldsymbol{v}')
D(\boldsymbol{v}',\boldsymbol{v})
-
D(\boldsymbol{x}_1,\boldsymbol{v}')
D(\boldsymbol{v}',\boldsymbol{x}_2)
-
D(\boldsymbol{v},\boldsymbol{x}_{1'})
D(\boldsymbol{x}_{2'},\boldsymbol{v})
\Bigg\rangle
\nonumber\\
-
\frac{1}{2C_FN_c}
\Bigg\langle
Q(\boldsymbol{x}_1,\boldsymbol{x}_2,
  \boldsymbol{x}_{2'},\boldsymbol{x}_{1'})
+1
-D(\boldsymbol{x}_1,\boldsymbol{x}_2)
-D(\boldsymbol{x}_{2'},\boldsymbol{x}_{1'})
\Bigg\rangle .
\label{Eq:SExactDipoleQuadrupole}
\end{gather}
As explained in the main text, the function $S$ vanishes when either $\boldsymbol{u}$ or
$\boldsymbol{u}'$ vanishes, its leading contribution is
\begin{equation}
    S
    =
    u^i u^{\prime j}
    \left.
    \frac{\partial}{\partial u^i}
    \frac{\partial}{\partial u^{\prime j}}
    S
    \right|_{\boldsymbol{u}=\boldsymbol{u}'=0}
    +
    \mathcal{O}(u^2u',uu'^2) = u^i u^{\prime j}
    \mathcal{G}^{ij}
    +
    \mathcal{O}(u^2u',uu'^2).
\label{Eq:MixedTaylorExpansion}
\end{equation}
The three-point functions depend either on $\boldsymbol{u}$ or on
$\boldsymbol{u}'$, but never on both. The two-point function is
independent of both variables. Consequently, the mixed derivative can
only act on the four-point functions:
\begin{align}
\mathcal{G}_{ij}(\boldsymbol{v},\boldsymbol{v}')
={}&
\frac{N_c}{2C_F}
\frac{\partial}{\partial u^i}
    \frac{\partial}{\partial u^{\prime j}} \left.
\left\langle
D(\boldsymbol{x}_1,\boldsymbol{x}_{1'})
D(\boldsymbol{x}_{2'},\boldsymbol{x}_2)
-
\frac{1}{N_c^2}
Q(\boldsymbol{x}_1,\boldsymbol{x}_2,
  \boldsymbol{x}_{2'},\boldsymbol{x}_{1'})
\right\rangle
\right|_{\substack{
\boldsymbol{u} = 0  \\
\boldsymbol{u}'=0}} .
\label{Eq:GijCompactDerivative}
\end{align}
A straightforward calculation gives
\begin{gather}
\left.
\partial_{u^i}\partial_{u^{\prime j}}
D(\boldsymbol{x}_1,\boldsymbol{x}_{1'})
D(\boldsymbol{x}_{2'},\boldsymbol{x}_2)
\right|_{\boldsymbol{u}=\boldsymbol{u}'=0}
=
\bar z^2
\big[
\partial_{v^i}\partial_{v^{\prime j}}
D(\boldsymbol{v},\boldsymbol{v}')
\big]
D(\boldsymbol{v}',\boldsymbol{v})
 \nonumber \\ \hspace{1.1 cm}
+
z^2
D(\boldsymbol{v},\boldsymbol{v}')
\big[
\partial_{v^i}\partial_{v^{\prime j}}
D(\boldsymbol{v}',\boldsymbol{v})
\big]
-
2z\bar z\,
{\rm Re}
\left\{
\big[
\partial_{v^i}D(\boldsymbol{v},\boldsymbol{v}')
\big]
\big[
\partial_{v^{\prime j}}
D(\boldsymbol{v}',\boldsymbol{v})
\big]
\right\} \; ,
\label{Eq:DoubleDipoleExpansionReal}
\end{gather}
and 
\begin{align}
&\left.
\partial_{u^i}\partial_{u^{\prime j}}
Q(\boldsymbol{x}_1,\boldsymbol{x}_2,
  \boldsymbol{x}_{2'},\boldsymbol{x}_{1'})
\right|_{\boldsymbol{u}=\boldsymbol{u}'=0}
 =
-\frac{1}{N_c}
{\rm Tr}_F\left[
\big(\partial_i\mathcal{U}_F(\boldsymbol{v})\big)
\mathcal{U}_F^\dagger(\boldsymbol{v})
\big(\partial_j\mathcal{U}_F(\boldsymbol{v}')\big)
\mathcal{U}_F^\dagger(\boldsymbol{v}')
\right].
\label{Eq:QuadrupoleExpansion}
\end{align}
\noindent Combining Eqs.~(\ref{Eq:DoubleDipoleExpansionReal}) and
(\ref{Eq:QuadrupoleExpansion}), we obtain eq.~(\ref{Eq:GijFinal_MainTex}).

\subsection{Gluon TMDs: operator definitions and the MV-model description}
\label{Sec:TMDs&MV_model_description}

In this subsection, we collect the operator definitions of the gluon TMDs entering Eqs.~(\ref{Eq:UnpolarDistr}) and (\ref{Eq:LLpolarDistr}), and summarize their analytical expressions in the MV model. The three transverse tensors relevant to the gluon channel admit the following decompositions~\cite{Marquet:2017xwy}\footnote{Taking into account the different conventions for the Wilson line of
Ref.~\cite{Marquet:2017xwy}.}
\begin{gather}
\tilde{\mathcal{G}}_{gg}^{(1)ij}(\boldsymbol{q})
=  
\frac{4}{g^2N_c}
\int\frac{d^2\boldsymbol{v}\,d^2\boldsymbol{v}'}{(2\pi)^3}
e^{-i\boldsymbol{q}\cdot(\boldsymbol{v}-\boldsymbol{v}')}
\left\langle
{\rm Tr}_F\!\left[
 (\partial_i\mathcal{U}_F(\boldsymbol{v}))
 (\partial_j\mathcal{U}_F^\dagger(\boldsymbol{v}'))
\right]
{\rm Tr}_F\!\left[
 \mathcal{U}_F(\boldsymbol{v}')
 \mathcal{U}_F^\dagger(\boldsymbol{v})
\right]
\right\rangle \nonumber \\
= 
\frac{\delta^{ij}}{2}F_{gg}^{(1)}(\boldsymbol{q})
+
\left(
 \frac{q^iq^j}{\boldsymbol{q}^{2}}
 -\frac{\delta^{ij}}{2} 
\right)
H_{gg}^{(1)}(\boldsymbol{q})  \; ,
\label{Eq:TMDOperatorDefinition1}
\end{gather}
\begin{gather}
\tilde{\mathcal{G}}_{gg}^{(2)ij}(\boldsymbol{q})
=
-\frac{4}{g^2N_c}
\int\frac{d^2\boldsymbol{v}\,d^2\boldsymbol{v}'}{(2\pi)^3}
e^{-i\boldsymbol{q}\cdot(\boldsymbol{v}-\boldsymbol{v}')}
{\rm Re}\left\langle
{\rm Tr}_F\!\left[
 (\partial_i\mathcal{U}_F(\boldsymbol{v}))
 \mathcal{U}_F^\dagger(\boldsymbol{v}')
\right]
{\rm Tr}_F\!\left[
 (\partial_j\mathcal{U}_F(\boldsymbol{v}'))
 \mathcal{U}_F^\dagger(\boldsymbol{v})
\right]
\right\rangle
\nonumber \\
= 
\frac{\delta^{ij}}{2}F_{gg}^{(2)}(\boldsymbol{q})
+
\left(
 \frac{q^iq^j}{\boldsymbol{q}^{2}}
 -\frac{\delta^{ij}}{2}
\right)
H_{gg}^{(2)}(\boldsymbol{q}) \; ,
\label{Eq:TMDOperatorDefinition2}
\end{gather}
\begin{gather}
\tilde{\mathcal{G}}_{gg}^{(3)ij}(\boldsymbol{q})
=
-\frac{4}{g^2}
\int\frac{d^2\boldsymbol{v}\,d^2\boldsymbol{v}'}{(2\pi)^3}
e^{-i\boldsymbol{q}\cdot(\boldsymbol{v}-\boldsymbol{v}')}
\left\langle
{\rm Tr}_F\!\left[
 (\partial_i\mathcal{U}_F(\boldsymbol{v}))
 \mathcal{U}_F^\dagger(\boldsymbol{v})
 (\partial_j\mathcal{U}_F(\boldsymbol{v}'))
 \mathcal{U}_F^\dagger(\boldsymbol{v}')
\right]
\right\rangle
\nonumber\\
=
\frac{\delta^{ij}}{2}F_{gg}^{(3)}(\boldsymbol{q})
+
\left(
 \frac{q^iq^j}{\boldsymbol{q}^{2}}
 -\frac{\delta^{ij}}{2}
\right)
H_{gg}^{(3)}(\boldsymbol{q}) .
\label{Eq:TMDOperatorDefinition3}
\end{gather}
The individual distributions are obtained from the trace and traceless
projections,
\begin{equation}
F_{gg}^{(n)}(\boldsymbol{q})
=
\delta_{ij}\,
\tilde{\mathcal{G}}_{gg}^{(n)ij}(\boldsymbol{q}),
\qquad
H_{gg}^{(n)}(\boldsymbol{q})
=
\left(
\frac{2q_iq_j}{\boldsymbol{q}^2}
-\delta_{ij}
\right)
\tilde{\mathcal{G}}_{gg}^{(n)ij}(\boldsymbol{q}),
\qquad n=1,2,3.
\label{Eq:GluonTMDProjectors}
\end{equation}
The distributions $F_{gg}^{(3)}$ and $H_{gg}^{(3)}$ are,
respectively, the unpolarized and linearly polarized
Weizs\"acker--Williams gluon distributions. \\

It is useful to verify explicitly that these definitions reproduce the
normalization adopted in the main text. Using
Eqs.~(\ref{Eq:TMDOperatorDefinition1})--
(\ref{Eq:TMDOperatorDefinition3}), one obtains
\begin{gather*}
    \tilde{\mathcal{G}}_{gg}^{ij}(z,\boldsymbol{q}) = \int \frac{d^2 {\bm v} \;  d^2 {\bm v}'}{(2 \pi)^4} \mathcal{G}^{ij}( {\bm v} , {\bm v}') e^{-i {\bm q} ( {\bm v} -{\bm v}')} \nonumber \\ = \frac{\alpha_s}{4 C_F} \left[ (z^2 + \bar z^2 ) \tilde{\mathcal{G}}_{gg}^{(1)ij}(\boldsymbol{q}) + 2 z \bar z \tilde{\mathcal{G}}_{gg}^{(2)ij}(\boldsymbol{q}) - \frac{1}{N_c^2} \tilde{\mathcal{G}}_{gg}^{(3)ij}(\boldsymbol{q}) \right] \; ,
\end{gather*}
from which eq.~(\ref{EqGtildeDecom}) immediately follows. \\   

For a heavy nucleus at moderately small $x$, the six gluon
distributions can be evaluated analytically in the MV model.  In this latter, the small-$x$ field is generated by a Gaussian distribution of static color sources. With this choice, the parametrizations derived in
Ref.~\cite{Marquet:2017xwy} read
\begin{align}
F_{gg}^{(1)}(\boldsymbol{q})
=
\mathcal{N}_{\rm MV}
\int_0^\infty\frac{dr}{r}\,
J_0(|\boldsymbol{q}|r)e^{-\frac{N_c}{2}\Gamma(r)}
\Bigg[
64\left(e^{\frac{N_c}{2}\Gamma(r)}-1\right)
-N_c^2Q_{sg}^4r^4K_r^2
+8(N_c^2-2)Q_{sg}^2r^2L_r
\Bigg],
\label{Eq:F1MV}
\end{align}

\begin{align}
H_{gg}^{(1)}(\boldsymbol{q})
=
\mathcal{N}_{\rm MV}
\int_0^\infty\frac{dr}{r}\,
J_2(|\boldsymbol{q}|r)e^{-\frac{N_c}{2}\Gamma(r)}
\Bigg[
\frac{64}{L_r}
\left(e^{\frac{N_c}{2}\Gamma(r)}-1\right)
+N_c^2Q_{sg}^4r^4K_r^2
+8(N_c^2-2)Q_{sg}^2r^2
\Bigg],
\label{Eq:H1MV}
\end{align}

\begin{align}
F_{gg}^{(2)}(\boldsymbol{q})
=
\mathcal{N}_{\rm MV}
\int_0^\infty\frac{dr}{r}\,
J_0(|\boldsymbol{q}|r)e^{-\frac{N_c}{2}\Gamma(r)}
\Bigg[
64\left(e^{\frac{N_c}{2}\Gamma(r)}-1\right)
+N_c^2Q_{sg}^4r^4K_r^2
-16Q_{sg}^2r^2L_r
\Bigg],
\label{Eq:F2MV}
\end{align}

\begin{align}
H_{gg}^{(2)}(\boldsymbol{q})
=
\mathcal{N}_{\rm MV}
\int_0^\infty\frac{dr}{r}\,
J_2(|\boldsymbol{q}|r)e^{-\frac{N_c}{2}\Gamma(r)}
\Bigg[
\frac{64}{L_r}
\left(e^{\frac{N_c}{2}\Gamma(r)}-1\right)
-N_c^2Q_{sg}^4r^4K_r^2
-16Q_{sg}^2r^2
\Bigg] \; ,
\label{Eq:H2MV}
\end{align}

\begin{equation}
F_{gg}^{(3)}(\boldsymbol{q})
=
32N_c^2\mathcal{N}_{\rm MV}
\int_0^\infty\frac{dr}{r}\,
J_0(|\boldsymbol{q}|r)
\left[
1-e^{-\frac{N_c}{2}\Gamma(r)}
\right],
\label{Eq:F3MV}
\end{equation}

\begin{equation}
H_{gg}^{(3)}(\boldsymbol{q})
=
32N_c^2\mathcal{N}_{\rm MV}
\int_0^\infty\frac{dr}{r}\,
\frac{J_2(|\boldsymbol{q}|r)}{L_r}
\left[
1-e^{-\frac{N_c}{2}\Gamma(r)}
\right].
\label{Eq:H3MV}
\end{equation}

The quantities entering the above expressions are
\begin{equation}
\Gamma(r)
=
\frac{r^2Q_{sg}^2}{2N_c}L_r,
\qquad
L_r
=
\ln\left(
 \frac{1}{r^2\Lambda^2}
\right),
\qquad
K_r
=
1-2\gamma_E+\ln4+L_r,
\label{Eq:MVCompactDefinitions}
\end{equation}
and
\begin{equation}
\mathcal{N}_{\rm MV}
=
\frac{S_\perp C_F}{\alpha_s\pi^3}
\frac{1}{32N_c^2}.
\label{Eq:MVNormalization}
\end{equation}
The common global normalization $\mathcal{N}_{\rm MV}$ can be omitted when computing the normalized
spin-correlation coefficients, since it cancels between numerator and
denominator. For the numerical illustrations, we set $Q_{sg}=Q_0$ and $\Lambda=Q_0/20$, and we shift the argument of the logarithm to $1 / r^2 \Lambda^2 + e $.\footnote{A consistent implementation of the shift in $L_r$ would require rederiving the six TMD expressions of Ref.~\cite{Marquet:2017xwy}, including the corresponding derivative terms. Since we seek an estimate rather than a precision prediction, we use the expressions above as a phenomenological benchmark.} We use the resulting infrared-regularized distributions as an MV-inspired phenomenological benchmark.

\section{Spin amplitudes in the light-front helicity basis and local spin rotations}
\label{HelicityBasis}
In this section, we give more details in support of the discussion of section \ref{sec:Coherent_Diffraction}.

\subsection{Spin amplitudes in the light-front helicity basis}

We begin by showing that, in the light-front helicity basis, the density matrix can be written as in eq.~(\ref{Eq:DensityAlaLightFront}). The impact factor in eq.~(\ref{Eq:IFmomRep}) has the following spinor structure
\begin{equation}
  \Phi^i_{\alpha \alpha'}(\boldsymbol{k})
 \propto
 \bar{u}_{\alpha} (p_q) \left[ (1- 2 z) k^i + i \epsilon^{ij} k^j \gamma^5 - m \gamma^i  \right] \gamma^+  v_{\alpha'} (p_{\bar{q}}) \; .
 \label{eq:transverse_numerator_helicity}
\end{equation}
We consider two circular polarizations~\cite{Lappi:2020ufv}
\begin{gather}
    \varepsilon_{+} = \frac{(-1,-i)}{\sqrt{2}} \; , \qquad \qquad \varepsilon_{-} = \frac{(1,-i)}{\sqrt{2}} ,
\end{gather}
and define $k_{\pm} \equiv k_x \pm i k_y $\footnote{In this section we use the subscript $\pm$ to denotes this specific combination of transverse components. They should not be confused with the superscript $\pm$ denoting the longitudinal light-cone components.}. In the light-front helicity basis, we can write
\begin{gather}
    \Phi^x_{\perp \alpha \alpha'}(\boldsymbol{k}) = 2 p_g^+ \sqrt{z \bar z} [ (1-2z) k_x \sigma^1_{\alpha \alpha'} + k_y \sigma^2_{\alpha \alpha'} - m \sigma^3_{\alpha \alpha'} ] \nonumber \\ = 2 p_g^+ \sqrt{z \bar z} 
\begin{pmatrix}
-m & (1-2z)k_x - i k_y \\ 
(1-2z)k_x + i k_y & m
\end{pmatrix} \; , 
\end{gather}
\begin{gather}
 \Phi^y_{\perp \alpha \alpha'}(\boldsymbol{k}) = 2 p_g^+ \sqrt{z \bar z} [ (1-2z) k_y \sigma^1_{\alpha \alpha'} - k_x \sigma^2_{\alpha \alpha'} + i m 1_{\alpha \alpha'} ] \nonumber \\ = 2 p_g^+ \sqrt{z \bar z} 
        \begin{pmatrix}
im & (1-2z)k_y + i k_x \\
(1-2z)k_y - i k_x & im
\end{pmatrix} \; .
\end{gather}
Therefore
\begin{gather}
    \Phi_{\perp \alpha \alpha' , +}(\boldsymbol{k}) \equiv \Phi^i_{\alpha \alpha'}(\boldsymbol{k})  \varepsilon_{+}^i = - \frac{\Phi^x_{\perp \alpha \alpha'}(\boldsymbol{k}) + i \Phi^y_{\perp \alpha \alpha'}(\boldsymbol{k}) }{ \sqrt{2} } = \sqrt{2 z \bar z  (p_g^+)^2 }  
    \begin{pmatrix}
    0 & z k_{+} \\
    - \bar z k_{+} & 0
   \end{pmatrix} 
   \label{Eq:PlusCom}
\end{gather}
\begin{gather}
    \Phi_{\perp \alpha \alpha' , -}(\boldsymbol{k}) \equiv \Phi^i_{\alpha \alpha'}(\boldsymbol{k})  \varepsilon_{-}^i = \frac{\Phi^x_{\perp \alpha \alpha'}(\boldsymbol{k}) - i \Phi^y_{\perp \alpha \alpha'}(\boldsymbol{k}) }{ \sqrt{2} } =  \sqrt{2 z \bar z  (p_g^+)^2 }  
    \begin{pmatrix}
    0 & \bar z k_{-} \\
    - z k_{-} & 0
   \end{pmatrix} 
   \label{Eq:MinusCom}
\end{gather}
In the basis $\{ | \uparrow \uparrow \rangle, | \downarrow \downarrow \rangle, | \uparrow \downarrow \rangle, | \downarrow \uparrow \rangle \} $, we can represent the two components of the impact factor as
\begin{gather}
    | \tilde{\Psi}_+ \rangle = m  | \uparrow \uparrow \rangle +   z k_+ | \uparrow \downarrow \rangle - \bar z k_+ | \downarrow \uparrow \rangle  \; ,  
\end{gather}
\begin{gather}
    | \tilde{\Psi}_- \rangle = m  | \downarrow \downarrow \rangle +  \bar z k_- | \uparrow \downarrow \rangle -  z k_- | \downarrow \uparrow \rangle  \; .
\end{gather}
Following the same reasoning as Ref.~\cite{Fucilla:2025kit}, it is easy to see that performing the convolution with target amounts to the replacements
\begin{gather}
    m  \longrightarrow m T_1 (\boldsymbol{P}) \; , \qquad \qquad  k_{\pm } \longrightarrow |\boldsymbol{P}| (T_1(\boldsymbol{P})+T_2(\boldsymbol{P})) \; ,
\end{gather}
giving the two states $| \Psi_+ \rangle$ and $| \Psi_- \rangle$ in eq.~(\ref{Eq:StatesInLightFront}).

\subsection{Spin-rotated density matrix}
We now work out the rotations giving $\rho_T^{\rm rot}$. We introduce the normalized states
\begin{gather}
   | \chi_{\pm} \rangle = \frac{| \Psi_+ \rangle \pm | \Psi_- \rangle }{\sqrt{2 (N \pm O)}} \; ,
\end{gather}
where $N$ and $O$ are defined in Eqs.~(\ref{Eq:NormalizationDensPsi}) and~(\ref{Eq:OverlapDensPsi}), respectively. The inverse transformations read
\begin{equation}
    | \Psi_+ \rangle = \sqrt{\frac{N+O}{2}} | \chi_+ \rangle + \sqrt{\frac{N-O}{2}} | \chi_- \rangle \; ,
\end{equation}
\begin{equation}
    | \Psi_- \rangle = \sqrt{\frac{N-O}{2}} | \chi_- \rangle - \sqrt{\frac{N+O}{2}} | \chi_+ \rangle \; .
\end{equation}
We can therefore write the density matrix $\rho_T$ in terms of these states as
\begin{equation}
    \rho_T = \frac{N+O}{2N} | \chi_{+} \rangle \langle \chi_{+} | + \frac{N-O}{2N} | \chi^{-} \rangle \langle \chi^{-} | = \frac{1+t}{2} | \chi_{+} \rangle \langle \chi_{+} | + \frac{1-t}{2} | \chi_{-} \rangle \langle \chi_{-} | \; ,
\end{equation}
where $t=O/N$ is explicitly given in eq.~(\ref{Eq:tDef}). By using the form of $| \Psi_+ \rangle $ and $| \Psi_- \rangle$ in the basis $\{ | \uparrow \uparrow \rangle, | \downarrow \downarrow \rangle, | \uparrow \downarrow \rangle, | \downarrow \uparrow \rangle \} $, eq.~(\ref{Eq:StatesInLightFront}), we can write
\begin{equation}
    | \chi_+ \rangle = \frac{1}{\sqrt{N+O}} \bigg( m^2 T_1 ^2 (\boldsymbol{P}) | \Phi^+ \rangle + \boldsymbol{P}^2 (T_1 (\boldsymbol{P}) + T_2 (\boldsymbol{P}) )^2 | \psi^- \rangle \bigg) \; ,
\end{equation}
\begin{equation}
    | \chi_- \rangle = \frac{1}{\sqrt{N-O}} \bigg( m^2 T_1^2 (\boldsymbol{P}) | \Phi^- \rangle - (1-2 z) \boldsymbol{P}^2 (T_1 (\boldsymbol{P}) + T_2 (\boldsymbol{P}) )^2 | \psi^+ \rangle \bigg) \; ,
\end{equation}
where we introduced the two Bell states
\begin{equation}
    | \psi^{\pm} \rangle = \frac{| \uparrow \downarrow \rangle \pm  | \downarrow \uparrow \rangle }{\sqrt{2}} \; .
\end{equation}
Let us observe that, introducing three parameters 
\begin{equation}
a=|\boldsymbol P|(T_1+T_2),\qquad
b=mT_1,\qquad
c= (z-\bar z) |\boldsymbol P|(T_1+T_2) ,
\end{equation}
we can write
\begin{gather}
     |\chi_+\rangle
=
\frac{b|\Phi^+\rangle+a|\psi^-\rangle}
     {\sqrt{b^2+a^2}},
\qquad
|\chi_-\rangle
=
\frac{b|\Phi^-\rangle+c|\psi^+\rangle}
     {\sqrt{b^2+c^2}}.
\end{gather}
The $|\chi_{\pm} \rangle$ are maximally entangled states that can be reached from the Bell states $|\Phi^{\pm} \rangle$ by local spin rotations~\cite{Lang:2010xtw} that preserve the concurrence, i.e.
\begin{equation}
|\chi_\pm\rangle
=(U_q (\alpha) \otimes U_{\bar q} (\beta) )|\Phi^\pm\rangle \; .
\label{Eq:RotExplicit}
\end{equation}
where
\begin{equation}
    U (\theta) = \begin{pmatrix}
\cos{\theta} & -\sin{\theta} \\ 
\sin{\theta} & \cos{\theta} 
\end{pmatrix} \; .
\end{equation}
Indeed, by applying the rotation explicitly
\begin{equation}
    (U_q (\alpha) \otimes U_{\bar q} (\beta) )|\Phi^\pm\rangle = \cos{(\alpha \mp \beta)} |\Phi^\pm\rangle + \sin{(\beta \mp \alpha)} |\psi^\mp\rangle
\end{equation}
and choosing
\begin{gather}
    \cos{(\alpha - \beta)} = \frac{b}{ \sqrt{b^2 + a^2}} \; , \qquad \qquad  \cos{(\alpha + \beta)} = \frac{b}{ \sqrt{b^2 + c^2}} \; ,
\end{gather}
\begin{gather}
    \sin{(\beta - \alpha)} = \frac{a}{ \sqrt{b^2 + a^2}} \; , \qquad \qquad  \sin{(\beta + \alpha)} = \frac{c}{ \sqrt{b^2 + c^2}} \; ,
\end{gather}
(\ref{Eq:RotExplicit}) immediately follows.
Consequently, in the rotated local spin bases, one has
\begin{equation}
\rho_T^{\rm rot}
=(U_q^\dagger\otimes U_{\bar q}^\dagger)
 \rho_T (U_q\otimes U_{\bar q}) =
\frac{1+t}{2} |\Phi^+\rangle\langle\Phi^+|
+ \frac{1-t}{2} |\Phi^-\rangle\langle\Phi^-|.
\end{equation}

\bibliographystyle{apsrev}
\bibliography{ref}

\end{document}